\documentclass[a4paper]{aa}
\usepackage[varg]{txfonts}
\usepackage{graphicx}
\usepackage{psfrag}
\usepackage{hyperref}
\usepackage{gensymb}
\usepackage{xcolor}
\usepackage{tabularx}
\usepackage{lipsum}
\usepackage{amsmath}

\usepackage{siunitx} 

\begin{document}

\bibpunct{(}{)}{;}{a}{}{,} 

\title{Optimal strategy for polarization modulation in the LSPE-SWIPE experiment} 
\subtitle{}
\author{A. Buzzelli\inst{1,4}
\and P. de Bernardis\inst{1,3}
\and S. Masi\inst{1,3}
\and N. Vittorio\inst{2,4}
\and G. de Gasperis\inst{2,4}}
\institute{Dipartimento di Fisica, Sapienza Universit\`a di Roma, piazzale Aldo Moro 5, I-00185, Roma, Italy \label{inst1}
 \and Dipartimento di Fisica, Universit\`a di Roma ``Tor~Vergata'', via della Ricerca Scientifica 1, I-00133, Roma, Italy \label{inst2}
 \and Sezione INFN Roma~1, piazzale Aldo Moro 5, I-00185, Roma, Italy \label{inst3}
 \and Sezione INFN Roma~2, via della Ricerca Scientifica 1, I-00133, Roma, Italy \label{inst4}}
\offprints{alessandro.buzzelli@roma2.infn.it}
\date{Received / Accepted}

\abstract
{Cosmic microwave background (CMB) B-mode experiments are required to control systematic effects with an unprecedented level of accuracy. Polarization modulation by a half wave plate (HWP) is a powerful technique able to mitigate a large number of the instrumental systematics.}
{Our goal is to optimize the polarization modulation strategy of the upcoming LSPE-SWIPE balloon-borne experiment, devoted to the accurate measurement of CMB polarization at large angular scales.}
{We depart from the nominal LSPE-SWIPE modulation strategy (HWP stepped every $60~\mathrm{s}$ with a telescope scanning at around $12~\mathrm{deg/s}$) and perform a thorough investigation of a wide range of possible HWP schemes (either in stepped or continuously spinning mode and at different azimuth telescope scan-speeds) in the frequency, map and angular power spectrum domain. In addition, we probe the effect of high-pass and band-pass filters of the data stream and explore the HWP response in the minimal case of one detector for one operation day (critical for the single-detector calibration process). We finally test the modulation performance against typical HWP-induced systematics.} 
{Our analysis shows that some stepped HWP schemes, either slowly rotating or combined with slow telescope modulations, represent poor choices. Moreover, our results point out that the nominal configuration may not be the most convenient choice. While a large class of spinning designs provides comparable results in terms of pixel angle coverage, map-making residuals and BB power spectrum standard deviations with respect to the nominal strategy, we find that some specific configurations (e.g., a rapidly spinning HWP with a slow gondola modulation) allow a more efficient polarization recovery in more general real-case situations.}
{Although our simulations are specific to the LSPE-SWIPE mission, the general outcomes of our analysis can be easily generalized to other CMB polarization experiments.}
\authorrunning{}

\titlerunning{Optimal strategy for polarization modulation in the LSPE-SWIPE experiment}


\maketitle

\section{Introduction}\label{intro}

Measurements of cosmic microwave background (CMB) temperature and polarization anisotropy allowed the establishment of a cosmological concordance scenario, the so-called $\Lambda CDM$ model, with very tight constraints on the parameters (see, e.g., Boomerang:~\citealt{2006ApJ...647..799M}; WMAP:~\citealt{2013ApJS..208...19H}; Planck:~\citealt{2015arXiv150201589P}).

The polarization field can be decomposed into a curl-free component, E-modes, and a curl component, B-modes \citep{1997PhRvL..78.2058K}. While E-modes have been widely detected (see, e.g., {DASI:~\citealt{2002Natur.420..772K}}; {WMAP:~\citealt{2003ApJS..148..175S}; Boomerang:~\citealt{2006ApJ...647..813M};
Planck:~\citealt{2015arXiv150702704P}), primordial B-modes are still hidden into foreground and noise contamination \citep[{see}][]{2015PhRvL.114j1301B}. 

The importance of a B-mode observation is twofold: on low multipoles, $\ell \lesssim 10$, a detection of the ``reionization bump'' in the BB angular power spectrum would allow to constrain some crucial aspects of the reionization epoch, which eventually moves part of the E-mode signal into B-modes at very large scales; on higher multipoles, $\ell \sim 80$, a measurement of the ``recombination bump'', i.e. the imprint of the tensor mode of primordial perturbations, would give a convincing confirmation to inflationary models \citep{1999PhR...314....1L}.

Beside primordial B-modes, gravitational lensing generated by growing matter inhomogeneities between us and the last scattering surface gives rise to a leakage from E to B modes at small scales \citep{1998PhRvD..58b3003Z}. Measurements of lensed B-modes have been recently claimed: 
see e.g., \citet{2014PhRvL.113b1301A, 2015arXiv151202882P, 2016arXiv160601968K}.

Several experiments have been designed or planned to detect primordial B-mode polarization: e.g., POLARBEAR \citep{2010SPIE.7741E..1EA}, SPIDER \citep{2010SPIE.7741E..1NF}, QUBIC \citep{2011APh....34..705Q}, COrE \citep{2011arXiv1102.2181T}, LSPE \citep{2012SPIE.8446E..7AA}, LiteBIRD \citep{2014JLTP..176..733M}, BICEP3 \citep{2016JLTP..184..765W}.  

Here, we focus on the SWIPE balloon-borne instrument \citep{2012SPIE.8452E..3FD}, that is part of the LSPE mission \citep{2012SPIE.8446E..7AA}, devoted to the accurate observation of CMB polarization at large angular scales. 

One of the most critical issues to face in the context of B-mode observation is the control and possibly removal of instrumental systematics, that are likely to severely degrade the performance of any B-mode experiment by introducing spurious contributions in general larger than the primordial signal (see, e.g., \citealt{2007MNRAS.376.1767O} and \citealt{2003PhRvD..67d3004H}). 

Modulating the incoming linear polarization by a half wave plate (HWP) is a powerful and widely employed technique to mitigate a large number of the intrumental systematics (see, e.g., \citealt{2004SPIE.5543..320O}, \citealt{2007ApJ...665...42J}, \citealt{2010SPIE.7741E..2BB}, \citealt{2015JLTP..tmp..100S}, \citealt{2016arXiv160707399H}). 

In particular, a HWP allows to (see, e.g., \citealt{2009MNRAS.397..634B, 2008ApJ...689..655M}): 
i) effectively mitigate calibration, beam and other instrumental systematics, as a HWP enables to perform the observation without differencing power from distinct orthogonal polarization-sensitive detectors and with no need to rotate the whole instrument; 
ii) reject the $1/f$ noise at the hardware level, as the polarization signal is shifted to higher frequencies ; 
iii) achieve a better angle coverage uniformity, since each pixel is observed over a wide range of orientations of the analyzer.

On the other hand, the presence of a HWP may introduce a large class of systematic effects of its own: mis-estimation of the HWP angle, differential transmittance of the two orthogonal states, leakage from temperature and E-modes to B-modes due to imperfect optical setup, etc. (see references above).

In this work, we depart from the nominal LSPE-SWIPE modulation strategy (HWP stepped every $60~\mathrm{s}$ with a telescope scanning at around $12~\mathrm{deg/s}$) and explore a wide range of possible HWP schemes, either in stepped or continuously spinning mode and at different azimuth telescope scan-speeds. See also \citealt{Buzzelli:2017mjw} for a preliminary discussion.

We investigate the HWP rotation designs in the frequency, map and angular power spectrum domains. In addition, we probe the effect of high-pass filtering, common practice to reject the $1/f$ noise, and band-pass filtering, which represent a more interesting possibility allowed by a spinning HWP scheme. 
We finally study the minimal observation case (one detector for one day of operation), which is an important test for the single-detector calibration process, and analyze the impact of typical HWP systematic effects of its own.

We adopt a robust noise model which, in addition to white noise, includes low-frequency $1/f$ noise, both self-correlated and cross-correlated among the different polarimeters.

The paper is organized as follows: in Section \ref{Sims} we introduce the scientific case of the LSPE-SWIPE experiment and outline the steps of the simulation pipeline followed in this work; in Section \ref{scan} we present and discuss our results; finally, we draw our conclusions in Section \ref{Conclusions}.

\section{Simulations and methodology}\label{Sims}

\subsection{The LSPE-SWIPE experiment}

LSPE is a next-generation CMB experiment \citep{2012SPIE.8446E..7AA}, aimed at detecting CMB polarization at large angular scales with the 
primary goal of constraining both the B-mode signal due to reionization (reionization bump) at very low multipoles and the imprint 
of inflationary tensor perturbations (recombination bump) at higher multipoles.

LSPE will survey the northern sky (the effective sky fraction will be around 30\%) with a coarse angular resolution of about 1.5 degrees FWHM.

The mission will consist of two instruments: a balloon-based array of bolometric polarimeters, the Short Wavelength Instrument for the 
Polarization Explorer (SWIPE, \citealt{2012SPIE.8452E..3FD}), that will observe the sky in three frequency bands centered at 
$140~\mathrm{GHz}$, $220~\mathrm{GHz}$ and $240~\mathrm{GHz}$, and a ground-based array of coherent polarimeters, the Survey TeneRIfe Polarimeter (STRIP, \citealt{2012SPIE.8446E..7CB}), that will scan the same region in two frequency bands centered at $43~\mathrm{GHz}$ and $90~\mathrm{GHz}$. This paper is specifically addressed to the SWIPE bolometric instrument. 

The SWIPE $140~\mathrm{GHz}$ band will be the main CMB channel, while measurements at $220-240~\mathrm{GHz}$ and at $43-90~\mathrm{GHz}$ will be devoted to monitor the thermal dust contamination and the synchrotron emission, respectively. 

The first optical element of SWIPE is a large ($50~\mathrm{cm}$ diameter) rotating HWP, followed by a $50~\mathrm{cm}$ diameter lens, focusing sky radiation on a large tilted (45$\degree$) grid polarizer, followed by two identical focal planes (transmitted and reflected by the polarizer, with orthogonal polarizations). Each focal plane accommodates 165 multimode bolometric detectors, for a total of 8800 radiation modes.

The nominal HWP setup consists of a step-and-integrate mode scanning the range $0-78.75~\mathrm{deg}$ with 11.25 steps every minute. 

SWIPE will be launched in the 2018-2019 winter from the Svalbard islands and will operate for around two weeks during the Arctic night (at latitude around 78$\degree$N), to take advantage of optimal observation conditions . 

SWIPE will scan the sky by spinning around the local vertical, keeping the telescope elevation constant for long periods, in the range $35~\mathrm{deg}$ to $55~\mathrm{deg}$. 
According to the default instrumental baseline, the azimuth telescope scan-speed will be set around $12~\mathrm{deg/s}$. The nominal sample rate is $100~\mathrm{Hz}$.

The SWIPE detectors are spiderweb multimode TES bolometers criogenically cooled down to $0.3~\mathrm{K}$ \citep{2016JLTP..tmp....3G}. All the optical components will operate at around $2~\mathrm{K}$ to reduce background thermal emission.

\subsection{Signal and noise simulations}

We generate angular power spectra from the publicly available CAMB software \citep{2002PhRvD..66j3511L} according to the latest Planck release of 
cosmological parameters \citep{2015arXiv150201589P} with a B-mode polarization signal corresponding to a tensor-to-scalar ratio $r=0.09$. 

In this paper, if not stated otherwise, we consider a subset of 18 detectors (arranged in three triples sparsely located in each of the two focal planes) for 5 days of observation, with the telescope elevation ranging from 35$\degree$ to 55$\degree$ in 5$\degree$ steps every 24 hours. 

Our flight simulator provides the pointing (right ascension and celestial declination) and the polarization angles, according to the nominal SWIPE
scanning strategy and assuming a late-December launch from Svalbard islands. 

Thus, we produce a sky map from the CAMB spectra by the use of the Synfast facility of the HEALPix package\footnote{http://healpix.sourceforge.net} (at HEALPix $N_{side}=1024$; see \citealt{2005ApJ...622..759G}), that we convert into Time Ordered Data (TOD) according to the SWIPE observation strategy. Any detector collects $8.64 \times 10^7$ samples per day.  

We assume the noise spectrum of the single detector to be the sum of a high-frequency white noise component (with amplitude $w$) and a low-frequency $(f_k/f)^\alpha$ component, where $f_k$ is the knee frequency and $\alpha$ the spectral index. 
Moreover, we include a low-frequency cross-correlated noise component shared by all the focal plane. Hence, our model is:
\begin{align}
P_{ii}(f) &= w \left[ 1 + \left(\frac{f_k}{f} \right)^\alpha \right], &
P_{ij}(f) &= w \left(\frac{f_k}{f} \right)^\alpha,
\end{align}
where $P_{ii}$ and $P_{ij}$ are the auto noise spectrum of the detector $i$ and the cross noise spectrum of the detectors $i$ and $j$, respectively. The knee frequency and the spectral index values are set to $0.1~\mathrm{Hz}$ and 2.0, respectively. 
Our noise model is largely based on the BOOMERanG polarization sensitive experiment \citep{2006A&A...458..687M}. See also \citet{2016A&A...596A.107P}.

In this paper we focus on the $140~\mathrm{GHz}$ band, the main SWIPE CMB channel. The assumed white noise amplitude (noise equivalent temperature, NET) for each molti-mode detector in this band is $15~\mathrm{\mu K s^{1/2}}$.

\subsection{Maps and angular power spectra estimates}\label{maps_aps_est}

The first step in CMB data analysis is the projection of the observational data into the sky, that means to build a sky-map. 

In this work we use the ROMA MPI-parallel code \citep{2005A&A...436.1159D}, an optimal map-making algorithm based on the iterative Generalised Least Squares (GLS) approach. The code is extended to allow for a possible cross-correlated noise component among the detectors \citep{2016A&A...593A..15D} introduced mainly by low-frequency atmospheric and temperature fluctuations affecting simultaneously the whole focal plane. 

In \citet{2008ApJ...681..708P}, \citet{2016JPhCS.689a2003B} and \citet{2016A&A...593A..15D} the authors show that accounting for common-mode noise results in more accurate sky maps and more faithful angular power spectra at low multipoles. 
The extended map-making algorithm is therefore expected to be particularly helpful in the context of large-scale B-mode observations. 

All the output maps are at HEALPix $N_{side}=128$, i.e. the pixel size is 27.4', and smoothed to $1.5~\mathrm{deg}$ FWHM. In Fig.~\ref{hits} we show the coverage map in time units (i.e. total integration time over each 27.4' pixel) for the SWIPE instrumental setup assumed here.
\begin{figure}
\begin{center}
 \includegraphics[angle=90,width=0.37\textwidth]{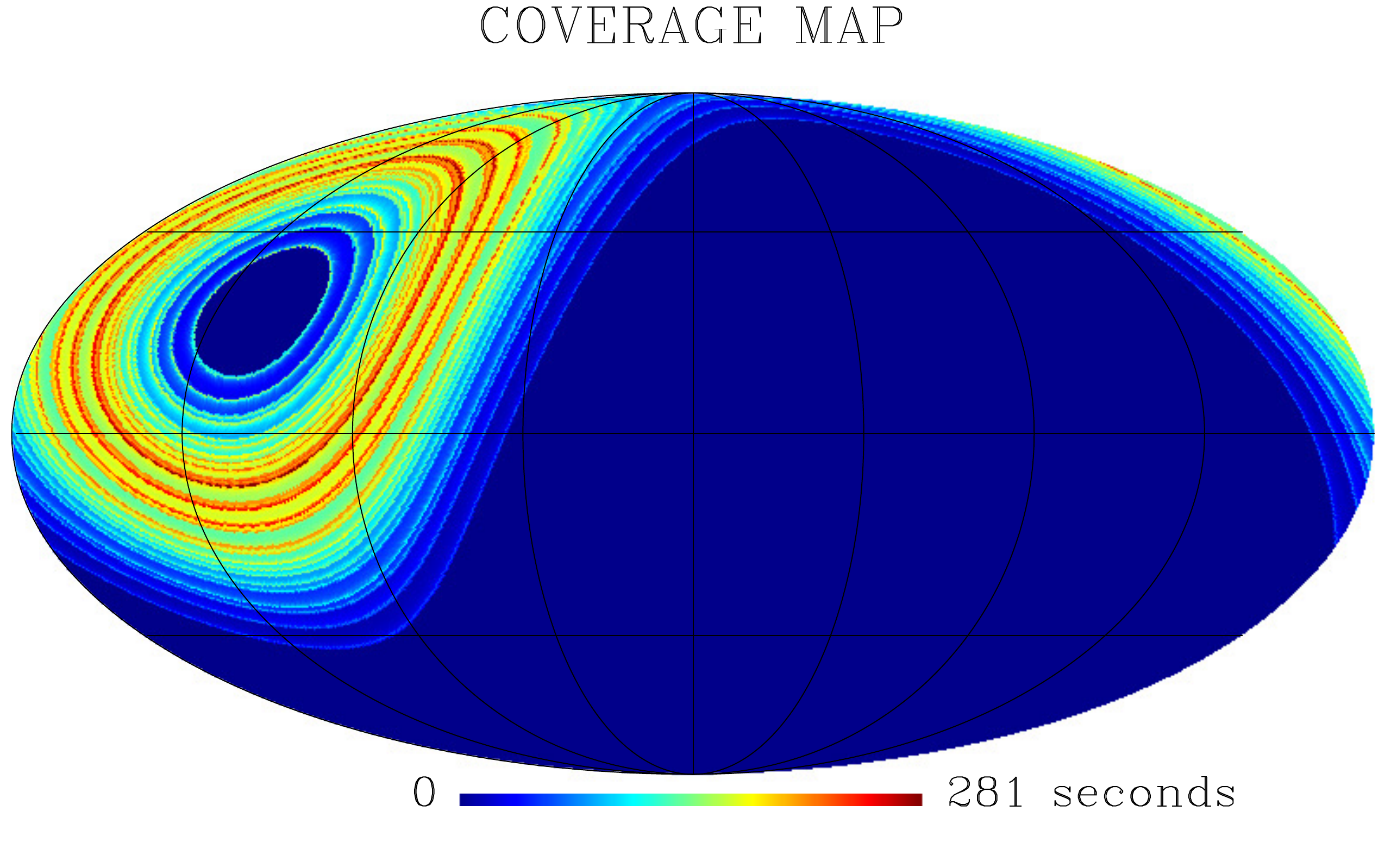}
 \caption{Coverage map in time units for the LSPE-SWIPE instrumental setup assumed in this work (18 detectors for 5 operation days, sampled at $100~\mathrm{Hz}$). Map is in Galactic coordinates and at resolution of HEALPix $N_{side}=128$.}\label{hits}
\end{center} 
\end{figure}

For the estimation of the angular power spectra, we follow the MASTER pseudo-$C_l$ estimator approach \citep{2002ApJ...567....2H}, that is a fast and accurate method that corrects for E/B mode and multipoles mixings due to the partial sky coverage. Nonetheless, as mentioned by \citet{2014MNRAS.440..957M}, at large angular scales the use of a quadratic maximum likelihood (QML) approach may be more convenient, even if much more demanding from a computational point of view. For the purposes of this work, we evaluated that the accuracy provided by the pseudo-$C_l$ algorithm is indeed sufficient.

For any of the setups assumed here, we use an indipendent sky mask calculated from the corresponding pixel inverse condition number map, taking the values $\gtrsim 10^{-2}$ (see Section \ref{invcond}). At this level, we did not use any mask apodisation. In addition, the power spectra are estimated according to a binning $\Delta \ell = 10$ (beside the first bin which is calculated from $\ell=2$), in such a way that the bin ranges are $2 < \ell < 9$, $10 < \ell < 19$, etc..

\section{Results}\label{scan}

In this Section we present and discuss our simulations aimed at optimizing the polarization modulation strategy of the SWIPE experiment. 
We test a variety of HWP configurations, either in stepped or spinning mode, exploring two different azimuth telescope scan-speeds. In detail, we investigate the following 12 schemes:

\begin{itemize}
 \item a HWP stepped every $1~\mathrm{s}$, $60~\mathrm{s}$ and $3600~\mathrm{s}$ with the telescope scanning at $12~\mathrm{deg/s}$ and $0.7~\mathrm{deg/s}$;
 \item a HWP spinning with mechanical frequency of $5~\mathrm{Hz}$, $2~\mathrm{Hz}$ and $0.5~\mathrm{Hz}$ with, again, the telescope scanning at $12~\mathrm{deg/s}$ and $0.7~\mathrm{deg/s}$.
\end{itemize}

The HWP setups listed above span the full range of the possible configurations that are currently considered experimentally feasible.
The maximum telescope rotation rate is given by the gondola pendulation, the detector noise and response time, while the maximum stepped or spinning HWP rotation rates are set by mechanical feasibility and preliminary tests on the heat generation, respectively.

\subsection{Periodograms}\label{periodograms}
\begin{figure*}
\psfrag{Frequency [Hz]}[c][][3]{Frequency $[Hz]$}
\psfrag{Power [uK^2/Hz]}[c][][3]{Power $[\mu K^{2}/Hz]$}
 \includegraphics[angle=90,width=0.33\textwidth]{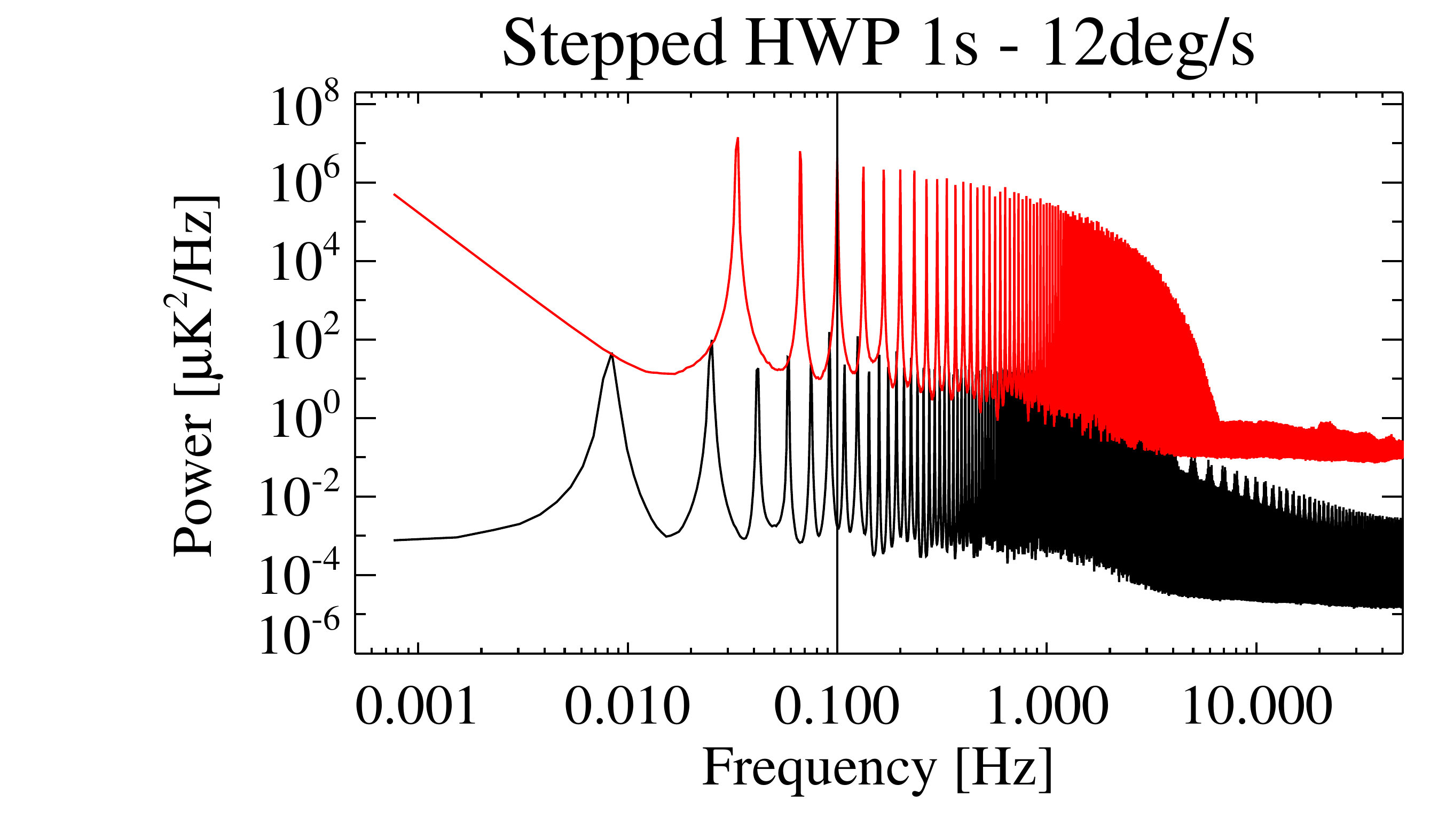} 
\hskip -1em  \includegraphics[angle=90,width=0.33\textwidth]{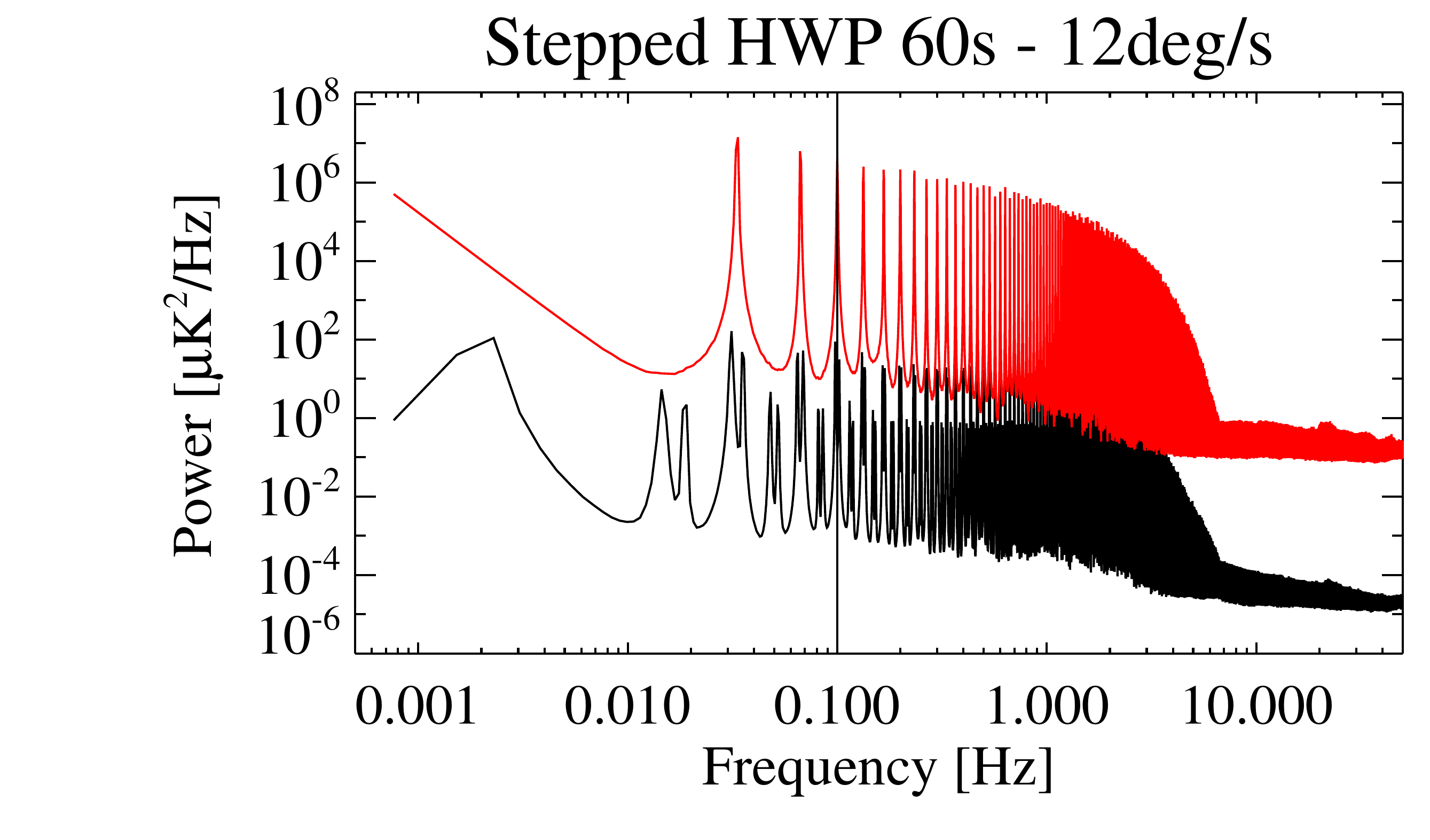}
\hskip -1em  \includegraphics[angle=90,width=0.33\textwidth]{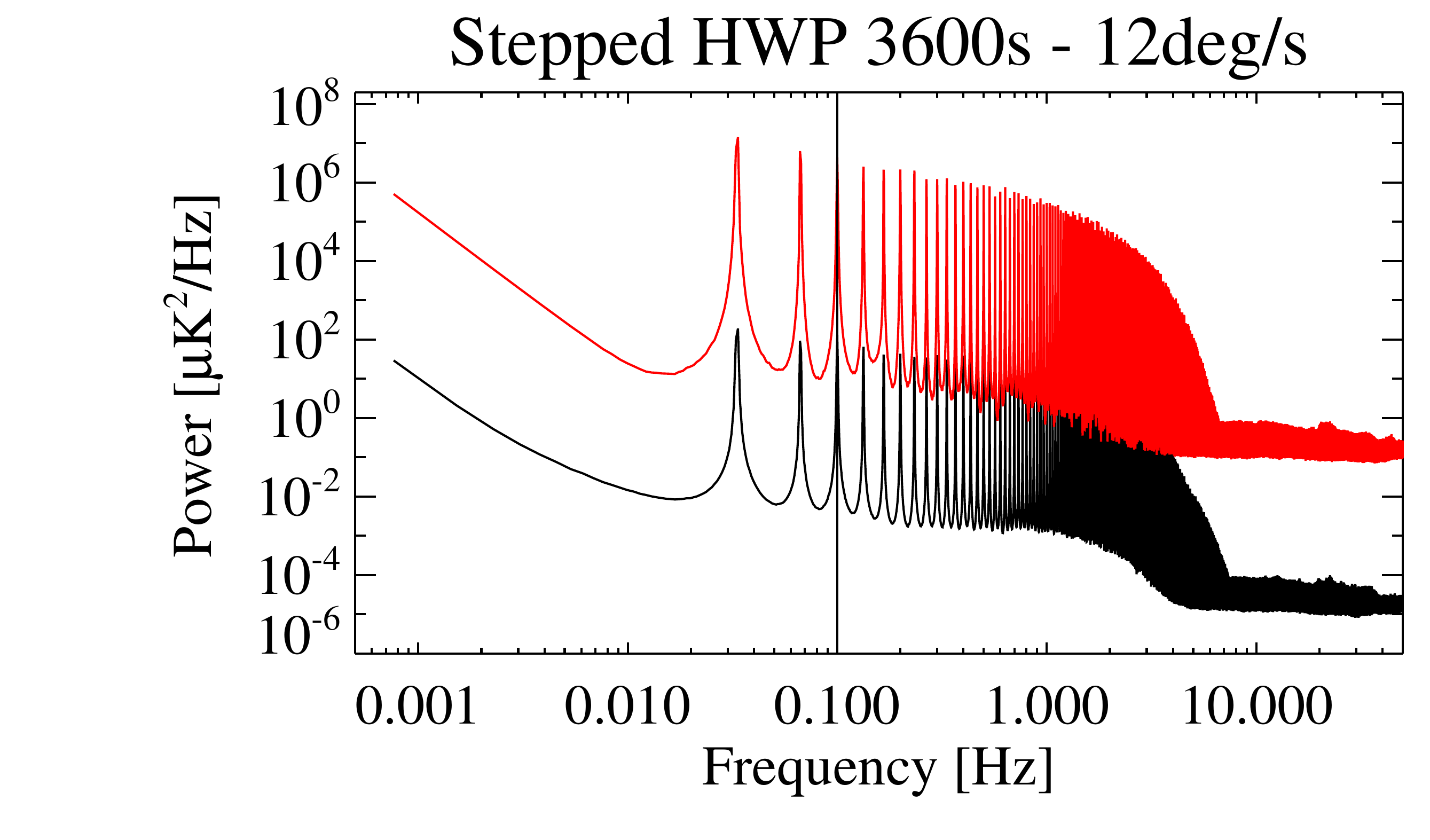}\\ \vskip -1.5em
  \includegraphics[angle=90,width=0.33\textwidth]{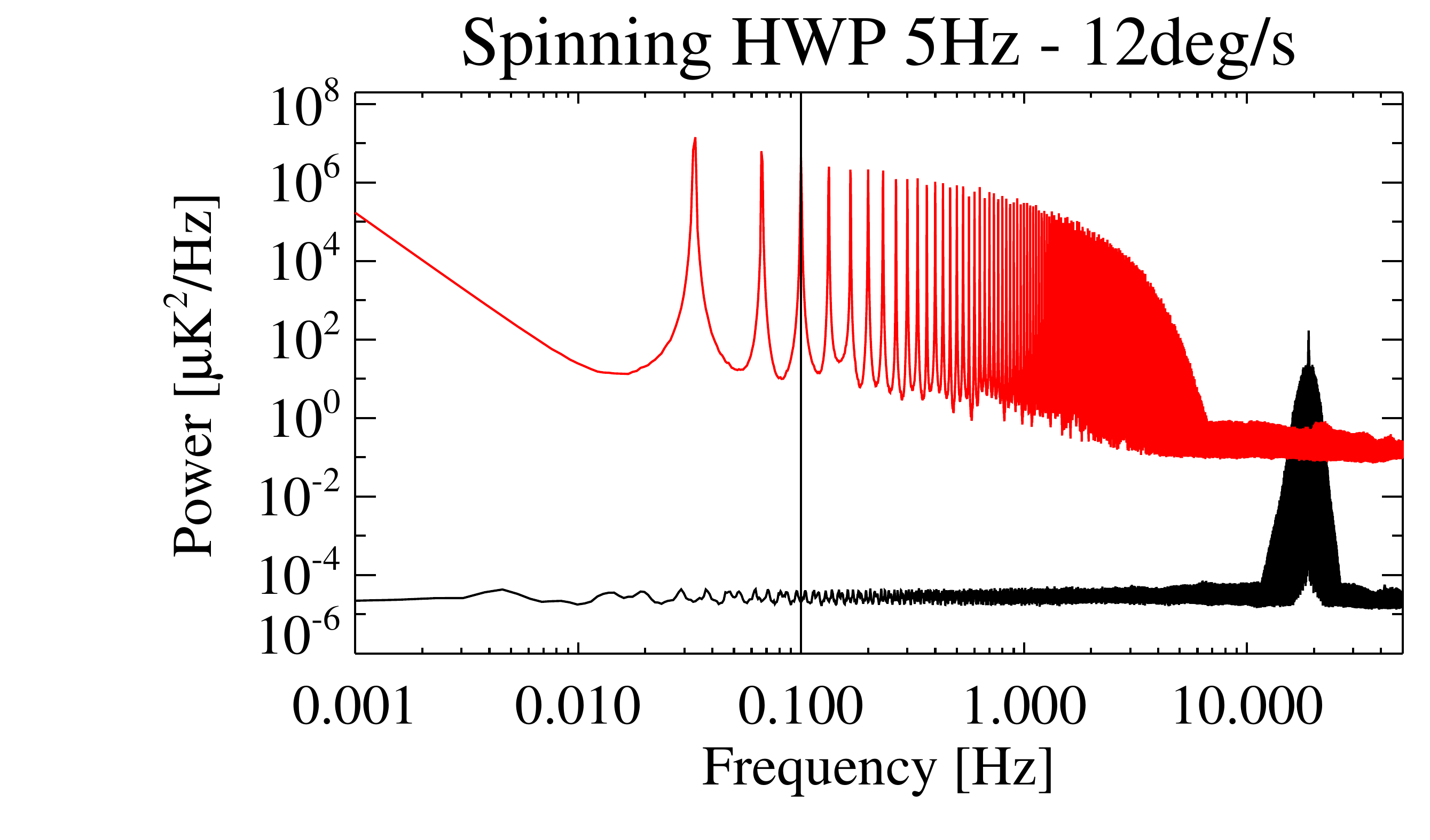}
\hskip -1em  \includegraphics[angle=90,width=0.33\textwidth]{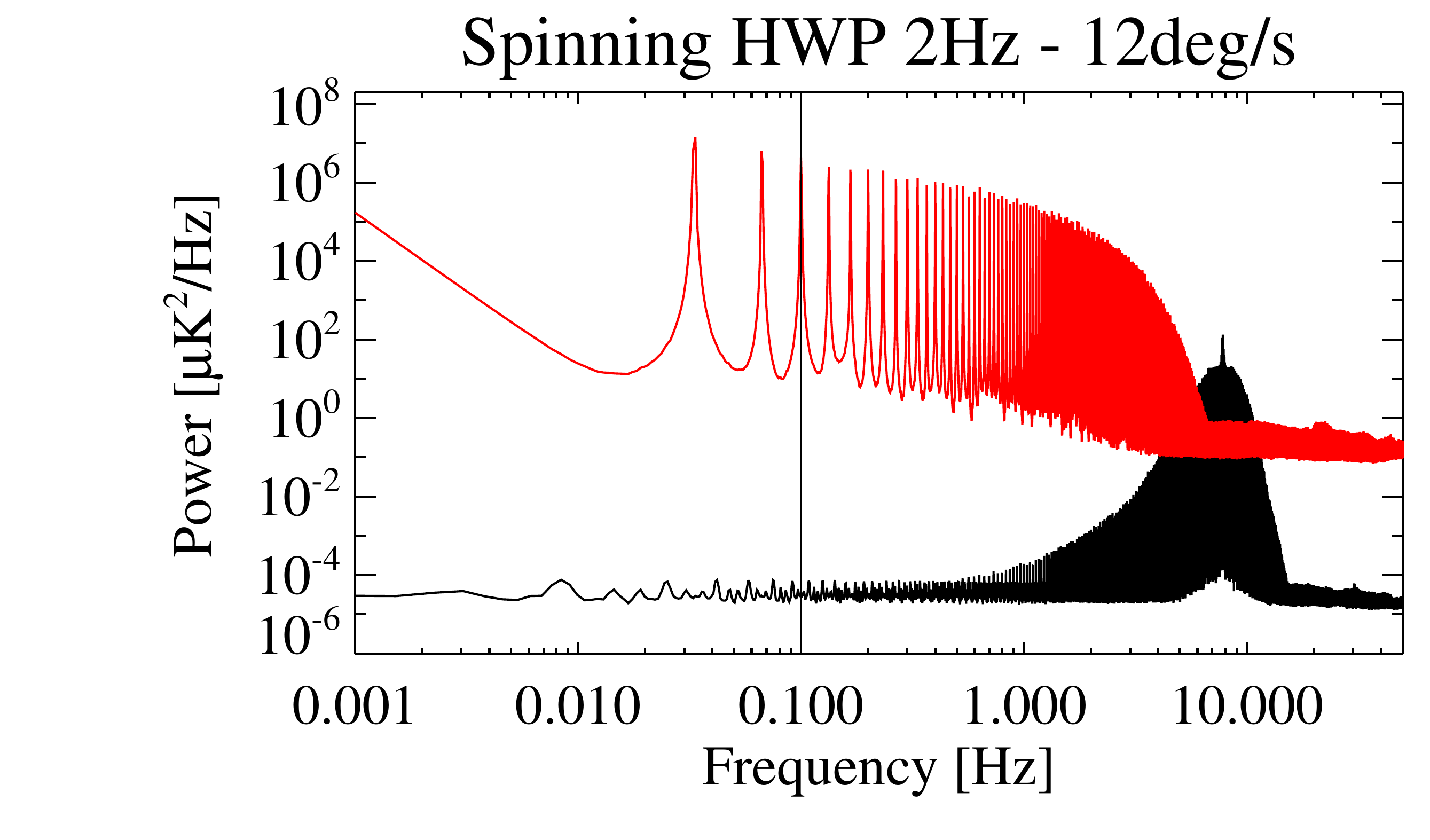}
\hskip -1em   \includegraphics[angle=90,width=0.33\textwidth]{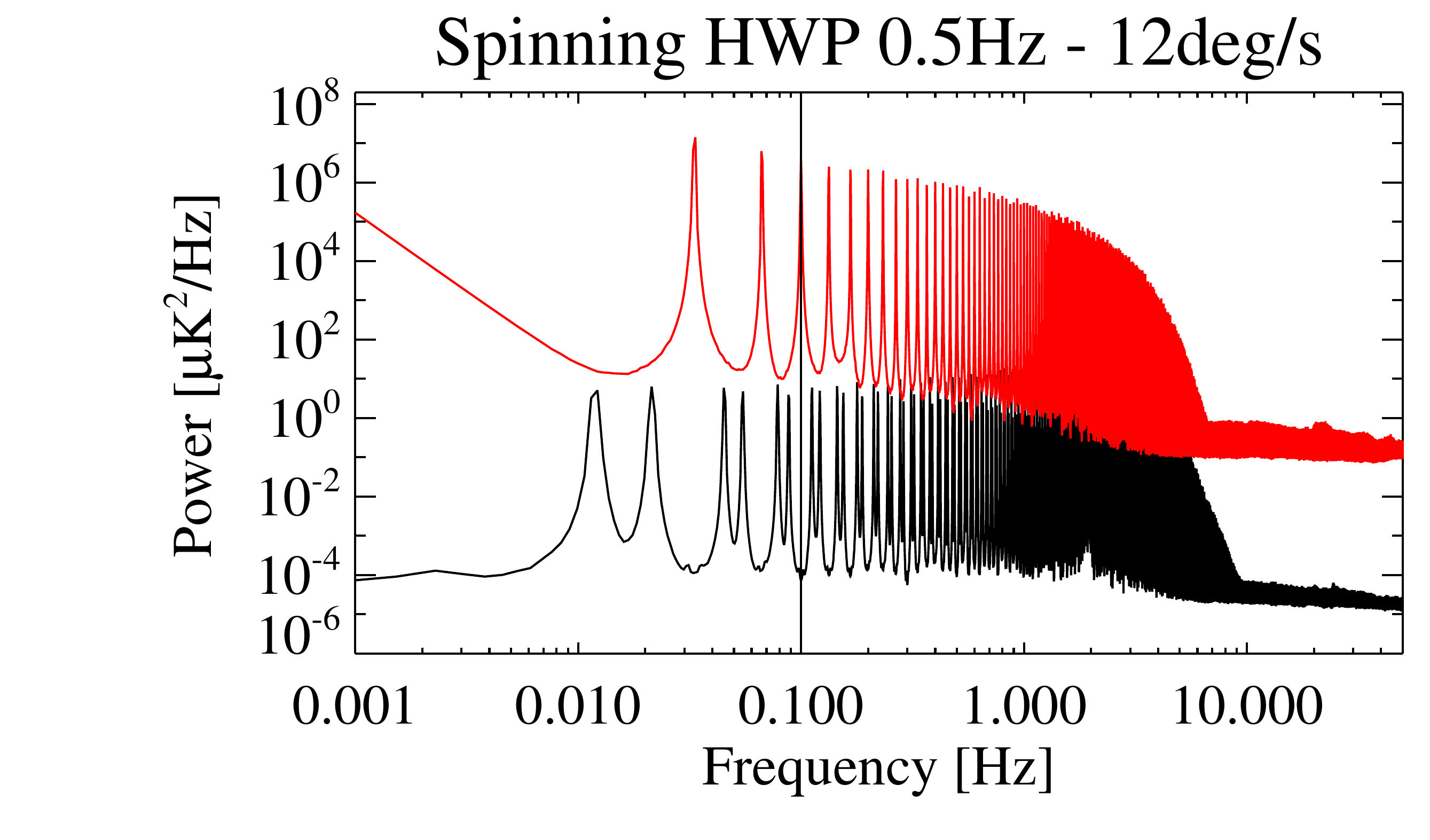}\\ \vskip -1.5em
  \includegraphics[angle=90,width=0.33\textwidth]{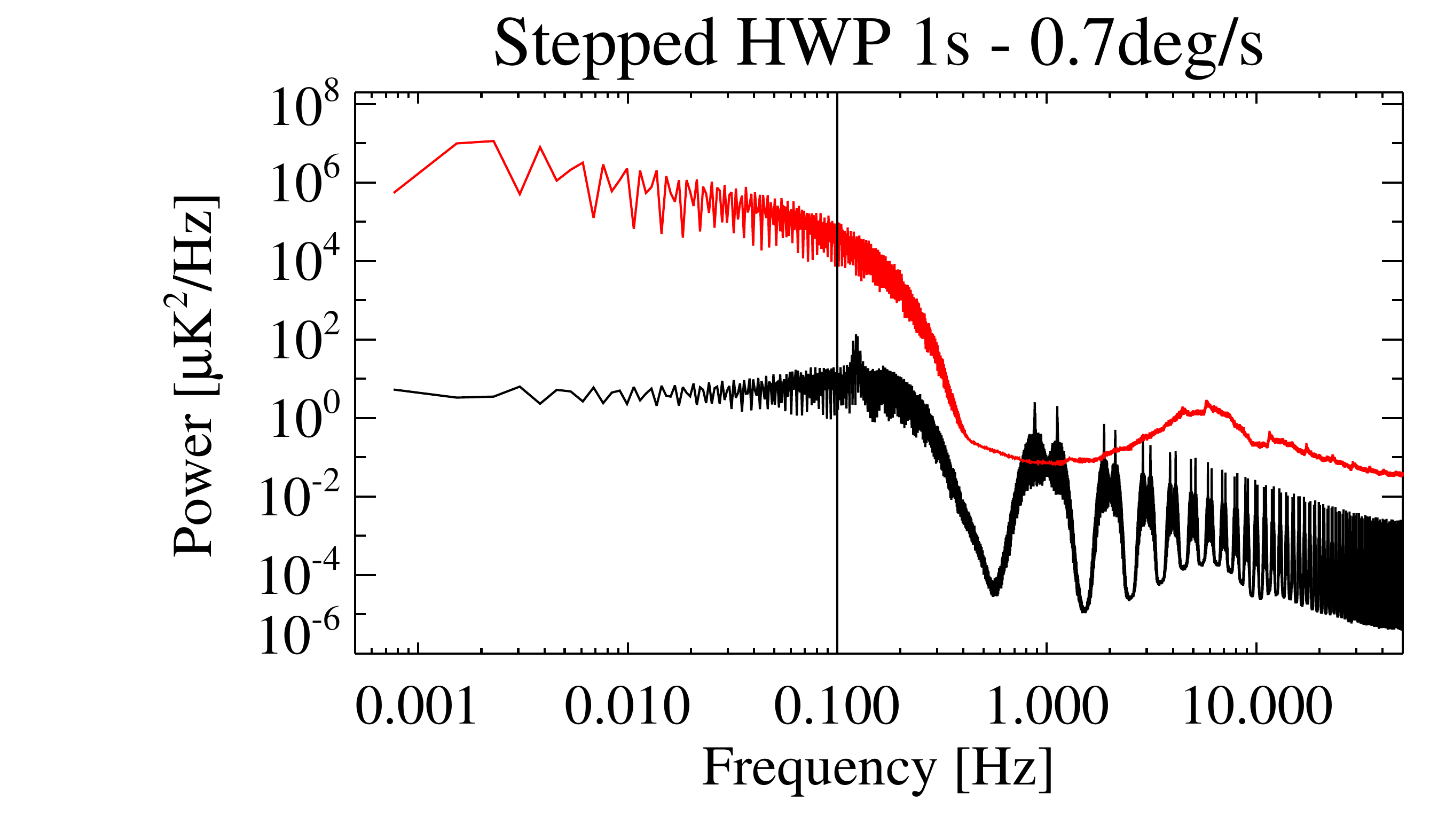}
\hskip -1em  \includegraphics[angle=90,width=0.33\textwidth]{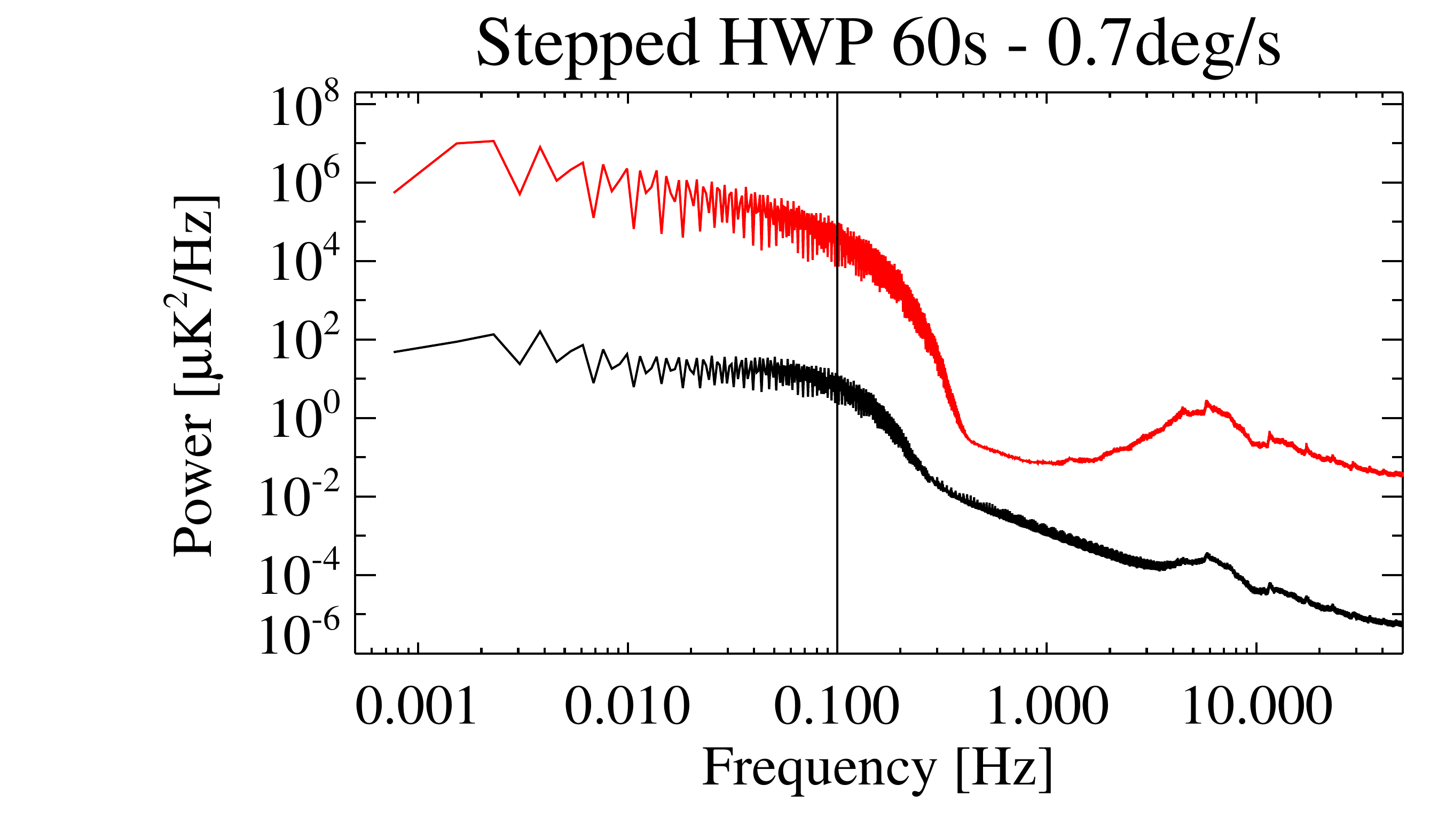}
\hskip -1em  \includegraphics[angle=90,width=0.33\textwidth]{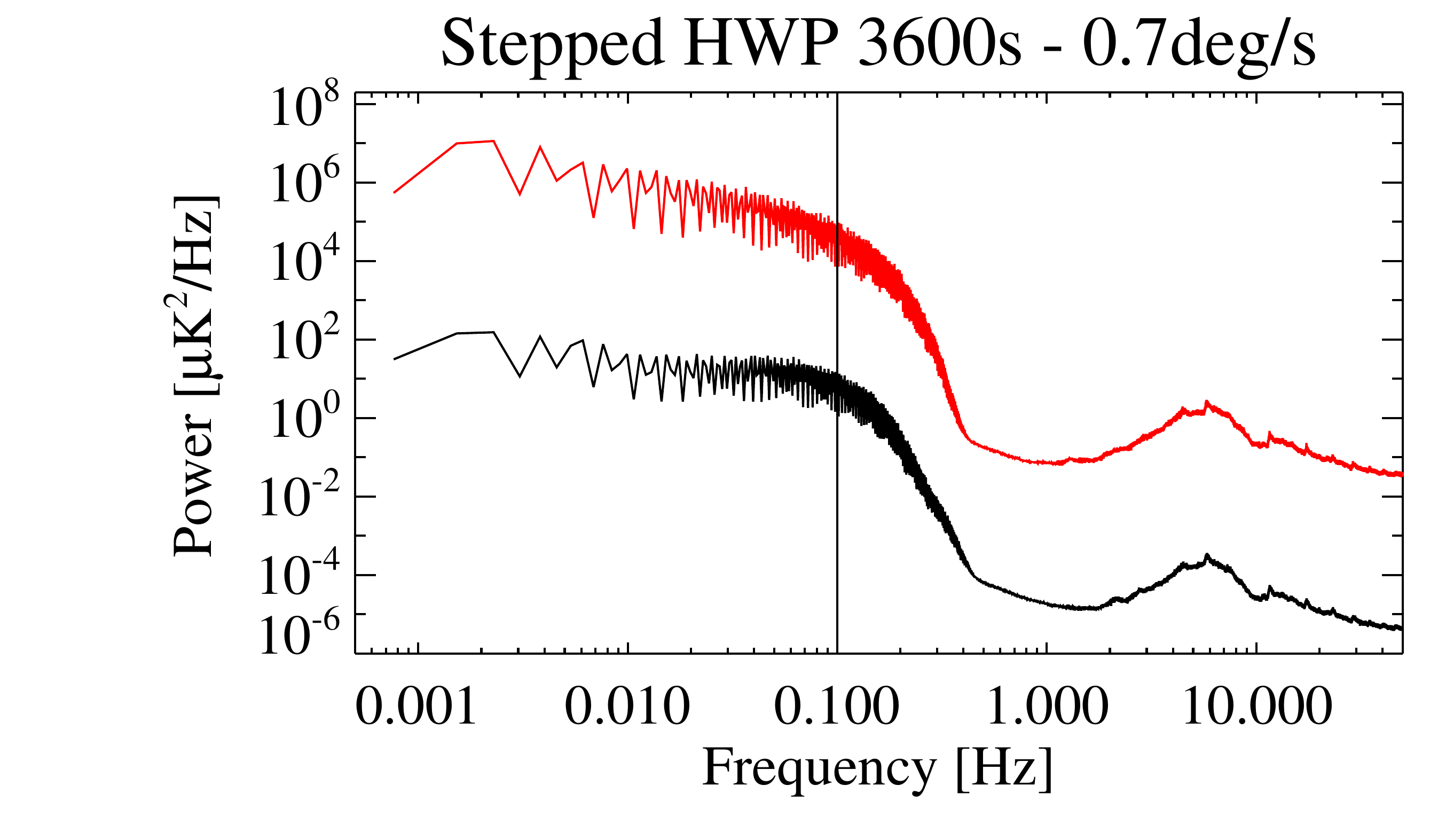}\\ \vskip -1.5em
  \includegraphics[angle=90,width=0.33\textwidth]{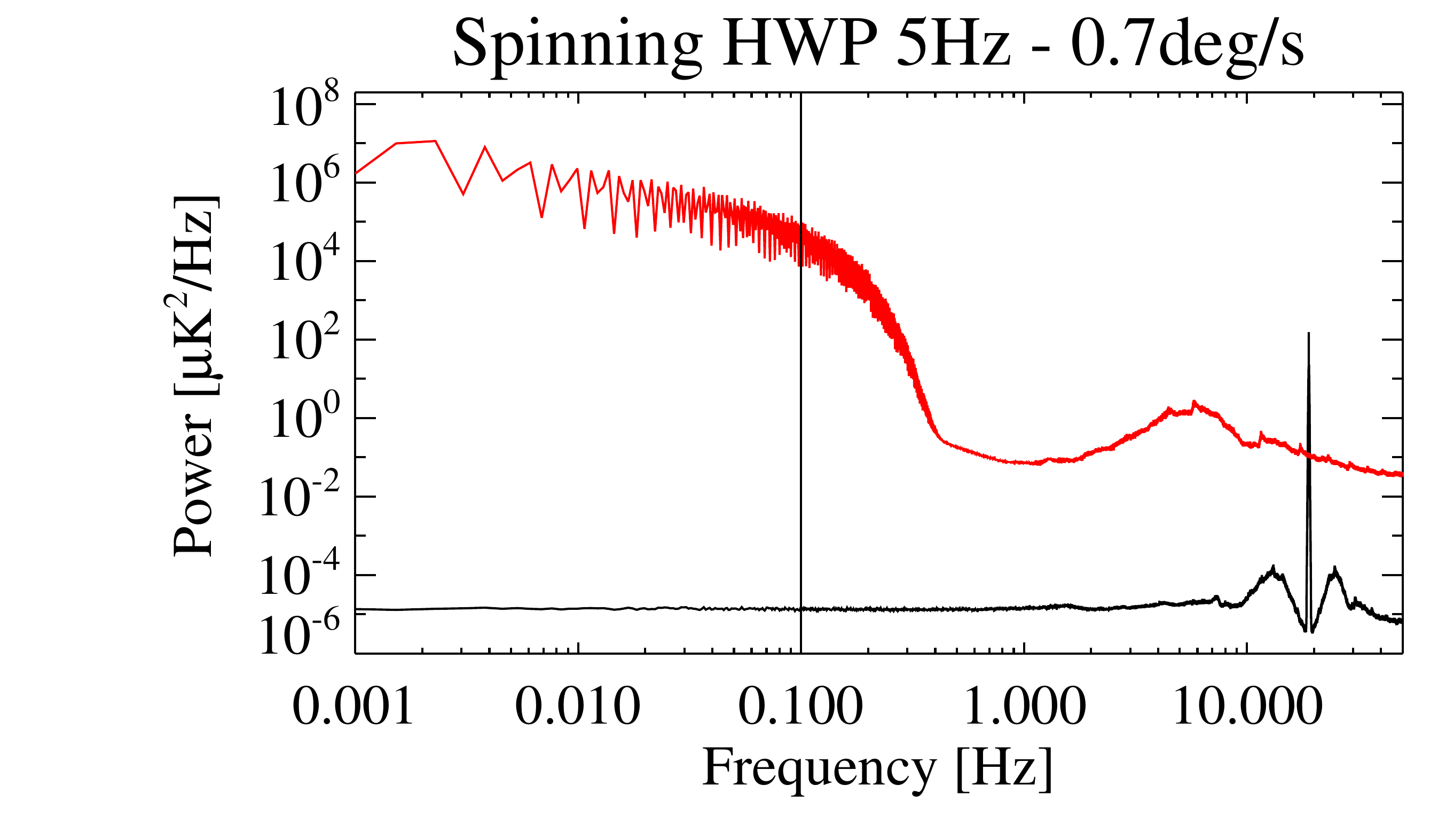}
\hskip -1em  \includegraphics[angle=90,width=0.33\textwidth]{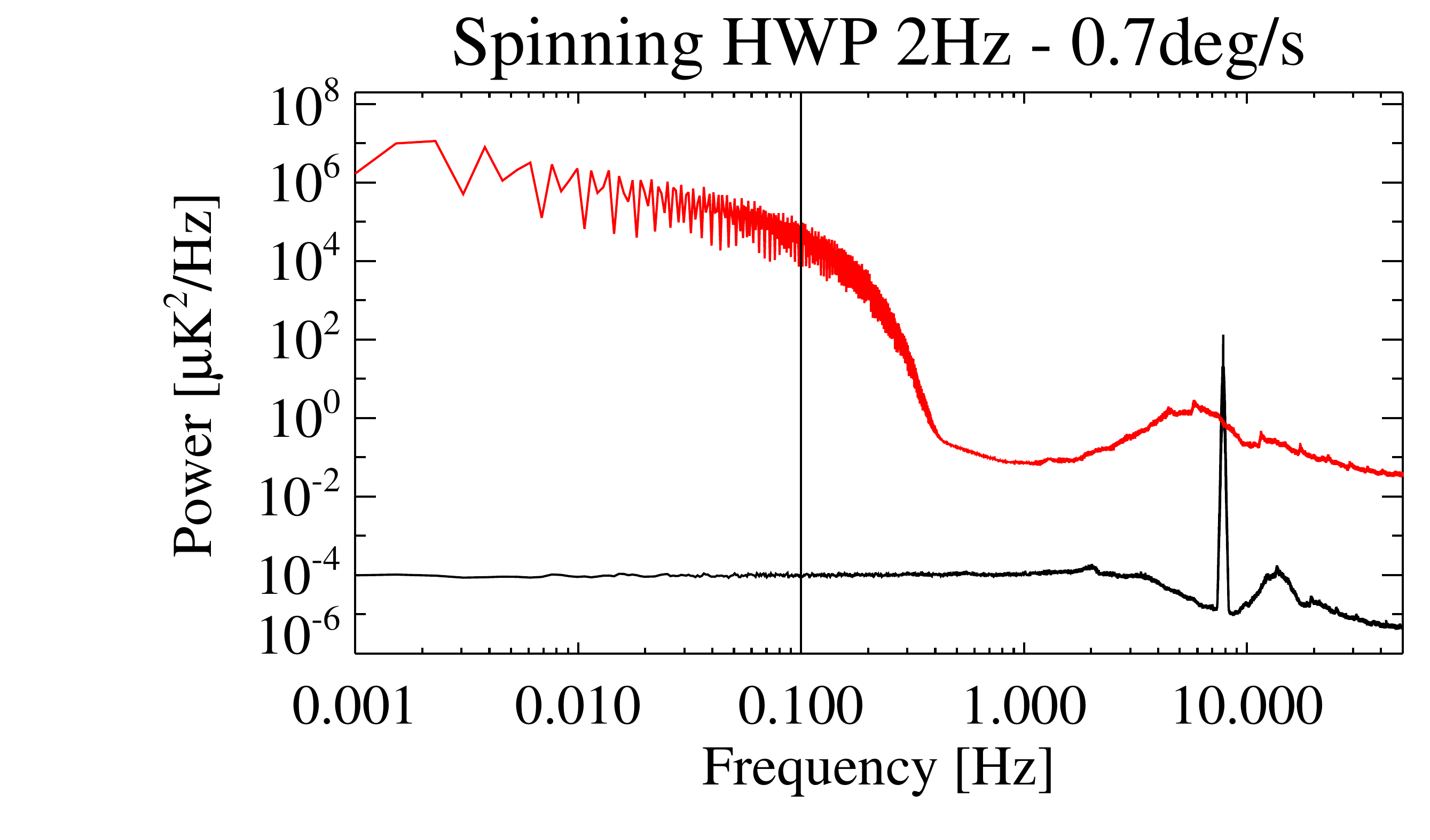}
\hskip -1em  \includegraphics[angle=90,width=0.33\textwidth]{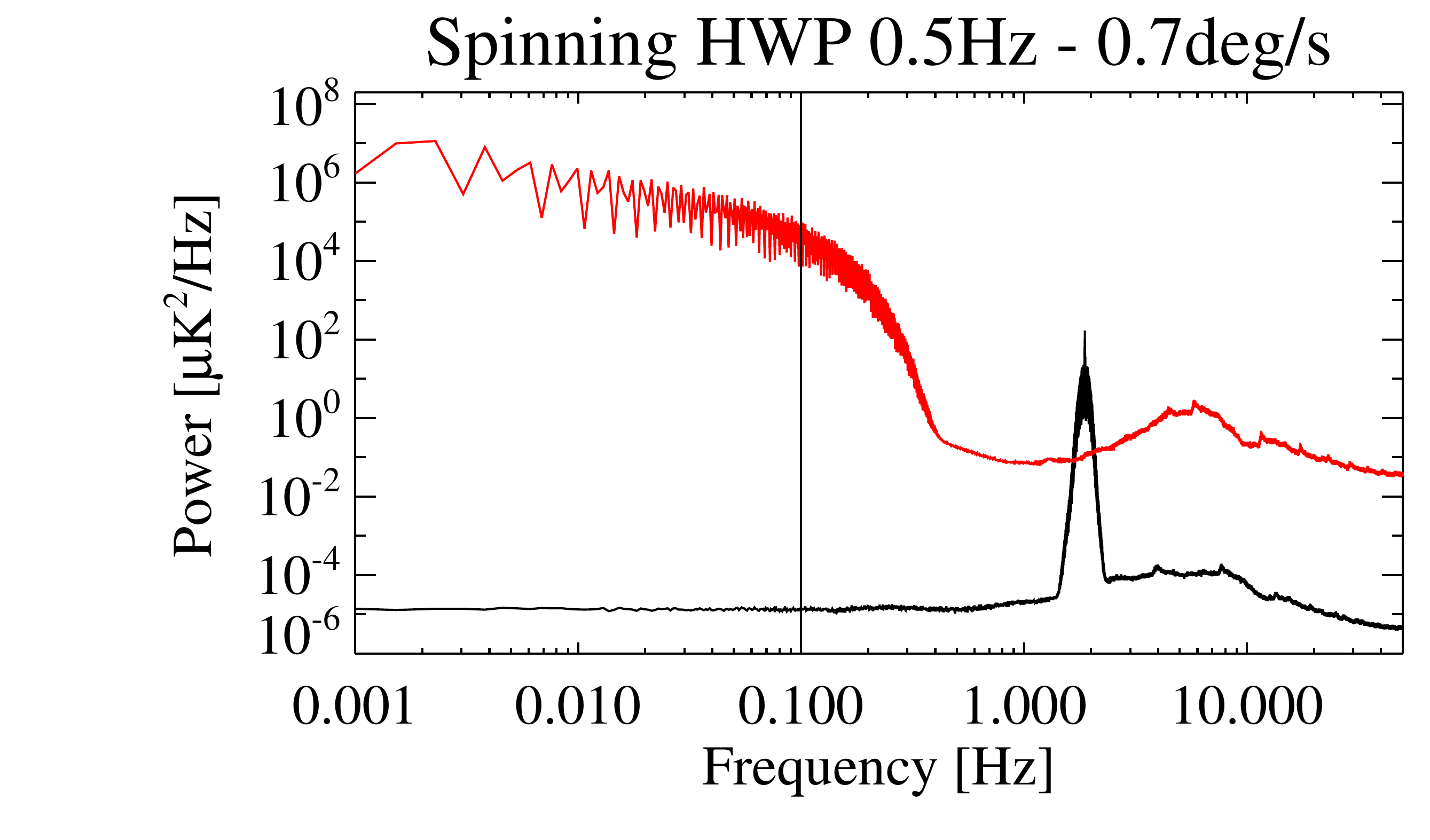}
  \caption{Periodograms, i.e. plots of temperature (in red) and polarization (in black) intensity power as function of frequency, for the HWP setups under consideration: stepping every $1~\mathrm{s}$, $60~\mathrm{s}$ and $3600~\mathrm{s}$ and spinning at $5~\mathrm{Hz}$, $2~\mathrm{Hz}$ and $0.5~\mathrm{Hz}$, for a telescope scanning at $12~\mathrm{deg/s}$ and $0.7~\mathrm{deg/s}$. Note that the telescope rotation affects both temperature and polarization, while the HWP modulates only polarization. For a spinning HWP design, the polarization signal is shifted to higher frequencies into a narrow band centered around a frequency $f=4f_r$, where $f_r$ is the rotation frequency.}\label{fig:periodo}
\end{figure*}
A natural consequence of employing a rotating HWP is the signal modulation. In Fig.~\ref{fig:periodo} we show the plots of temperature and polarization intensity power as function of frequency (periodograms) for the HWP rotation schemes under examination.

We highlight that a HWP modulates only the polarization signal, while the telescope scan-speed affects both intensity and polarization. Intensity modulation can therefore be achieved only via telescope scanning and the (small) amount of sky rotation \citep{2009MNRAS.397..634B}.

It is clearly desirable to move the polarization power away from the low frequency $1/f$ noise as far as possible. Furthermore, it would be preferable to filter out the least amount of intensity signal, since temperature anisotropy provides a viable source of calibration by direct comparison with Planck anisotropy maps.

To better understand the meaning of the periodograms, we point out that the first peak in the temperature plots is the fundamental mode corresponding to the gondola scanning frequency, followed by its harmonics. 

Note that the pattern of the polarization plots in the case of a slow stepped HWP is very similar to the corresponding temperature plots, as the HWP modulation contribution is subdominant with respect to the gondola spinning. It should be also pointed out that the high-frequency tail of the periodograms has no physical meaning since is due to pixelization effects. 

This analysis confirms the following general expectations:
i) the ability of shifting the polarization signal to higher frequencies is typical of the spinning HWP mode. In this case the polarization power is moved to a narrow band centered at frequency $f=4f_r$, where $f_r$ is the rotation frequency. The bandwidth depends on both the HWP and the telescope rotation rates: in particular, we find that a narrower bandwidth corresponds to slower telescope scanning rates;
ii) for a stepped HWP, the gain in the polarization signal modulation is less clear. The modulation performance is only slightly sensitive to the HWP rotation rate, while we find a moderate dependence on the telescope scan-speed.

In addition, the periodograms display an interesting feature: the ``doubling'' of the polarization peaks, due to the interaction of the HWP and telescope scanning strategies, which is particularly evident for some configurations (see, e.g., the stepped mode at $60~\mathrm{s}$ or the spinning mode at $0.5~\mathrm{Hz}$ at telescope scan-speed of $12~\mathrm{deg/s}$). 

\subsection{Pixel inverse condition number}\label{invcond}
\begin{figure*}
\hskip 3.5em  \includegraphics[angle=90, width=0.39\textwidth]{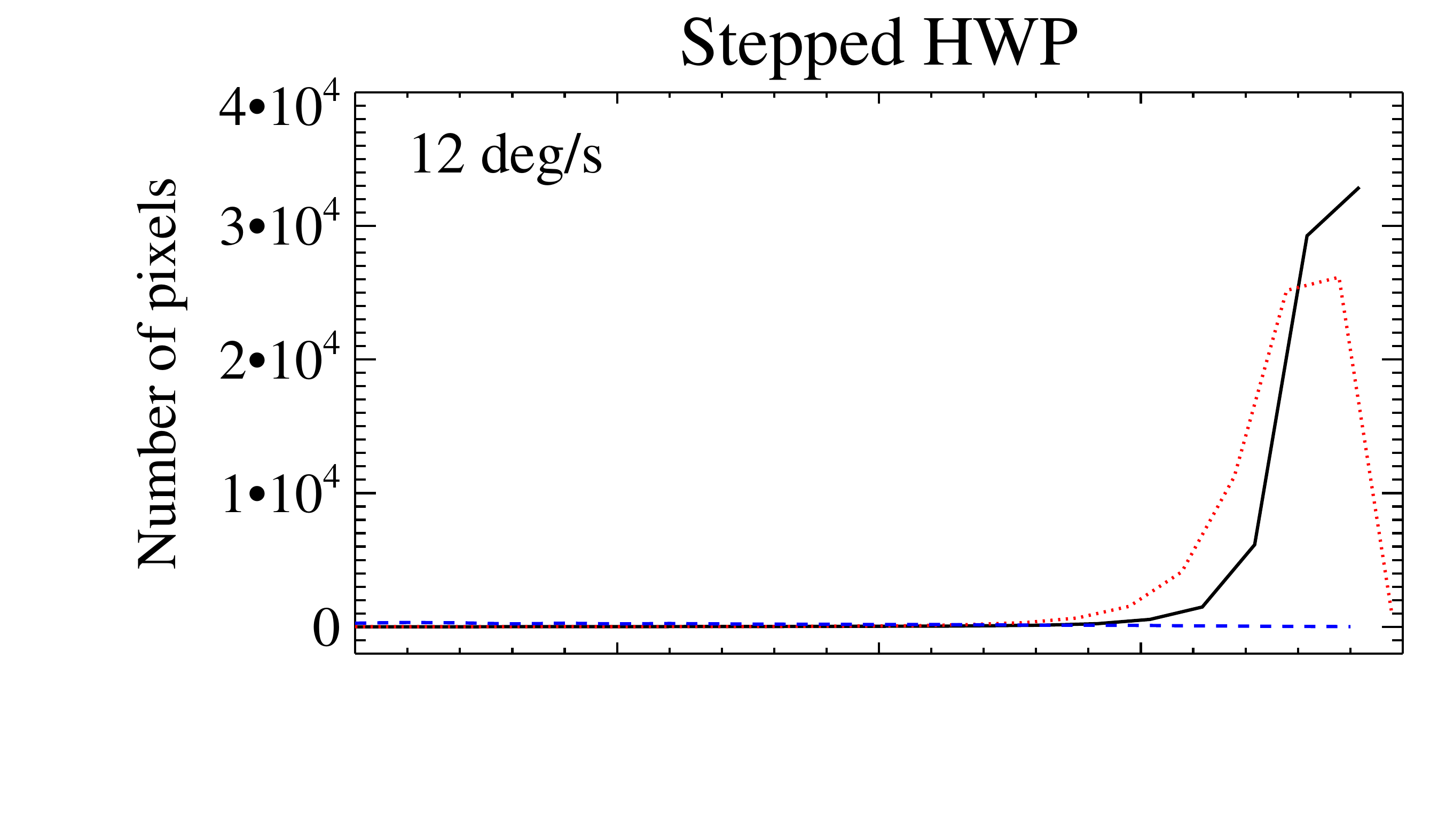}  
\hskip 3em  \includegraphics[angle=90, width=0.39\textwidth]{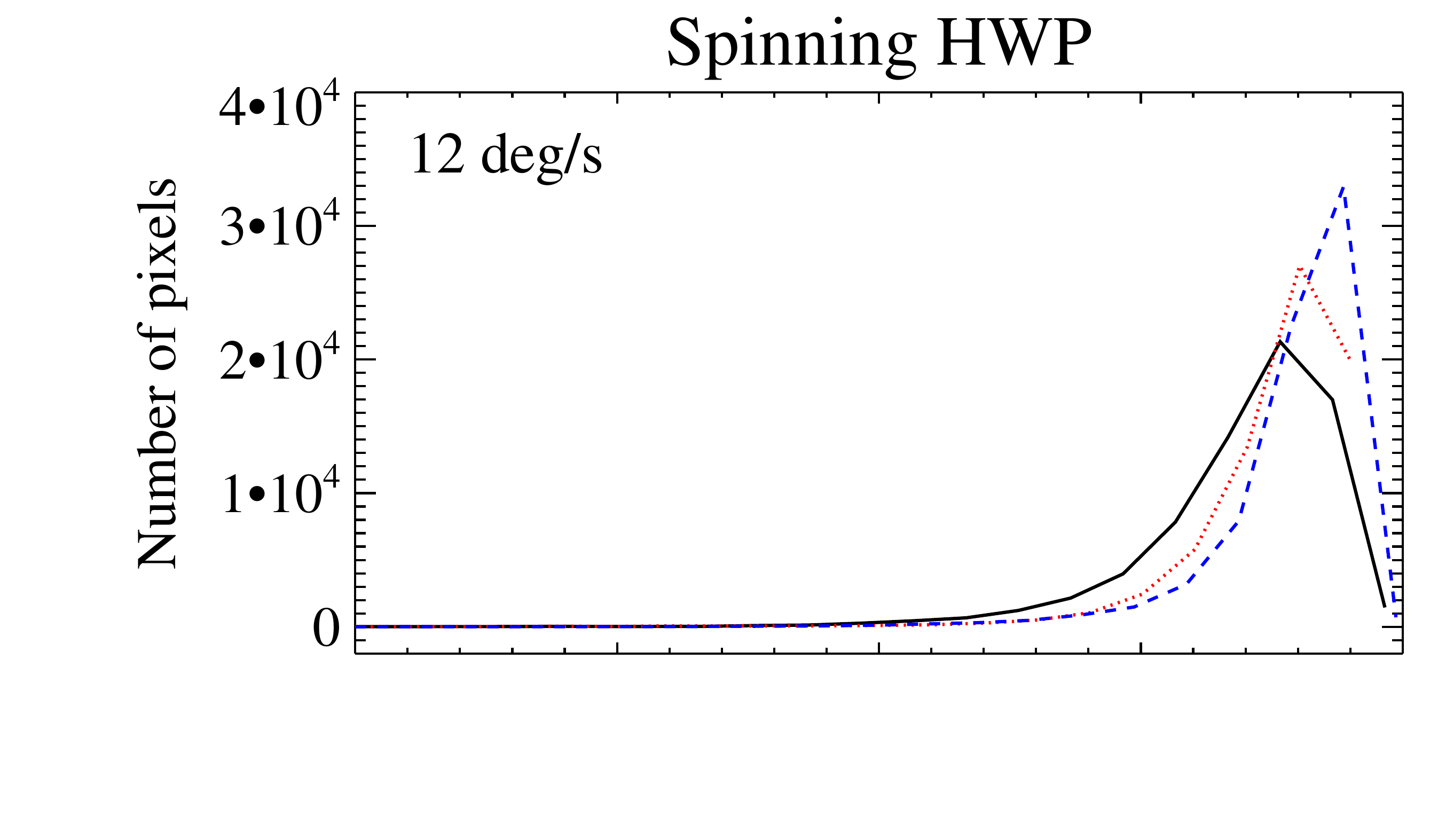} \\ \vskip -4.5em    
\hskip 3.5em  \includegraphics[angle=90, width=0.39\textwidth]{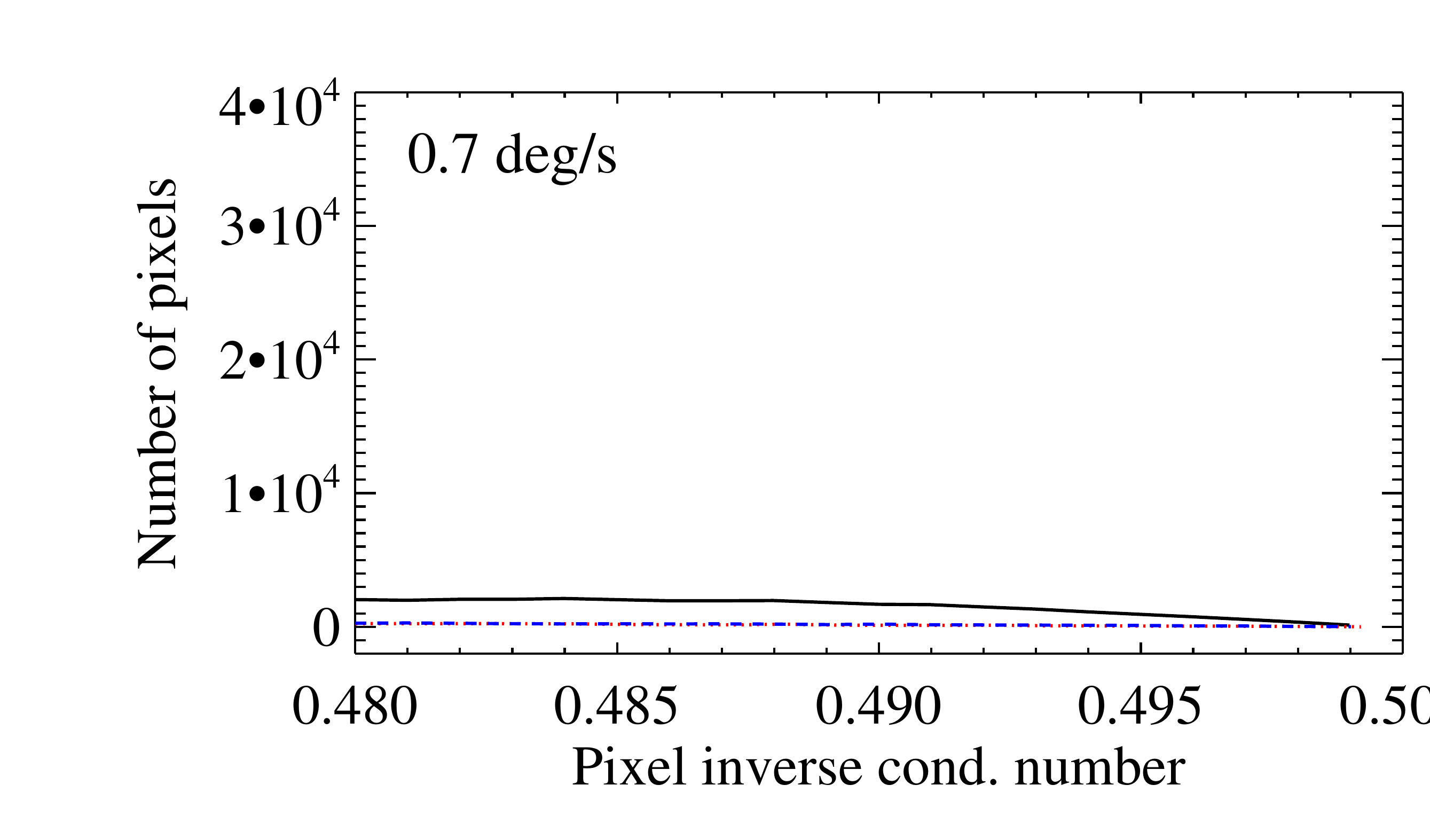} 
\hskip 3em  \includegraphics[angle=90, width=0.39\textwidth]{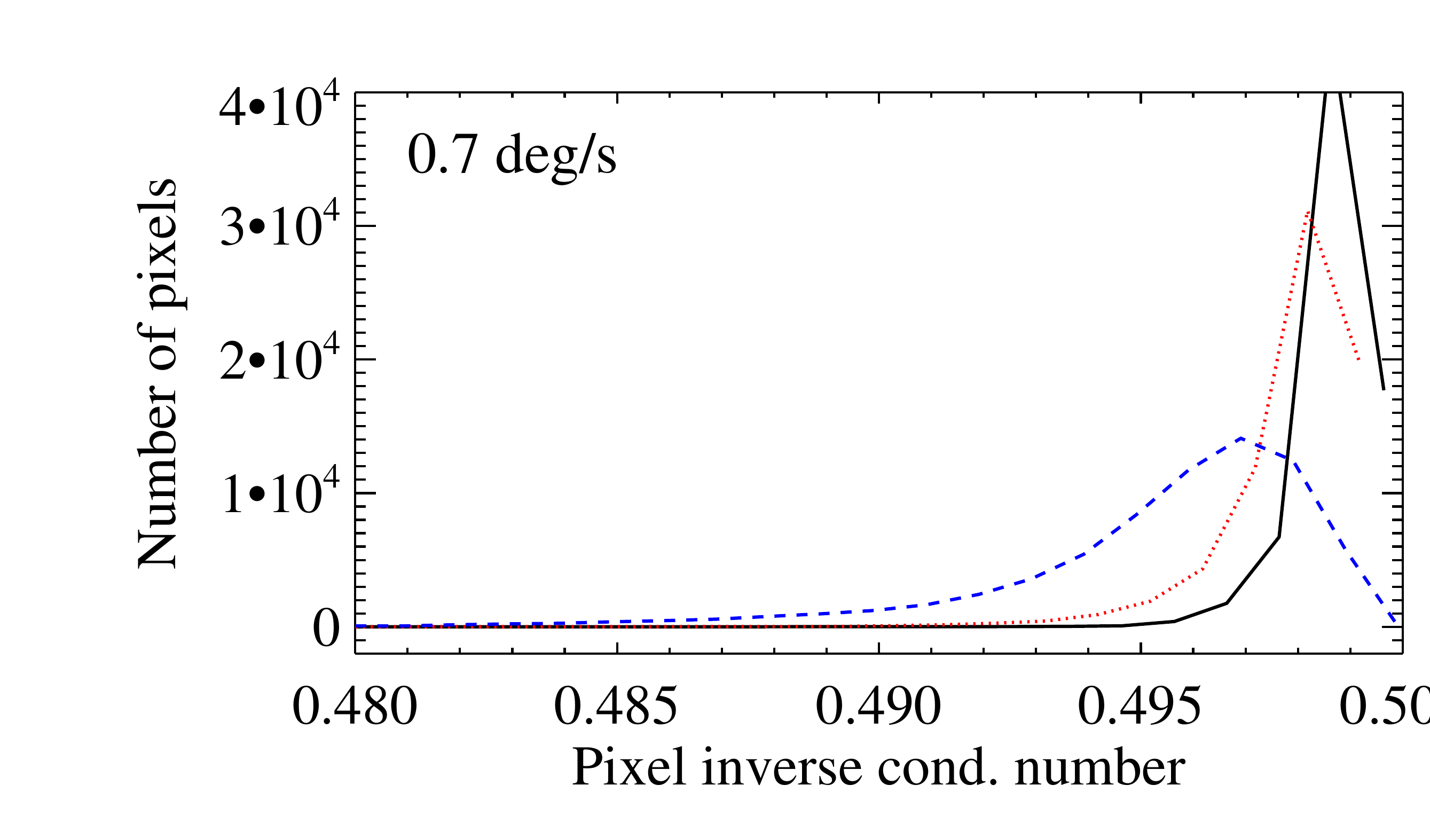}
  \caption{Histograms of the pixel inverse condition number for the HWP setups under consideration. First column: HWP stepped every $1~\mathrm{s}$ (in solid black line), $60~\mathrm{s}$ (in dotted red line) and $3600~\mathrm{s}$ (in dashed blue line), for a telescope scanning at $12~\mathrm{deg/s}$ (top) and $0.7~\mathrm{deg/s}$ (bottom). Second column: HWP spinning at $5~\mathrm{Hz}$ (in solid black line), $2~\mathrm{Hz}$ (in dotted red line) and $0.5~\mathrm{Hz}$ (in dashed blue line), for a telescope scanning at $12~\mathrm{deg/s}$ (top) and $0.7~\mathrm{deg/s}$ (bottom). In our convention, the condition number $R_{cond}$ has a value running from 0 to 0.5, where $R_{cond}=0.5$ means perfect angle coverage.} \label{fig:hist}
\end{figure*}
The pixel inverse condition number (hereafter, $R_{cond}$) is a useful tool to quantify the angle coverage uniformity of the observation on a given pixel. 
In our definition, $R_{cond}$ is the ratio of the absolute values of the smallest and largest eigenvalues for each block of the preconditioner matrix 
employed in the map-making algorithm. In this formalism, $R_{cond}$ has a value running from zero to $1/2$, with $R_{cond}=1/2$ in case of perfect 
angle coverage (see, e.g., \citealt{2009A&A...506.1511K} for more details). Notice that the condition number has a pure geometrical meaning, being essentially independent from the map-making algorithm employed. It has been usually assumed that a value $R_{cond} \le 10^{-2}$ means that the polarization cannot be solved and hence those pixels must be removed from the analysis.

The use of a HWP modulator clearly improves the pixel angle coverage during the observation, since each pixel is observed from more directions. 

In Fig.~\ref{fig:hist} we show the histograms of the pixel inverse condition number for the HWP configurations under examination. We find that:
i) for a fast telescope rotation ($12~\mathrm{deg/s}$), all the HWP configurations provide very good pixel angle coverage, but the slowest stepped mode ($3600~\mathrm{s}$). The performance of a stepped HWP increases as its rotation rate does, while the opposite happens to a spinning HWP; 
ii) for a slow telescope rotation ($0.7~\mathrm{deg/s}$), the performance of a stepped HWP dramatically decreases, while a spinning HWP still provides a very good pixel angle coverage. As opposite to the fast telescope rotation case, now a rapidly spinning HWP provides a better coverage than a slowly spinning one.

We conclude that a spinning scheme offers a very good pixel angle coverage for any HWP rotation frequency and any gondola scanning rate. A slow stepped HWP ($3600~\mathrm{s}$) is never effective to provide a good coverage, while a fast stepping HWP ($1~\mathrm{s}$ and $60~\mathrm{s}$) provide a good coverage only when combined to a fast telescope scan-speed.

\subsection{Map-making residuals}\label{residuals}

In this Section we investigate the impact of the HWP modulation on the map reconstruction. The GLS maps are the primary output of our map-making code, but we will rather work in terms of residual maps, i.e. the difference between the output optimal and the input maps. Hence, smaller residuals imply smaller map-making errors. As output maps, we consider both the signal-only (S) and signal plus noise (SN) cases. Note that poorly observed pixels are removed from the analysis by setting a filter on the pixel inverse condition number ($R_{cond}>10^{-2}$).

We evaluate the residual map distribution over the angular scales by producing the corresponding BB angular power spectra (see Fig.~\ref{fig:BB_res_N}).  

In both the S and SN cases, the slowest HWP stepped mode ($3600~\mathrm{s}$) provides higher residuals at any telescope scan-speed. Moreover, we find that the performance of a HWP stepped every $60~\mathrm{s}$ worsens when a slow telescope scan-speed ($0.7~\mathrm{deg/s}$) is performed.
On the contrary, the residuals corresponding to the spinning HWP configurations do not significantly change against variations of both the HWP and gondola rotation rates.

The residuals corresponding to the default SWIPE configuration are comparable to those of a generic spinning scheme.

\begin{figure*}
\psfrag{Cl BB res [uK^2]}[c][][3]{Residual $C_{\ell}^{BB}$ $[\si{\mu K^2}]$}
\hskip 3.5em  \includegraphics[angle=90, width=0.39\textwidth]{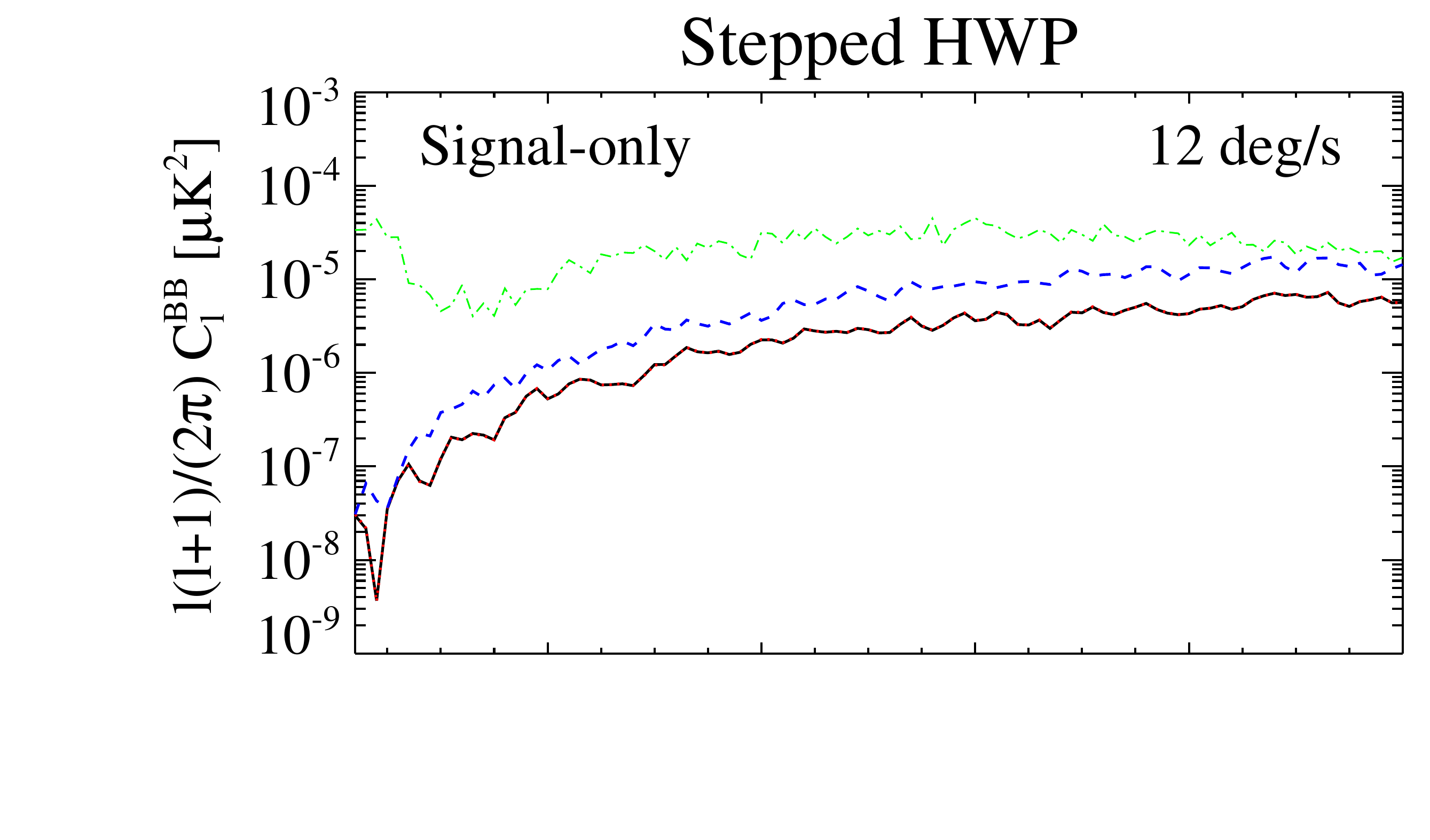}
\hskip 3em  \includegraphics[angle=90, width=0.39\textwidth]{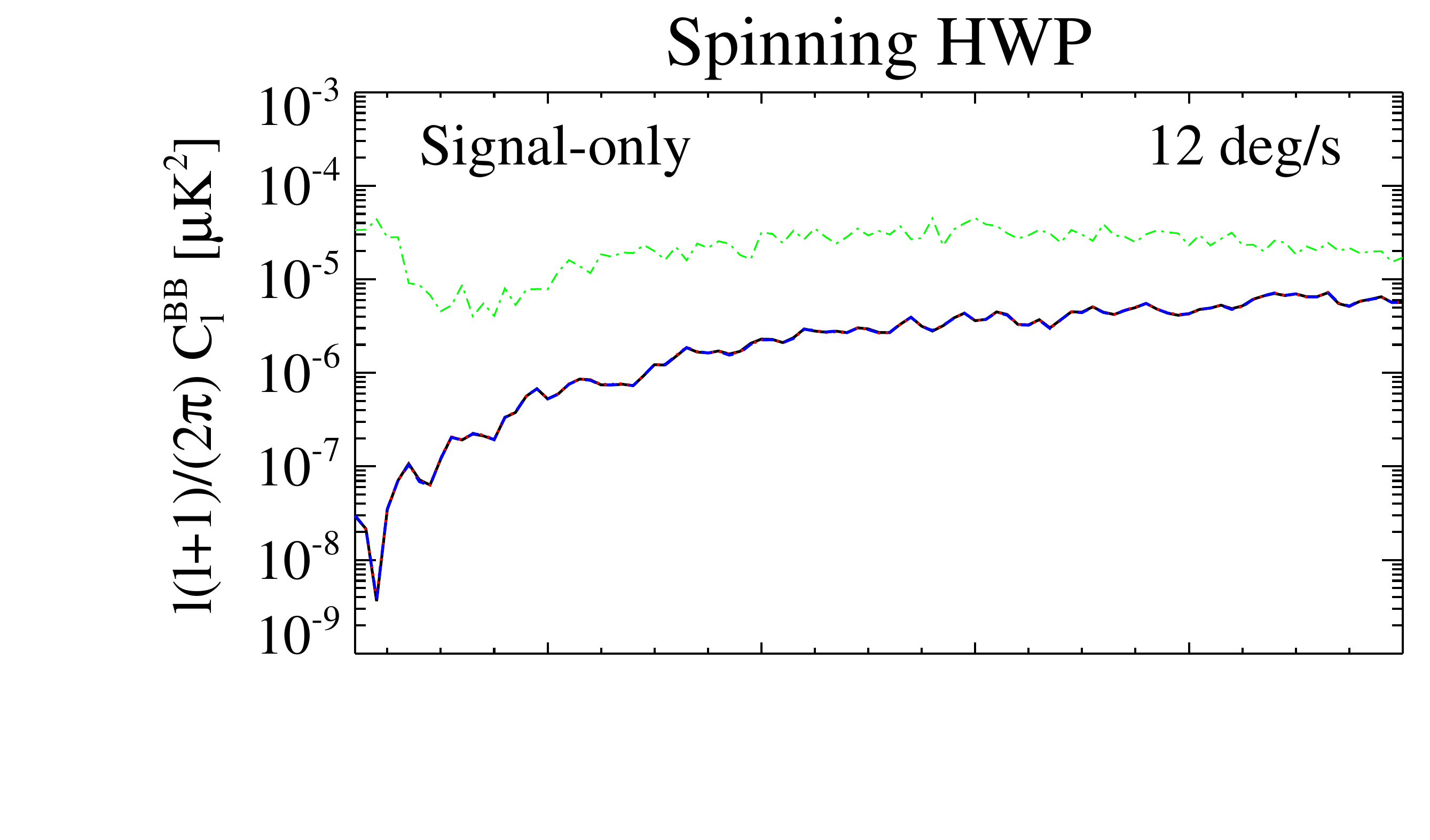}\\ \vskip -4.5em 
\hskip 3.5em  \includegraphics[angle=90, width=0.39\textwidth]{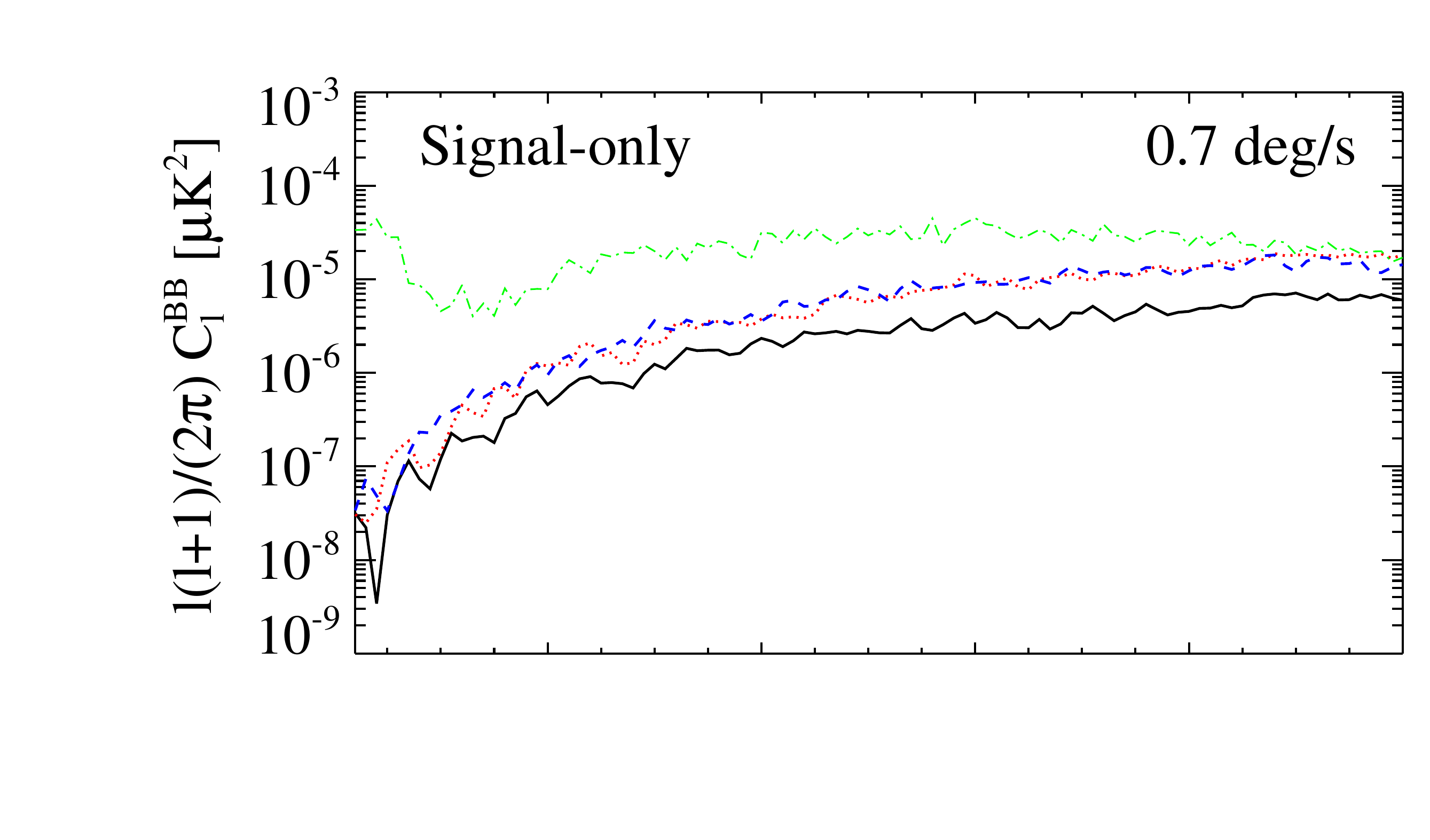}
\hskip 3em  \includegraphics[angle=90, width=0.39\textwidth]{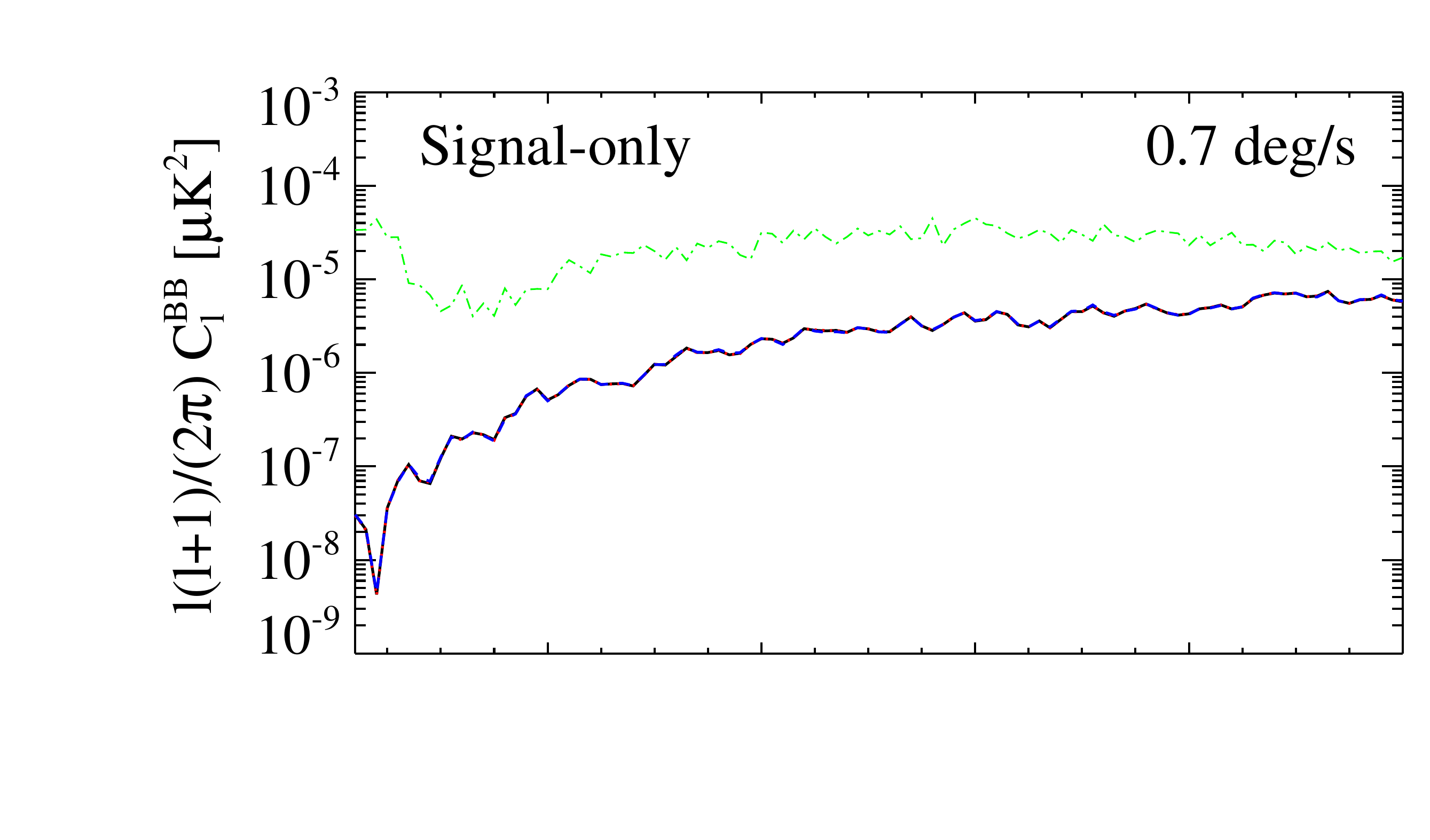}\\ \vskip -4.5em 
  \hskip 3.5em  \includegraphics[angle=90, width=0.39\textwidth]{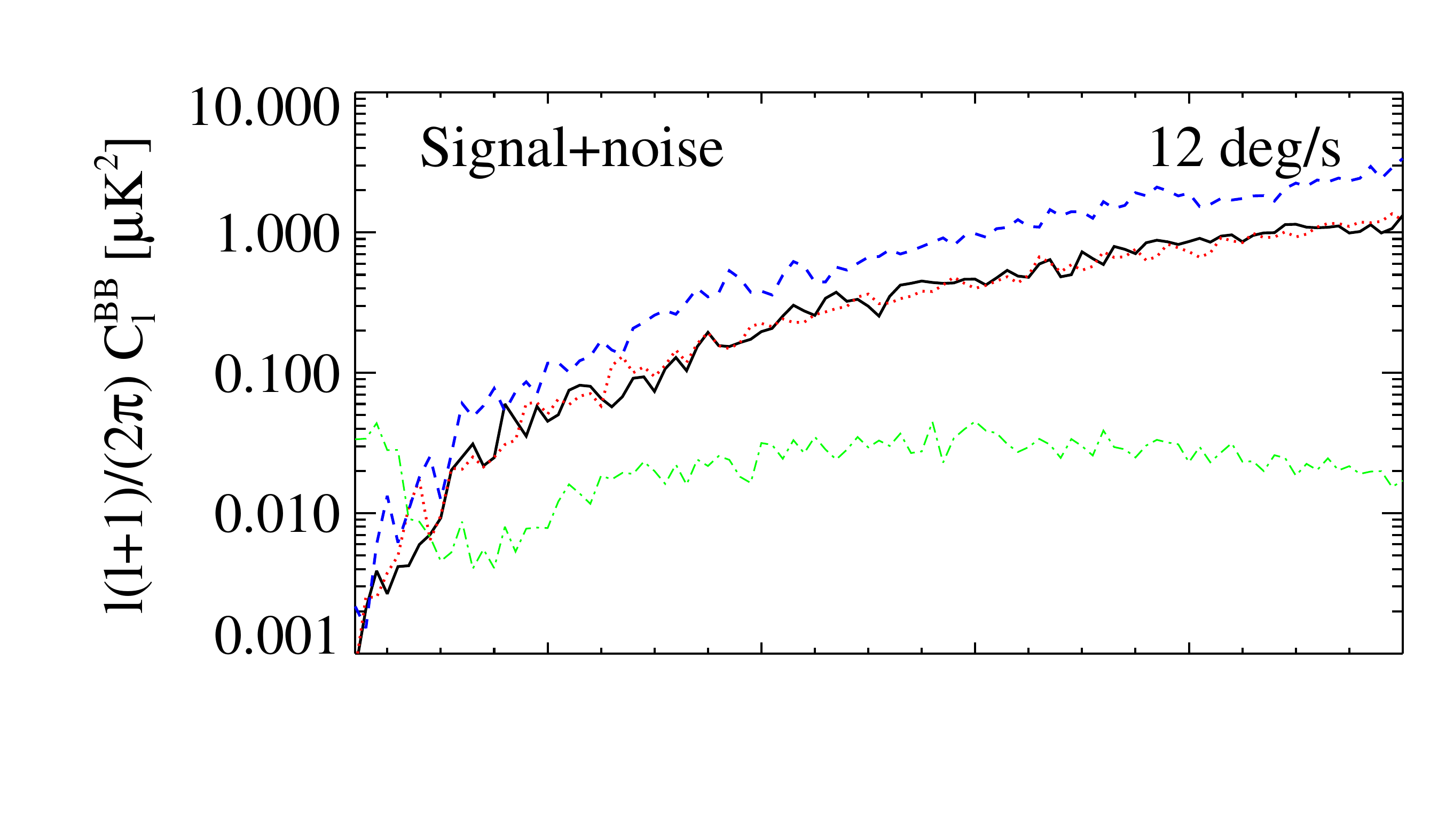}
\hskip 3em  \includegraphics[angle=90, width=0.39\textwidth]{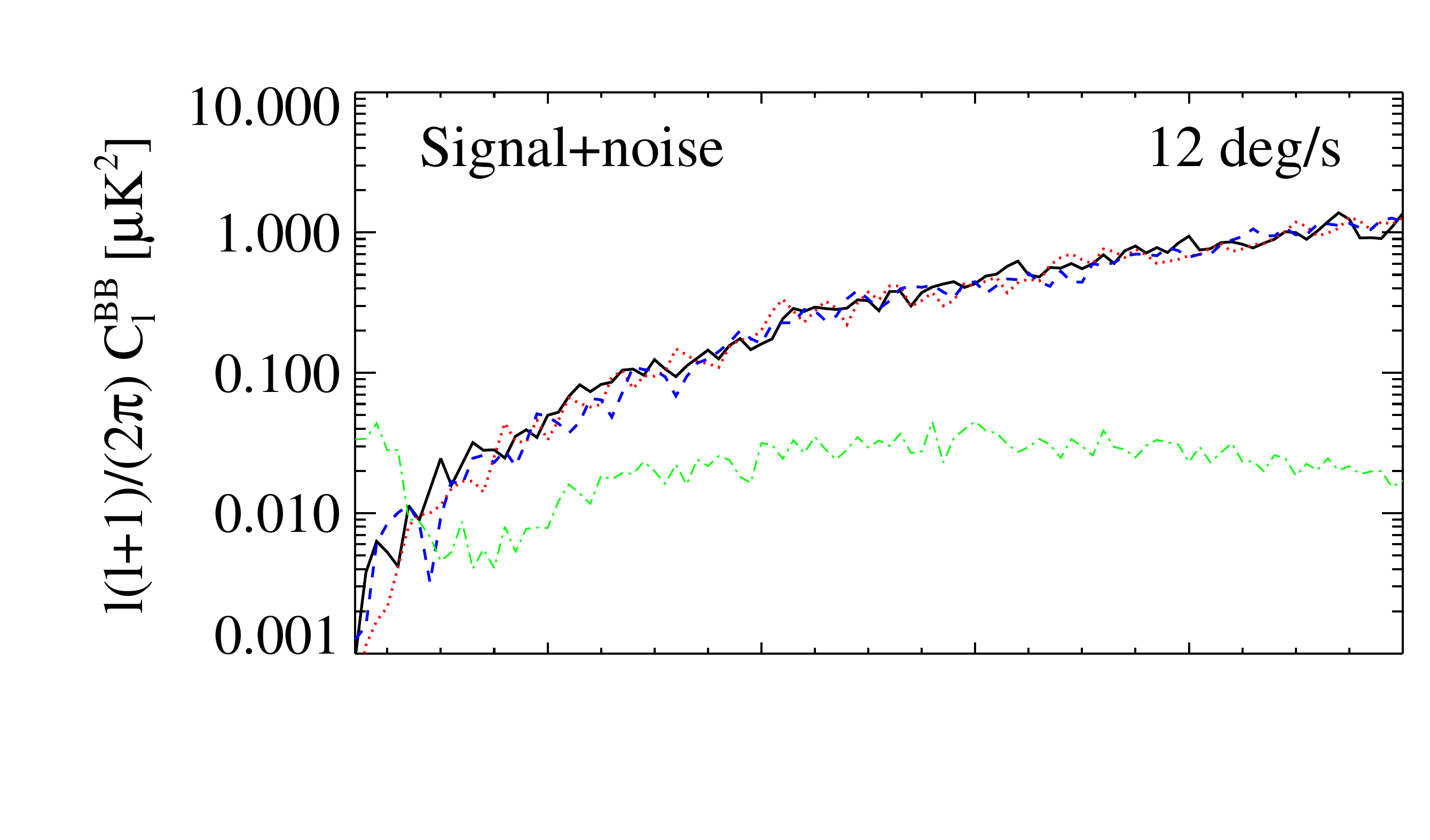}\\ \vskip -4.5em 
\hskip 3.5em  \includegraphics[angle=90, width=0.39\textwidth]{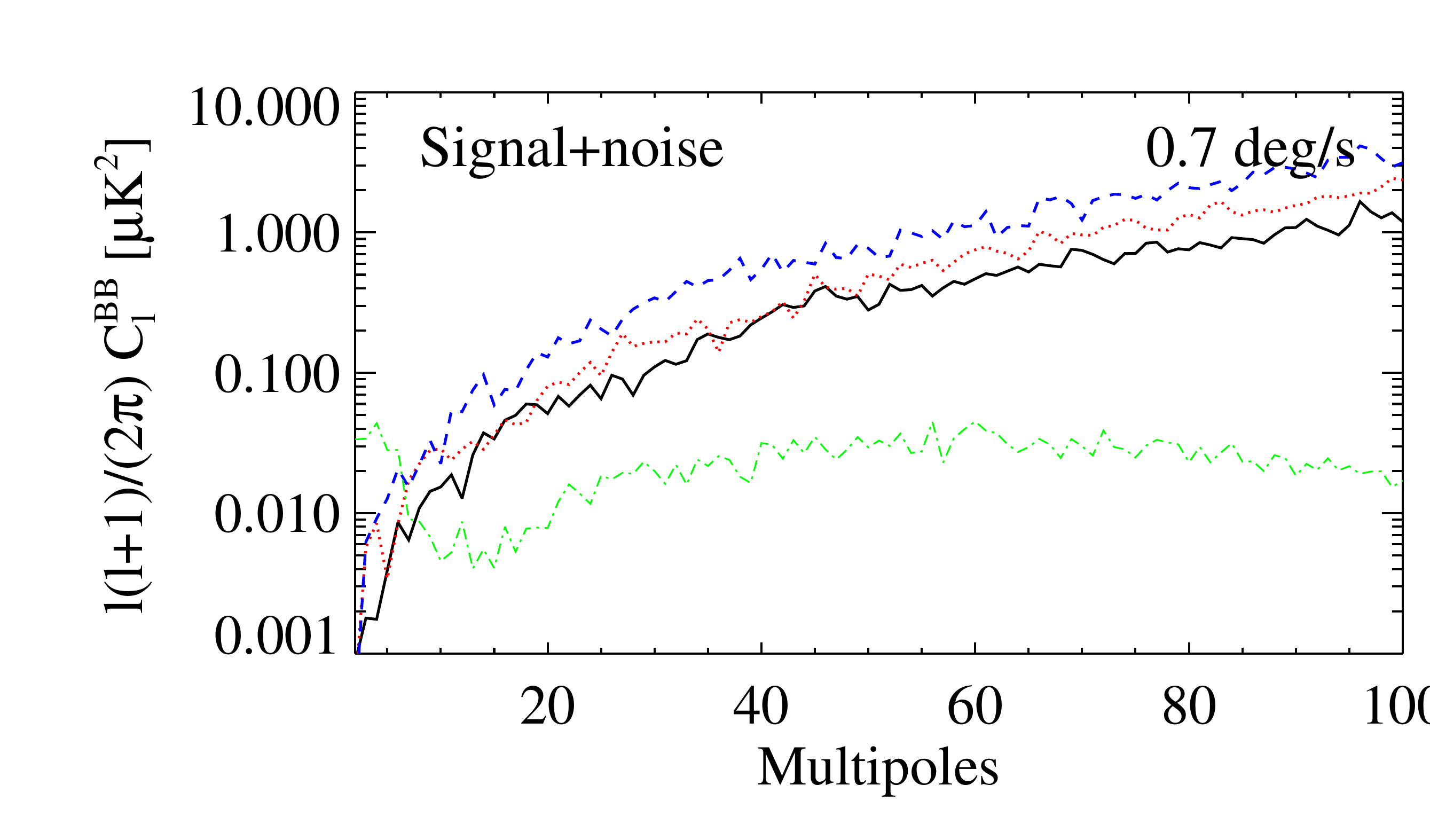}
\hskip 3em  \includegraphics[angle=90, width=0.39\textwidth]{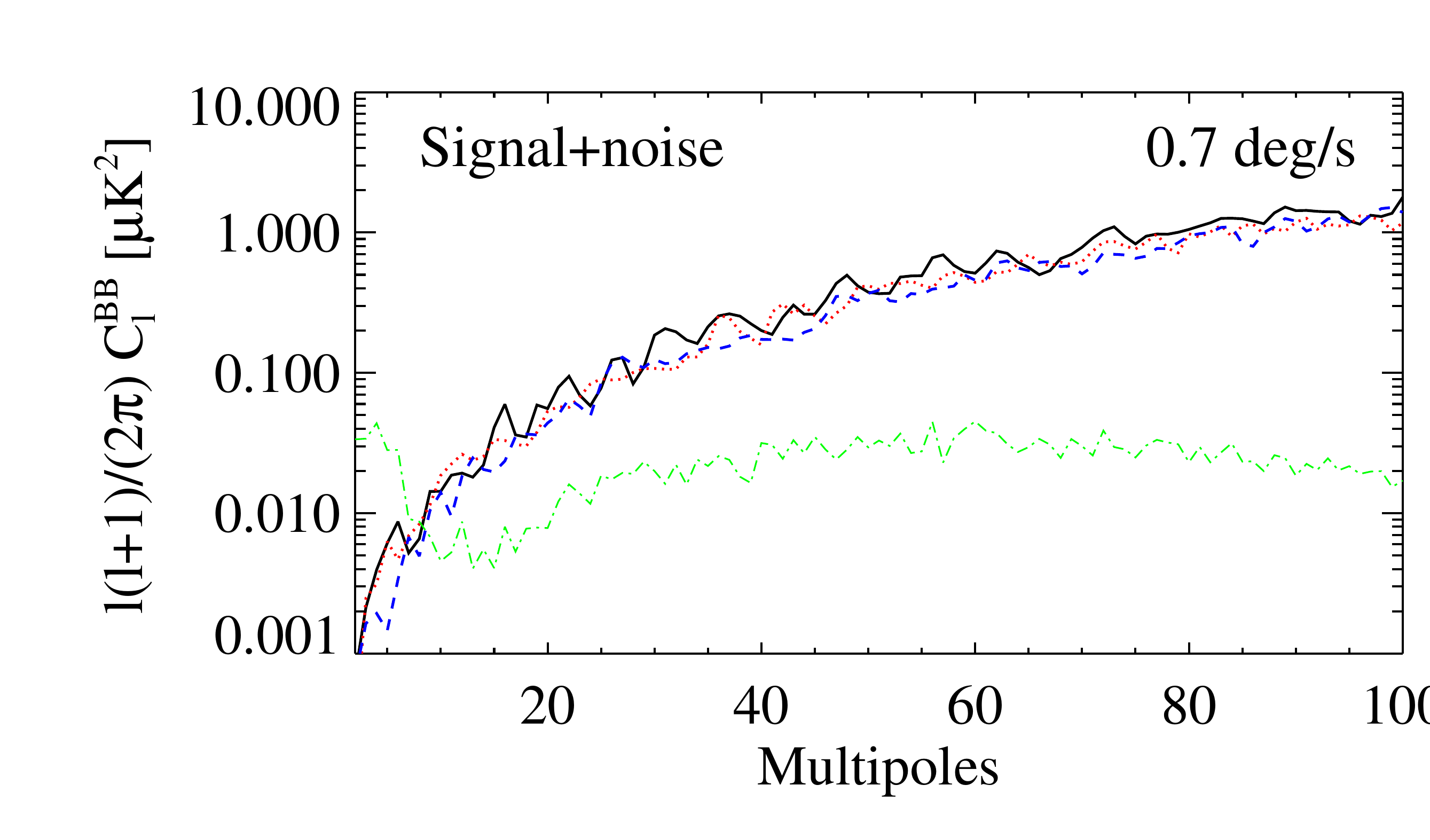}
  \caption{BB angular power spectra from signal-only (S, first two rows) and signal plus noise (SN, last two rows) residual maps for the HWP setups under consideration. First column: HWP stepped every $1~\mathrm{s}$ (in solid black line), $60~\mathrm{s}$ (in dotted red line) and $3600~\mathrm{s}$ (in dashed blue line), for a telescope scanning at $12~\mathrm{deg/s}$ and $0.7~\mathrm{deg/s}$. Second column: HWP spinning at $5~\mathrm{Hz}$ (in solid black line), $2~\mathrm{Hz}$ (in dotted red line) and $0.5~\mathrm{Hz}$ (in dashed blue line), for a telescope scanning at $12~\mathrm{deg/s}$ and $0.7~\mathrm{deg/s}$. The BB spectrum from the input map is shown for comparison in green dot-dashed line (in the signal-only case the spectrum is rescaled down by a factor $10^4$). The assumed noise level corresponds to 5 days and 18 detectors.}\label{fig:BB_res_N}
\end{figure*}
\subsection{Angular power spectra}\label{BB_ps}
\begin{figure*}
\psfrag{Standard deviations}[c][][3]{Standard deviations $[\si{\micro  \kelvin \squared}]$}
\hskip 3.5em  \includegraphics[angle=90, width=0.39\textwidth]{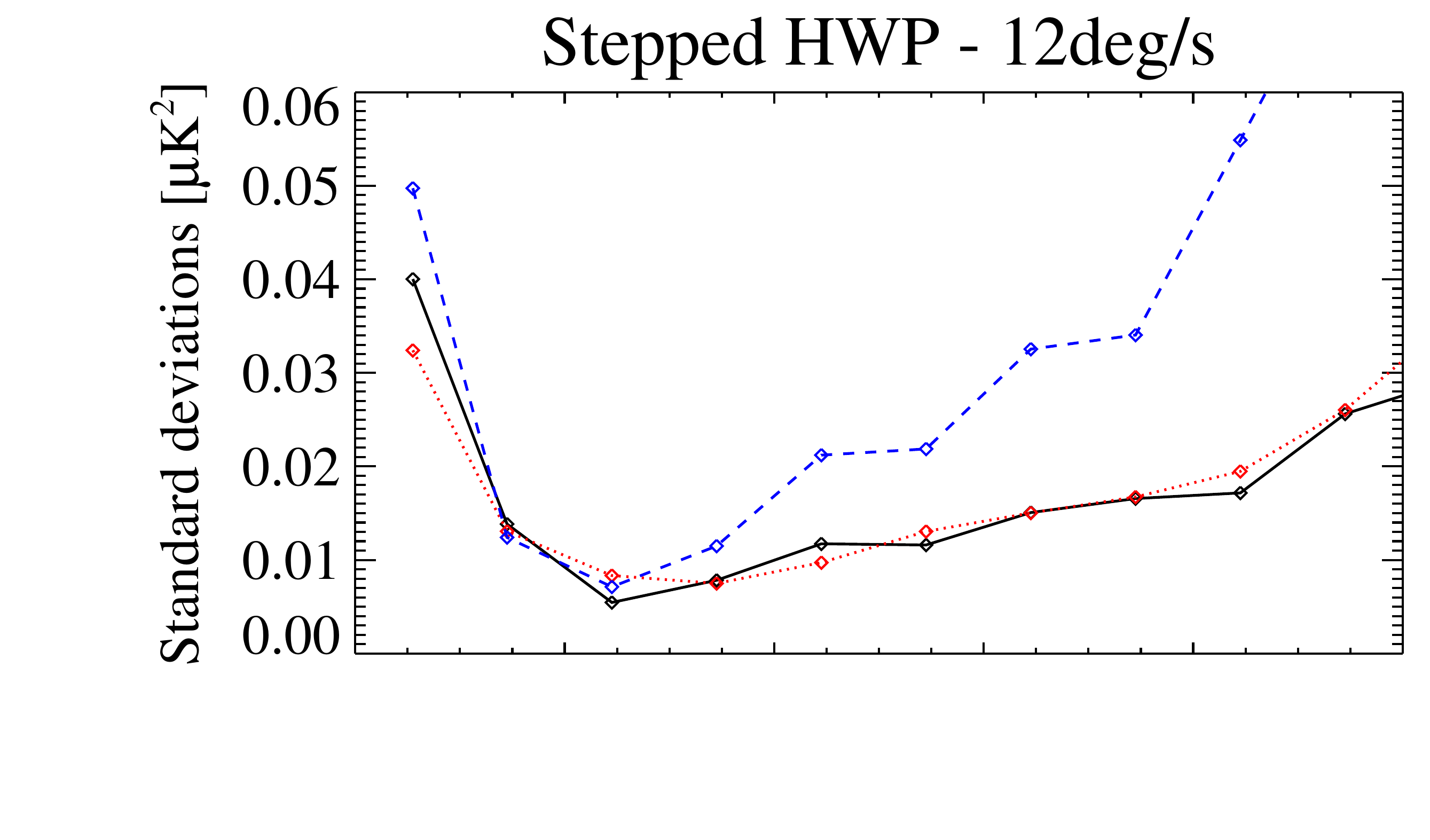}
\hskip 3em   \includegraphics[angle=90, width=0.39\textwidth]{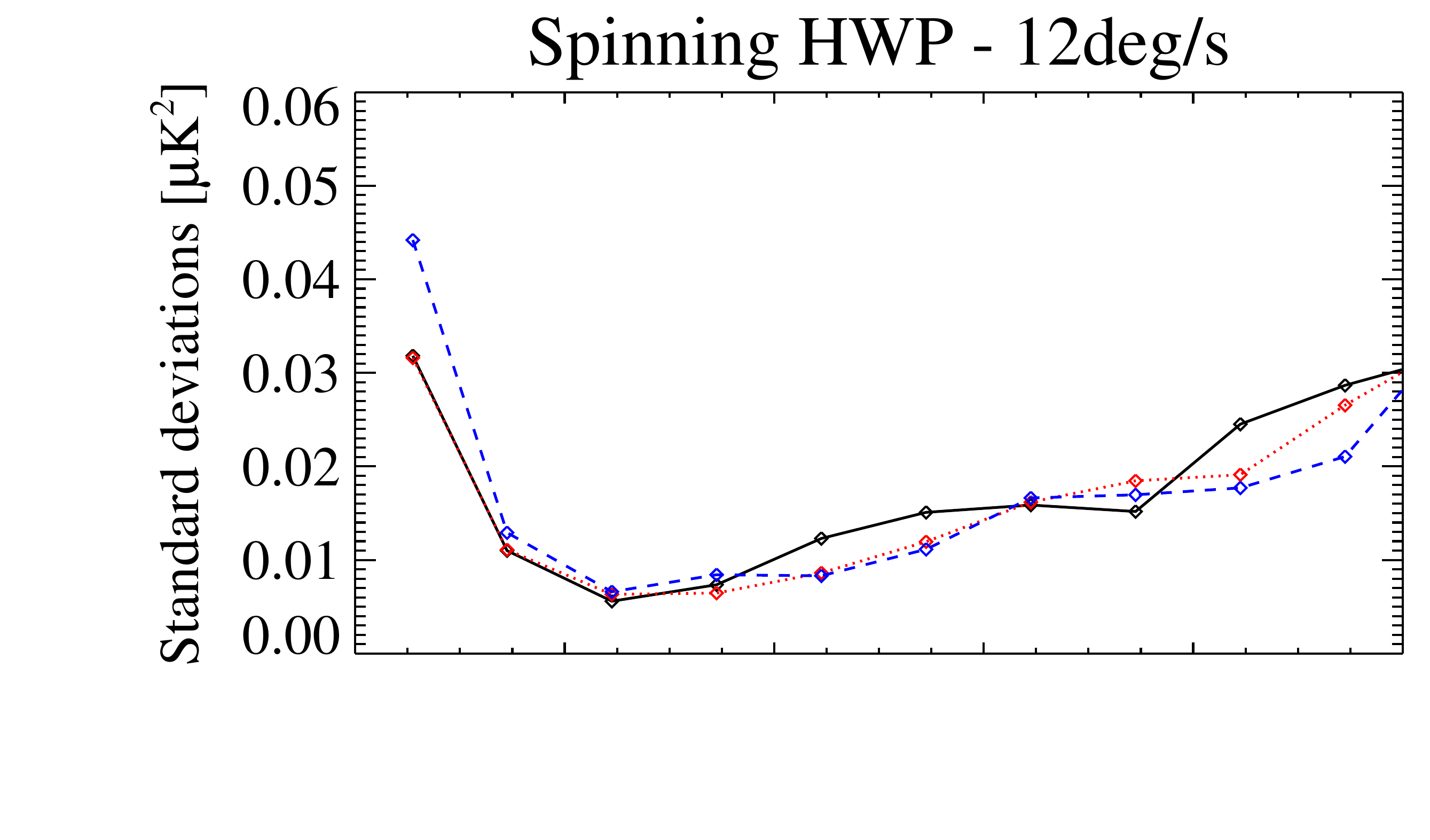}\\ \vskip -4.5em
\hskip 3.5em   \includegraphics[angle=90, width=0.39\textwidth]{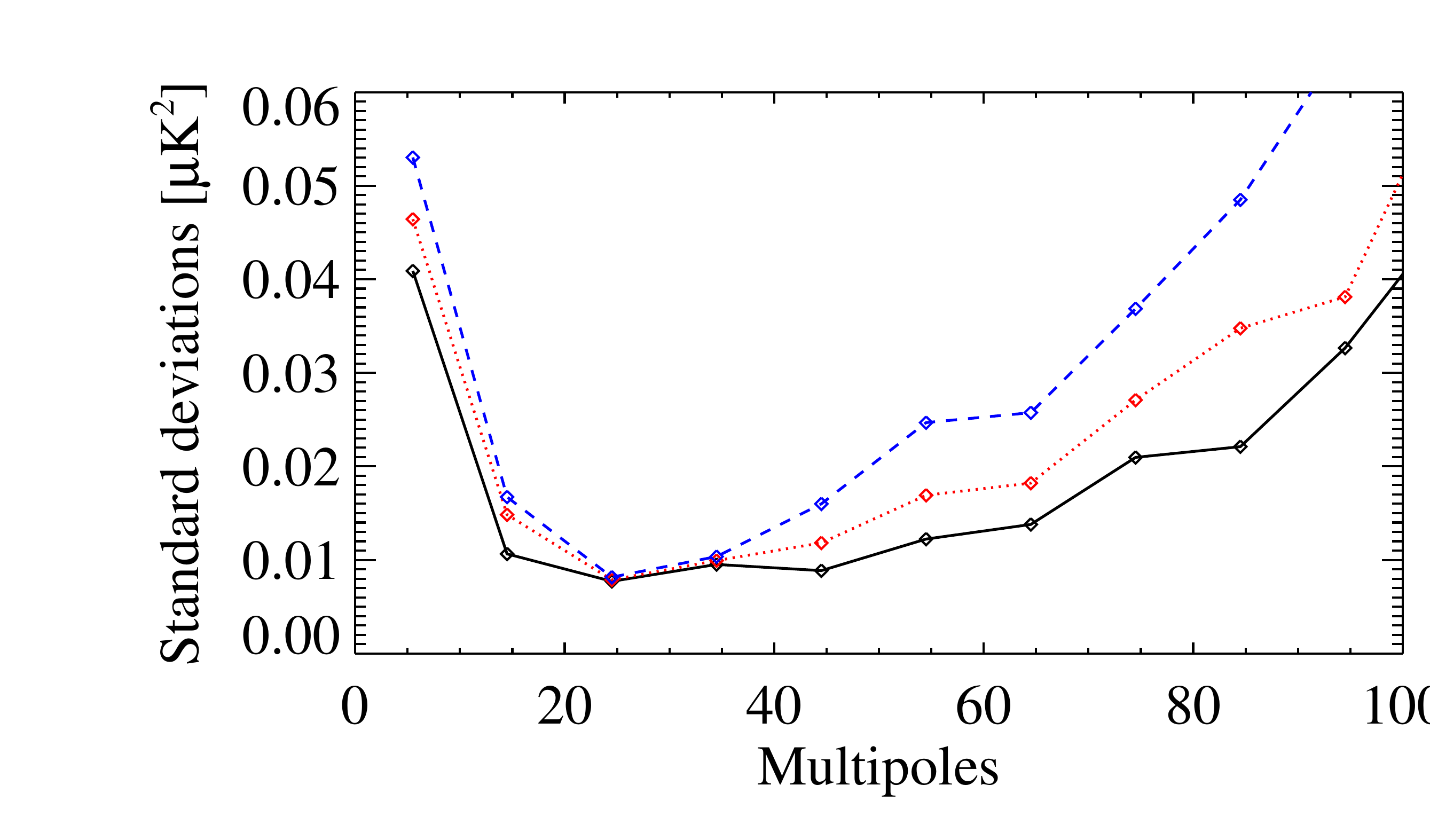}
\hskip 3em   \includegraphics[angle=90, width=0.39\textwidth]{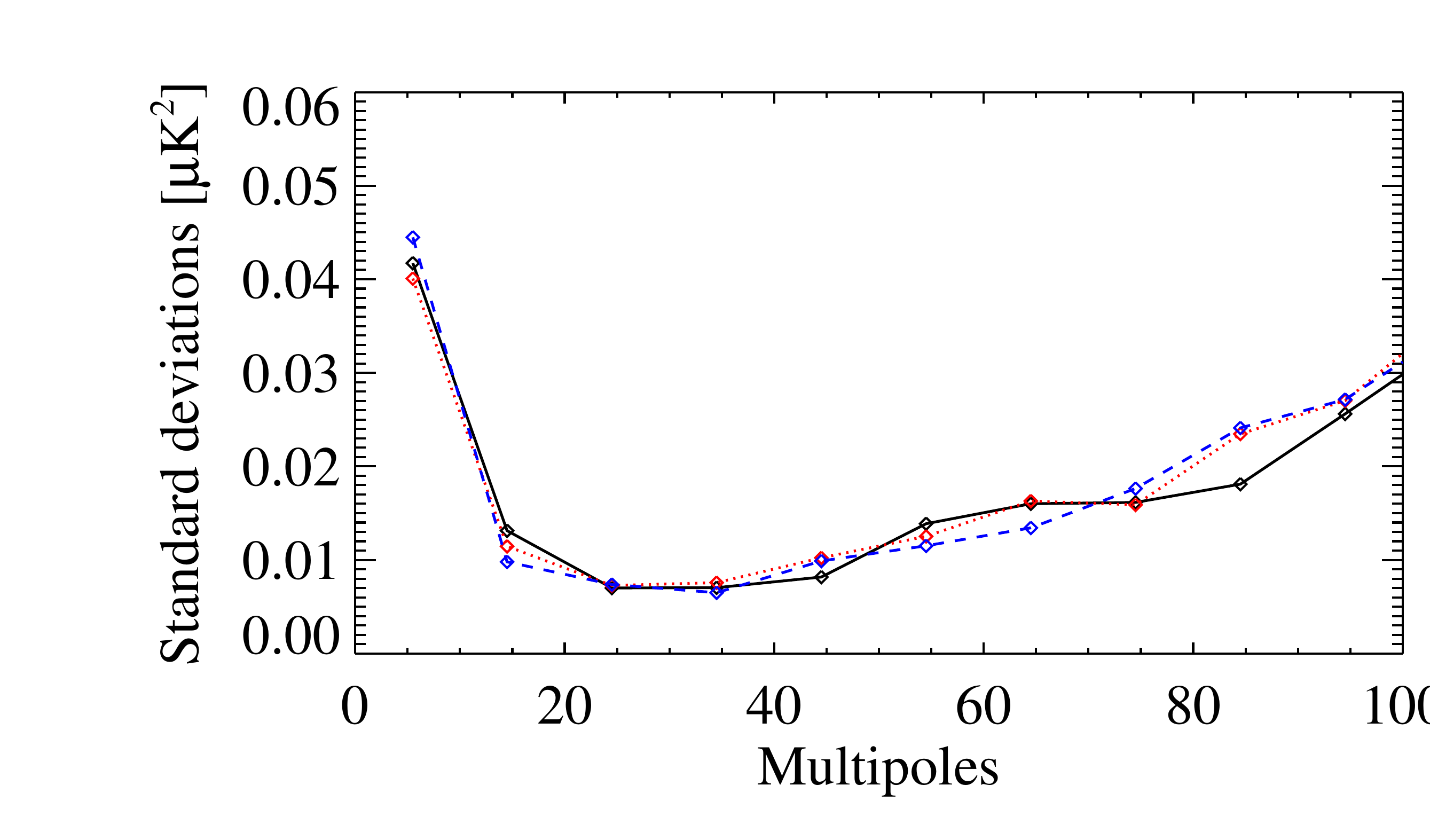}
  \caption{Comparison of BB angular power spectrum error bars (in $\mu K^2$) for the HWP setups under consideration. The spectra have been estimated by an implementation of the MASTER pseudo-$C_l$ method from 50 signal-only, noise-only and signal plus noise Monte Carlo maps. The error bars correspond to the standard deviation of the realizations (i.e. the dispersion of the simulations). The spectra have been produced assuming a couple of detectors for one day of operation.  Note that the noise-only maps have been rescaled by a factor $\sim28.7$, as (optimistically) expected by extending the observation to 110 detectors and 15 days. First column: HWP stepped every $1~\mathrm{s}$ (in solid black line), $60~\mathrm{s}$ (in dotted red line) and $3600~\mathrm{s}$ (in dashed blue line), for a telescope scanning at $12~\mathrm{deg/s}$ (top) and $0.7~\mathrm{deg/s}$ (bottom). Second column: HWP spinning at $5~\mathrm{Hz}$ (in solid black line), $2~\mathrm{Hz}$ (in dotted red line) and $0.5~\mathrm{Hz}$ (in dashed blue line), for a telescope scanning at $12~\mathrm{deg/s}$ (top) and $0.7~\mathrm{deg/s}$ (bottom).}\label{fig:devstand_BB}
\end{figure*}
We now extend to the power spectrum domain the results derived at the map-making stage. 

We generate 50 noise-only, signal-only and signal plus noise Monte Carlo maps for each HWP scheme under consideration and estimate the angular power spectra following the MASTER pseudo-$C_l$ estimator method.
To reduce the computational requirements, the maps are produced assuming a couple of detectors for one day of operation (at telescope elevation of $35~\mathrm{deg}$). 

Notice that the noise-only maps have been rescaled by a factor $\sim 28.7$, as (optimistically) expected by extending the observation to 110 detectors and 15 days

In Fig.~\ref{fig:devstand_BB} we compare the full (i.e. cosmic variance plus noise) BB spectrum error bars for the HWP configuration under analysis.

We find that, in agreement with the residual analysis, a slow stepped HWP provides larger spectrum standard deviations at any telescope rotation rate, and that the performance of a stepped configuration worsens by slowing down the gondola scan-speed. 

We stress that the low number of Monte Carlo maps considered here impacts on the accuracy of the spectrum error bars at the very large angular scales. The optimal estimate of the low-multipole power spectra is beyond the scope of this paper and will be addressed in a forthcoming work.

\subsection{High-pass filters}\label{filter}
\begin{figure*}
\psfrag{f_cut [Hz]}[c][][3]{$f_{cut}$ $[\si{\hertz}]$}
\hskip 3.5em   \includegraphics[angle=90, width=0.39\textwidth]{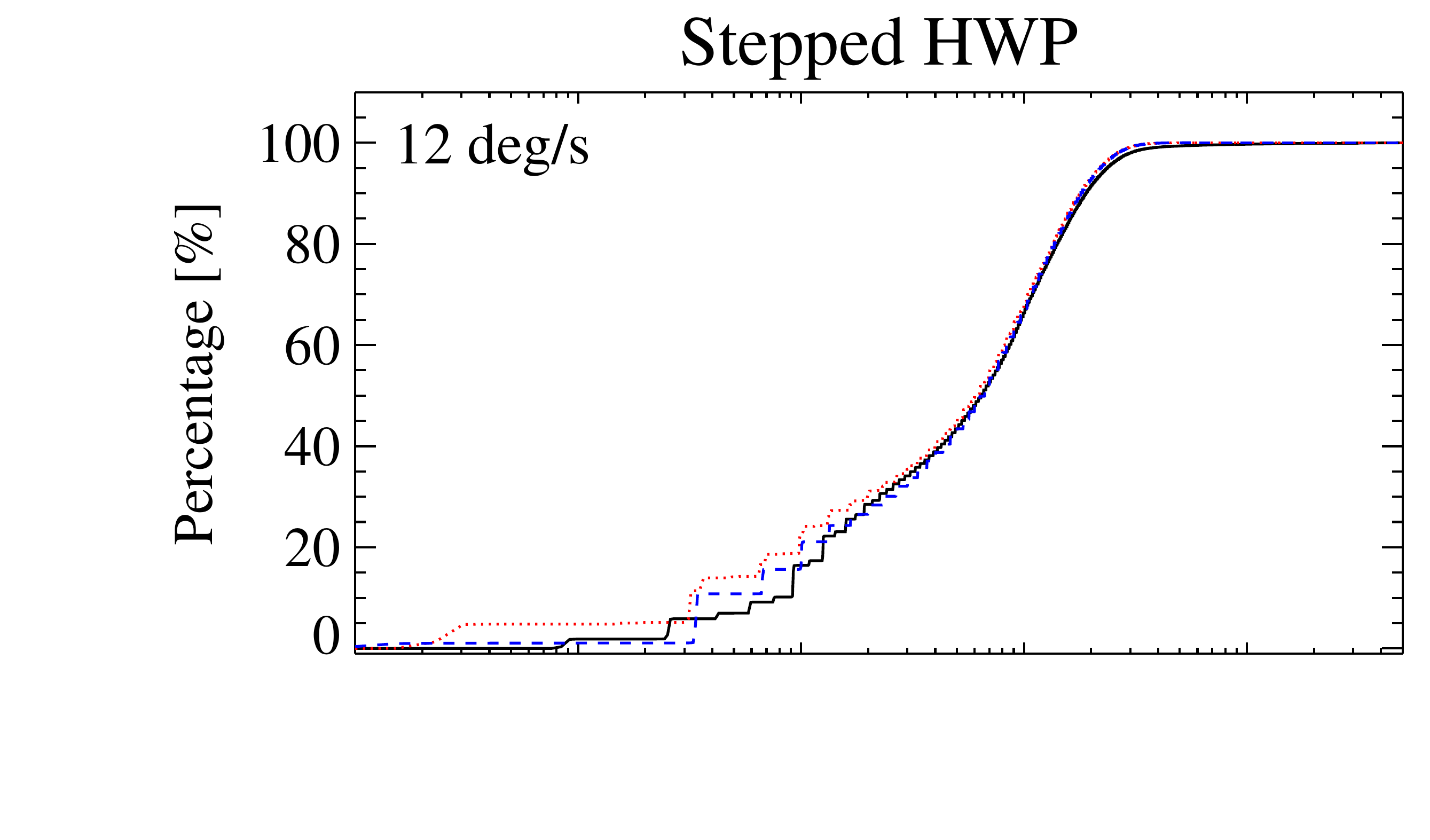}
\hskip 3em   \includegraphics[angle=90, width=0.39\textwidth]{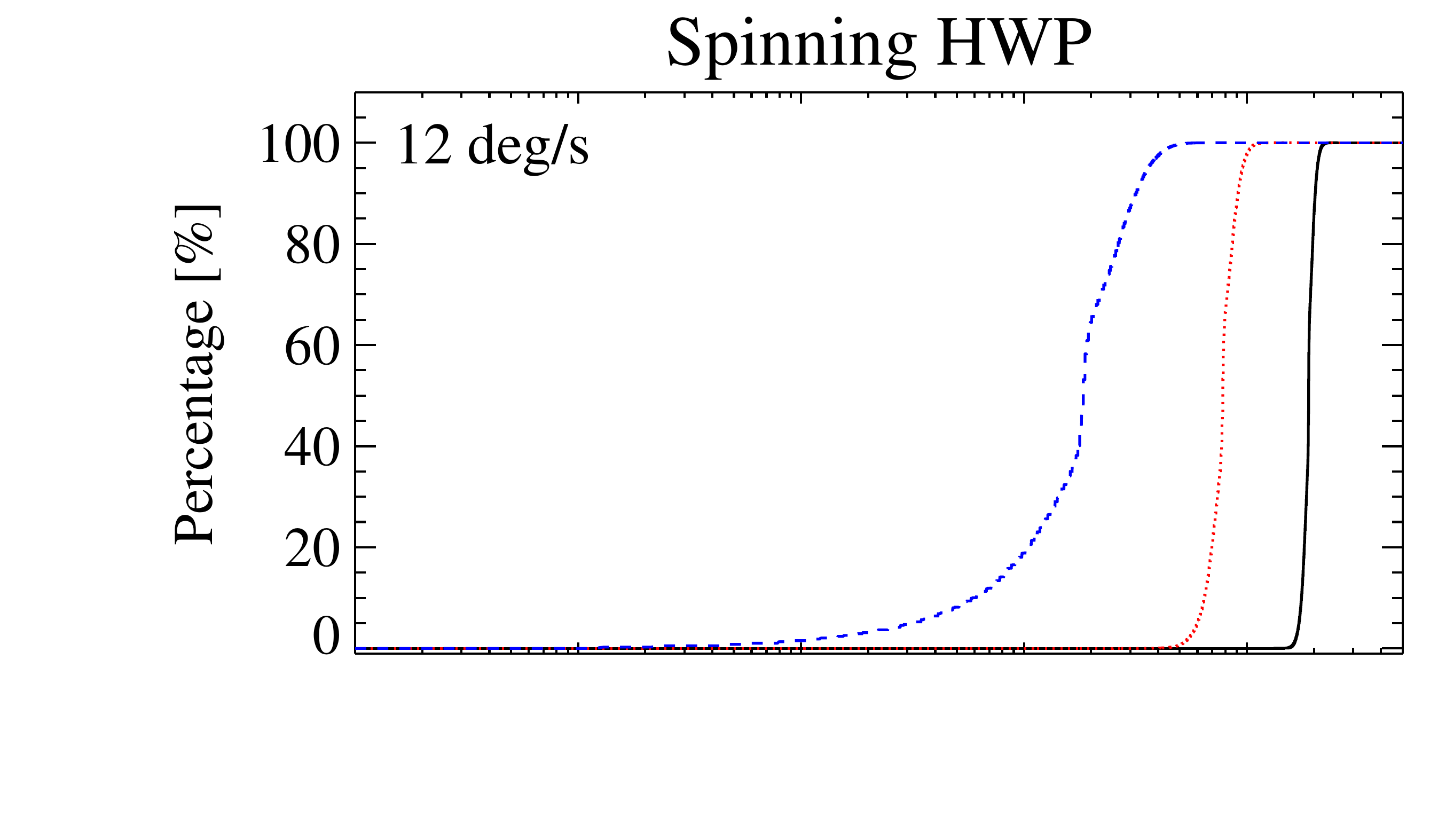}\\ \vskip -4.5em
\hskip 3.5em   \includegraphics[angle=90, width=0.39\textwidth]{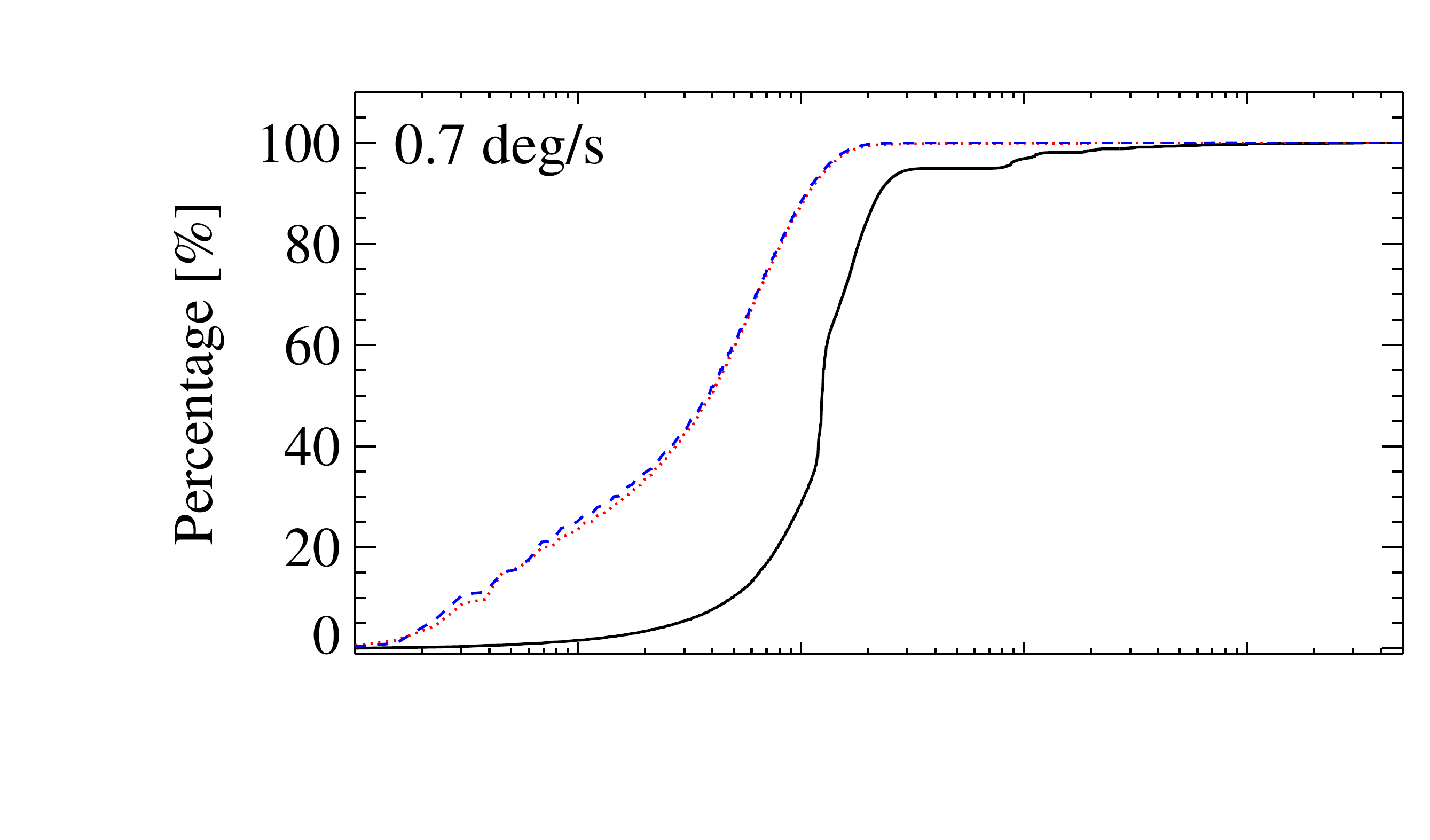}
\hskip 3em   \includegraphics[angle=90, width=0.39\textwidth]{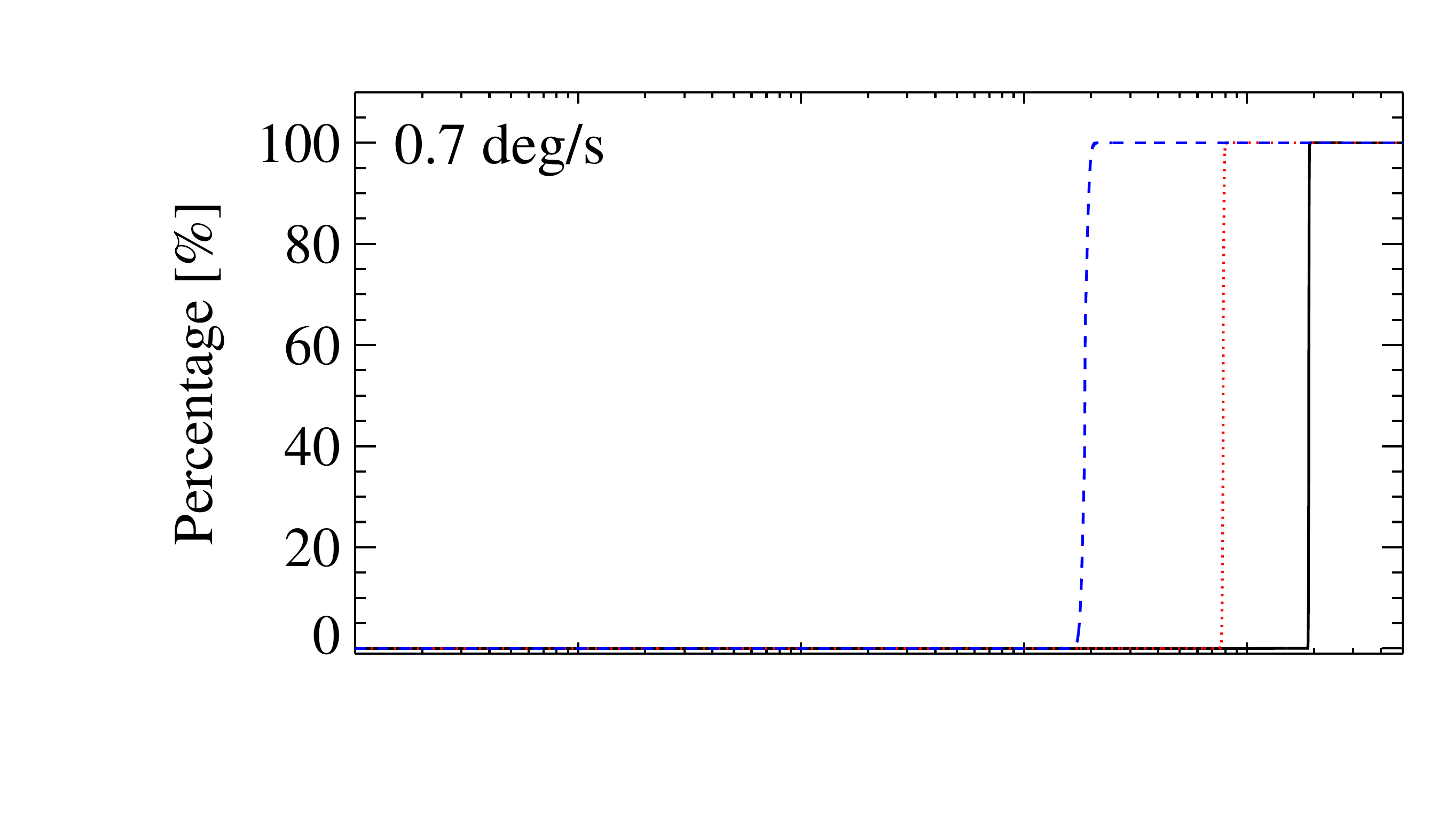}\\ \vskip -4.5em
\hskip 3.5em   \includegraphics[angle=90, width=0.39\textwidth]{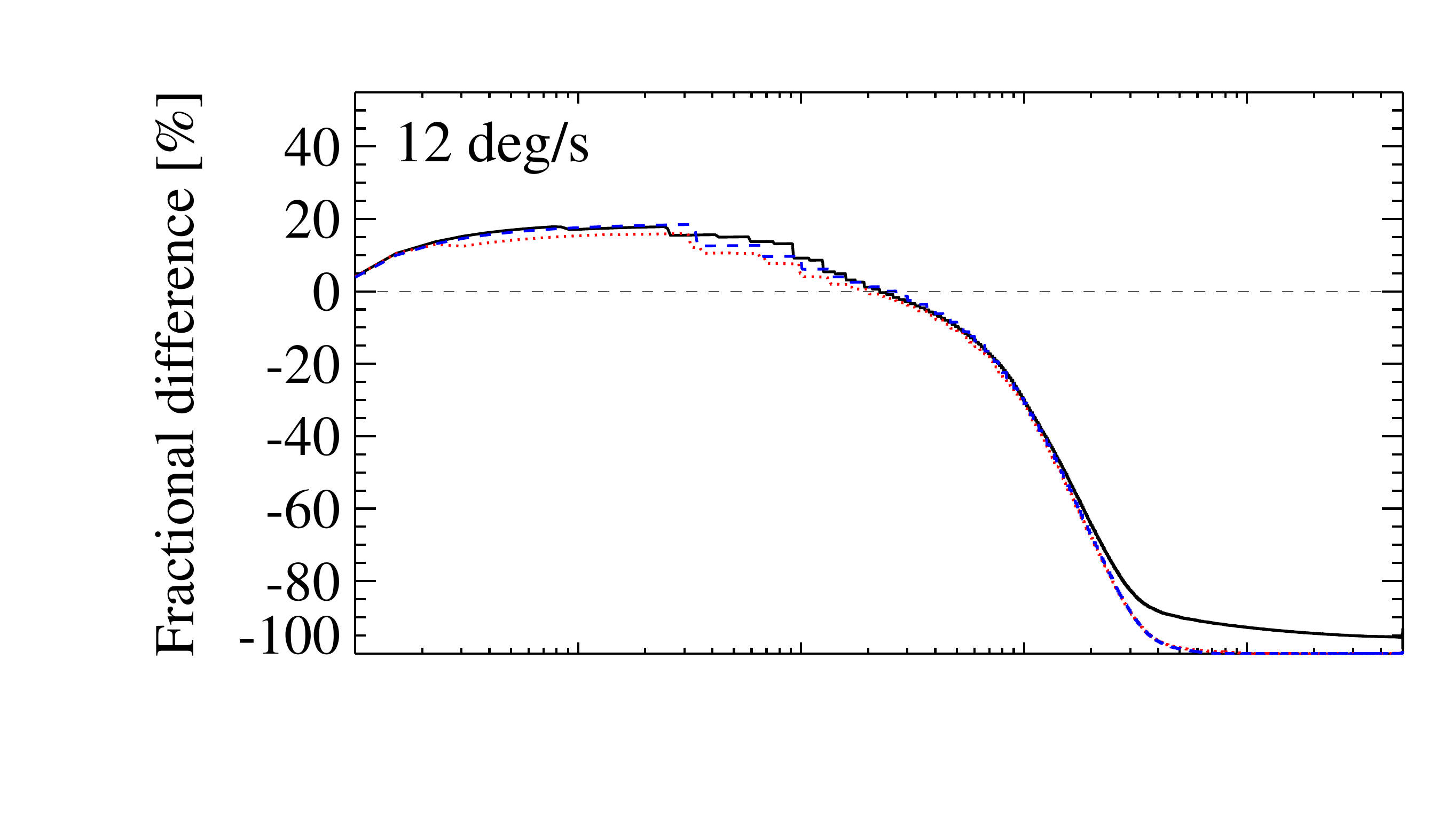}
\hskip 3em   \includegraphics[angle=90, width=0.39\textwidth]{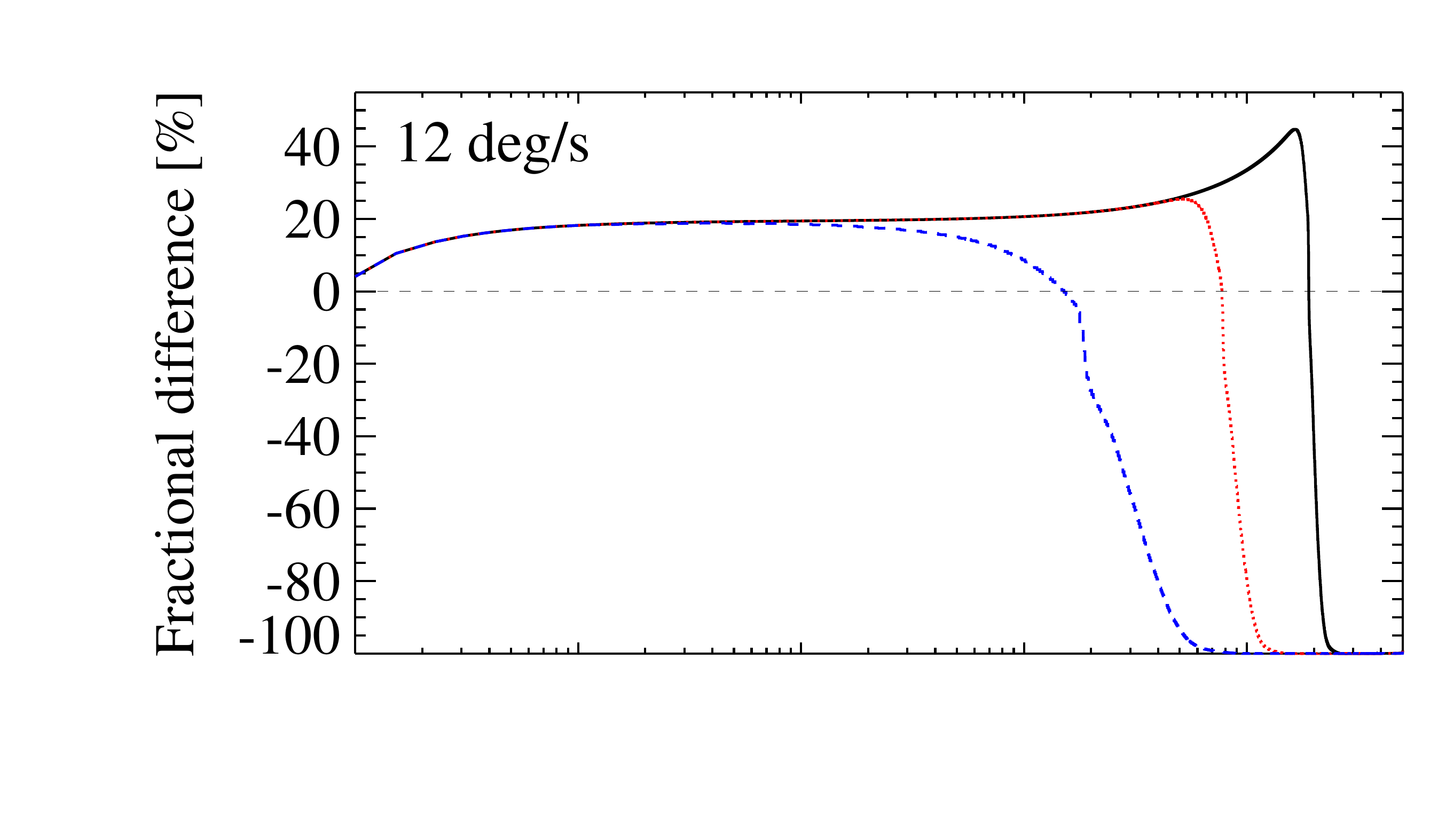}\\ \vskip -4.5em
\hskip 3.5em   \includegraphics[angle=90, width=0.39\textwidth]{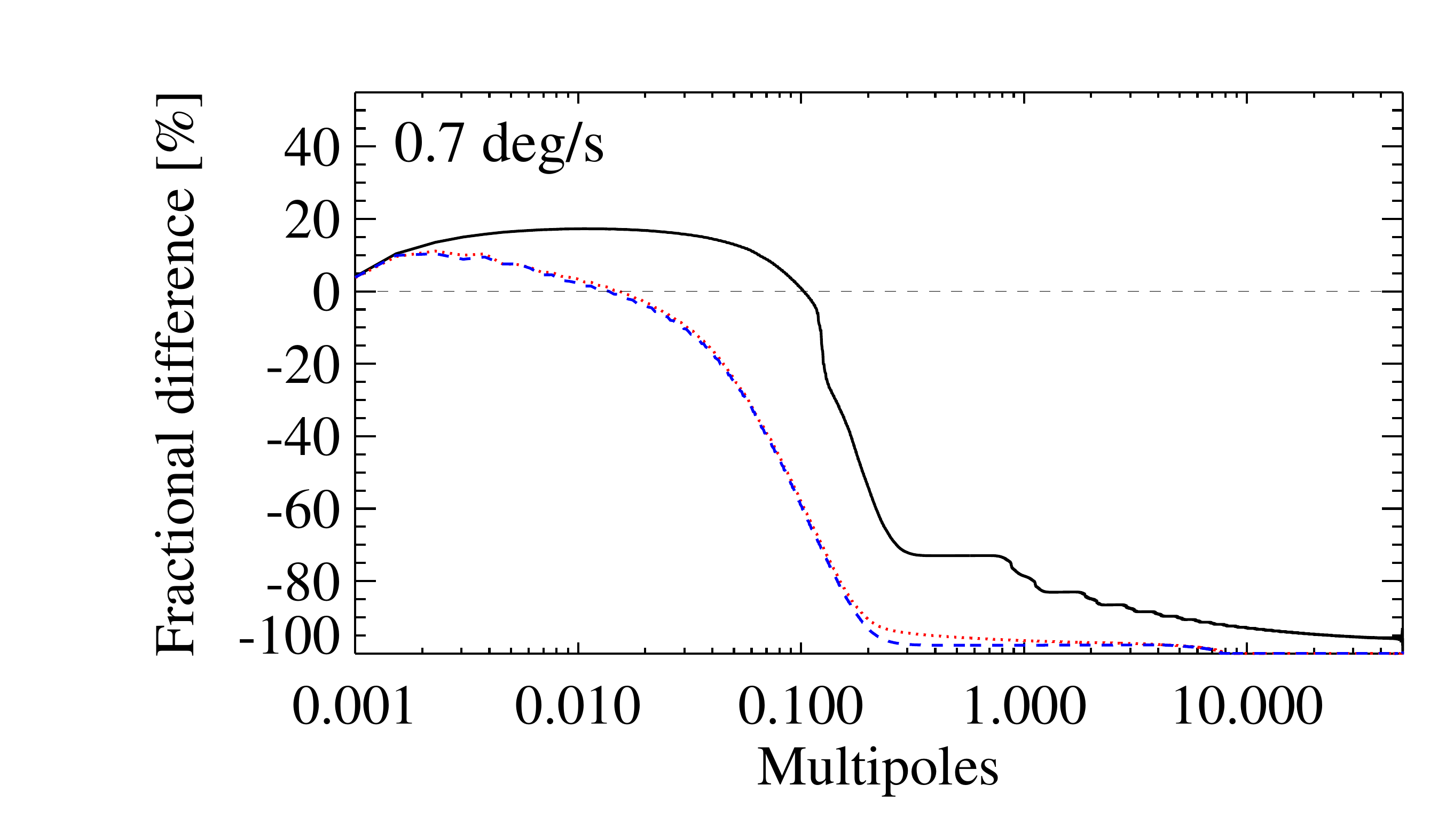}
\hskip 3em   \includegraphics[angle=90, width=0.39\textwidth]{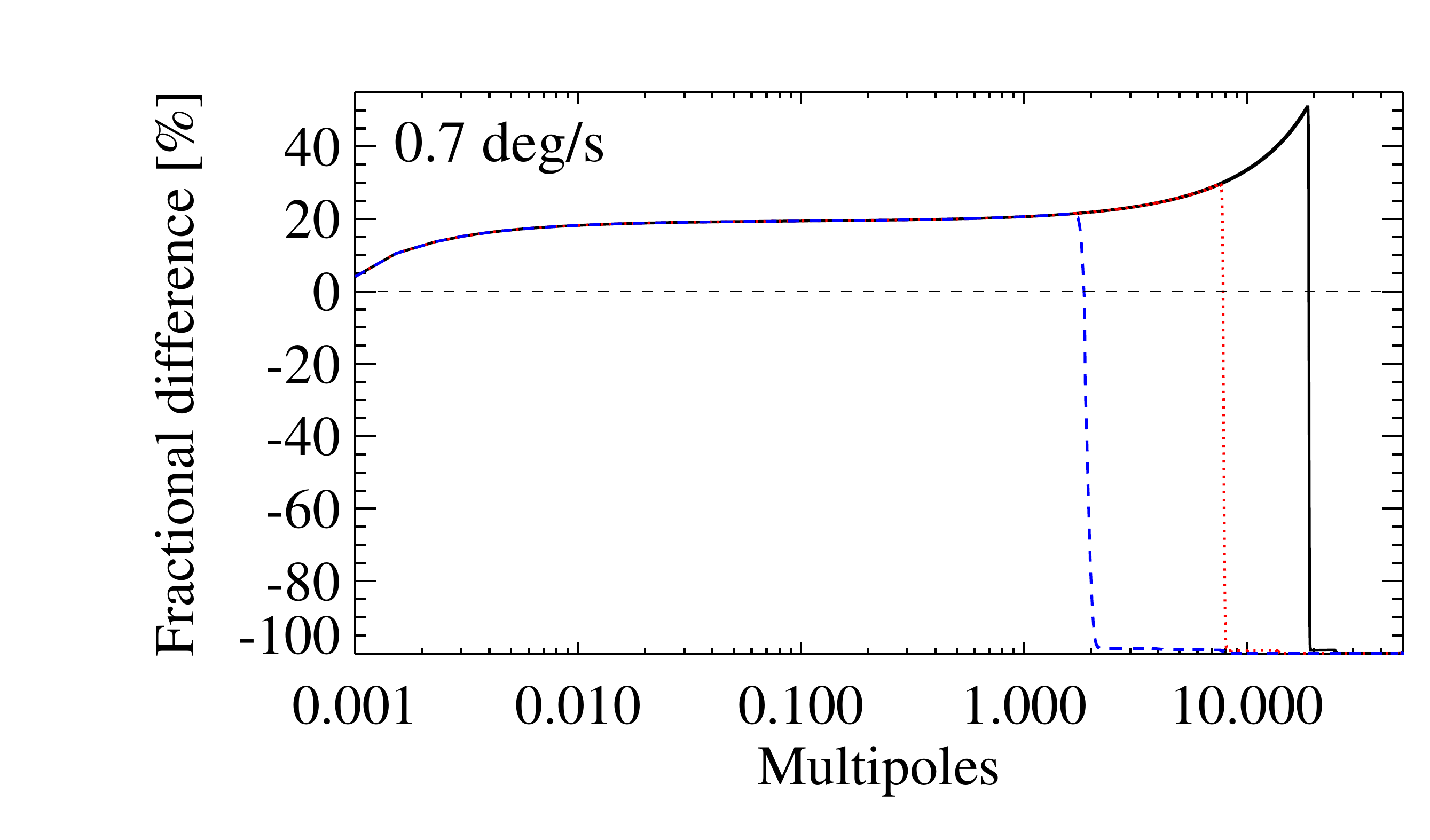}
  \caption{First two rows: percentage of polarization intensity power lost when different high-pass filters are applied to the TOD as function of the cut-on frequency $f_c$. Second two rows: fractional difference of the polarization signal-to-noise ratio (SNR) between the cases with and without high-pass filters as function of $f_c$. The fractional difference is calculated as $(SNR_{fil}-SNR_{tot})\times100/SNR_{tot}$, where $SNR_{fil}$ and $SNR_{tot}$ refer to the cases with and without filters. Note that the polarization signal intensity is extracted directly from the plots of Fig.~\ref{fig:periodo}. The cut-on frequency $f_c$ is set to vary from $0.001~\mathrm{Hz}$ to $50~\mathrm{Hz}$. First column: HWP stepping every $1~\mathrm{s}$ (in solid black line), $60~\mathrm{s}$ (in dotted red line) and $3600~\mathrm{s}$ (in dashed blue line), for a telescope scanning at $12~\mathrm{deg/s}$ and $0.7~\mathrm{deg/s}$. Second column: HWP spinning at $5~\mathrm{Hz}$ (in solid black line), $2~\mathrm{Hz}$ (in dotted red line) and $0.5~\mathrm{Hz}$ (in dashed blue line), for a telescope scanning at $12~\mathrm{deg/s}$ and $0.7~\mathrm{deg/s}$.} \label{fig:filter_perc_SN}
\end{figure*}

A very important issue is to test the HWP configurations against a possible filtering of the low-frequency data streams, which is a common practice to cut the $1/f$ part of the noise spectrum in order to maximize the signal-to-noise ratio (SNR). It is beyond the aim of this work to forecast a filtering strategy for the SWIPE experiment, nonetheless it is interesting to investigate how the polarization modulation could impact on the choice of a suitable data stream filter.

In choosing a proper filter, two parameters must be accounted for: the fraction of signal lost and the increase in the SNR.

In Fig.~\ref{fig:filter_perc_SN} we show the percentage of polarization intensity power that would be cut out from different high-pass filters, for the HWP schemes under consideration. The cut-on frequency $f_c$ is set to vary from $10^{-3}~\mathrm{Hz}$ to $50~\mathrm{Hz}$. It is also shown the fractional difference of the (polarization) SNR between the cases with and without high-pass filters. 

Due to the shift of the polarization signal to higher frequencies, a spinning HWP allows clearly much more freedom in the choice of the cut-on frequency: $f_c$ can be freely chosen sufficiently below the peak frequency with no loss in cosmological signal.
For instance, we find that, in order to preserve at least the $90\%$ of polarization signal, the maximum allowed cut-on frequency for the nominal configuration is around $f_{c}=0.03~\mathrm{Hz}$, while for a fast spinning scheme ($5~\mathrm{Hz}$) almost $f_{c}=20~\mathrm{Hz}$. 

This implies that, for the nominal modulation setup, the increase in SNR with respect to the no filter case cannot be more than a factor 20\%, while with a fast spinning HWP we can gain a factor 40-50\%.

We finally quantified the impact of a typical high-pass filter ($f_{c}=0.1~\mathrm{Hz}$; see, e.g., \citealt{2006A&A...458..687M}) on the polarization map recovery. We generate maps from the filtered TOD and calculate the residuals (for both the S and SN cases) and the corresponding BB angular power spectra. 
As expected, the spinning mode is not generally affected by the presence of this specific filter. On the contrary, the impact on the stepped HWP setups is very large: we find a raise in the BB residual spectrum up to six and three orders of magnitude for the S and SN cases, respectively. This increase is generally more prominent at large scales.

\subsection{Band-pass filters}\label{filter:bandpass}
\begin{figure*}
\psfrag{delta f [Hz]}[c][][3]{$\Delta f$ $[\si{\hertz}]$}
\hskip 3.5em  \includegraphics[angle=90, width=0.39\textwidth]{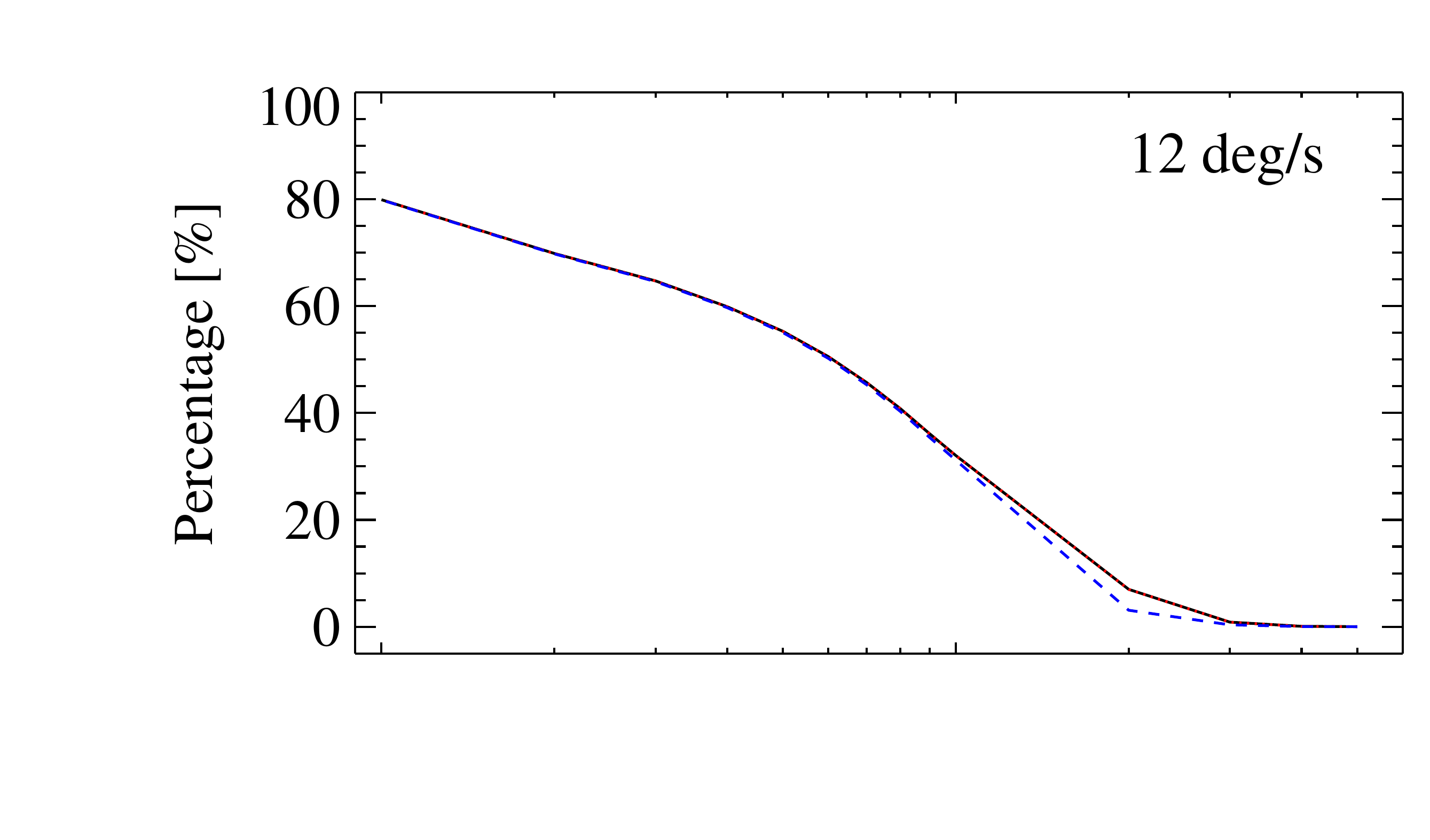}
\hskip 3em  \includegraphics[angle=90, width=0.39\textwidth]{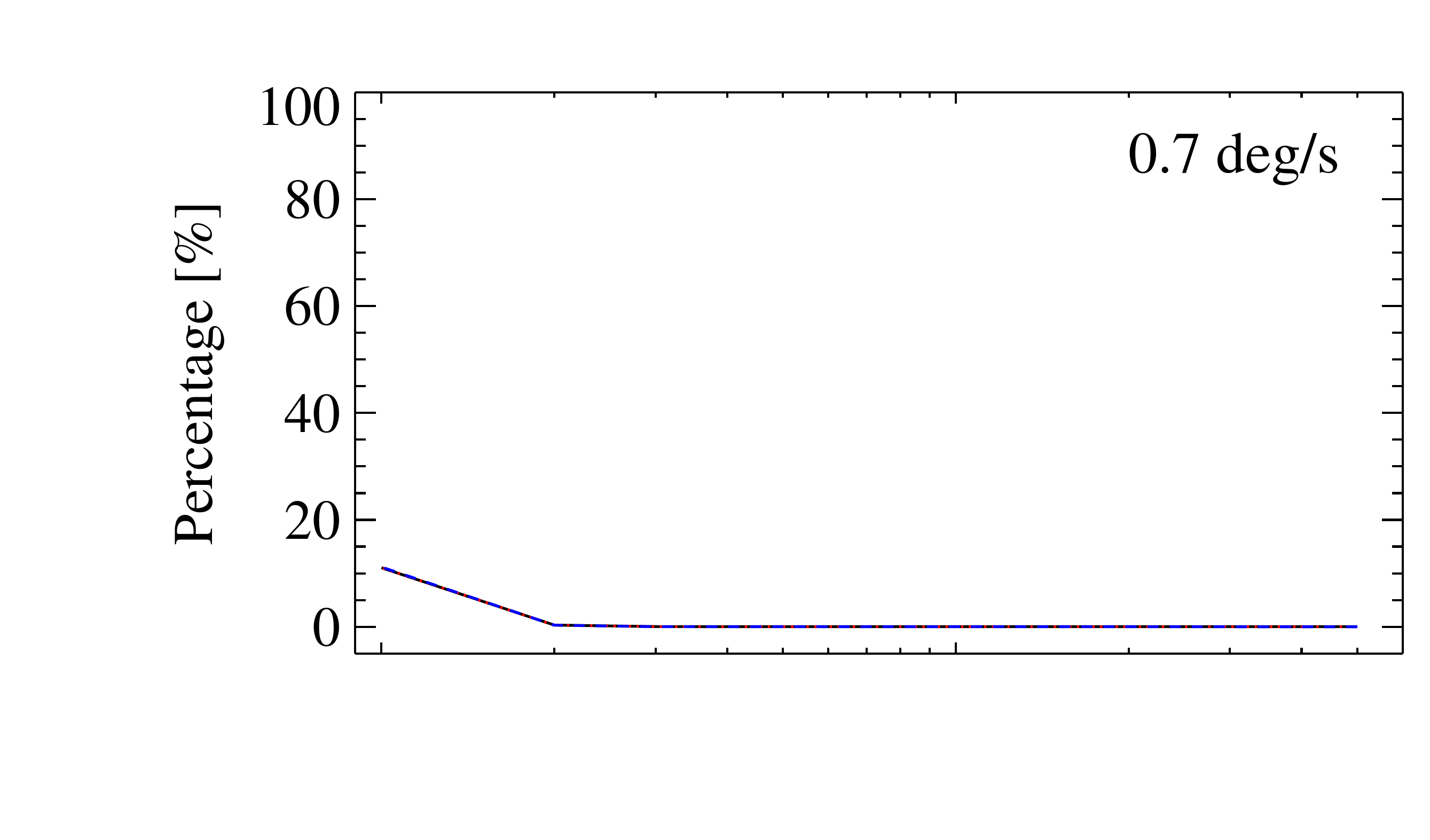}\\ \vskip -4.5em
\hskip 3.5em  \includegraphics[angle=90, width=0.39\textwidth]{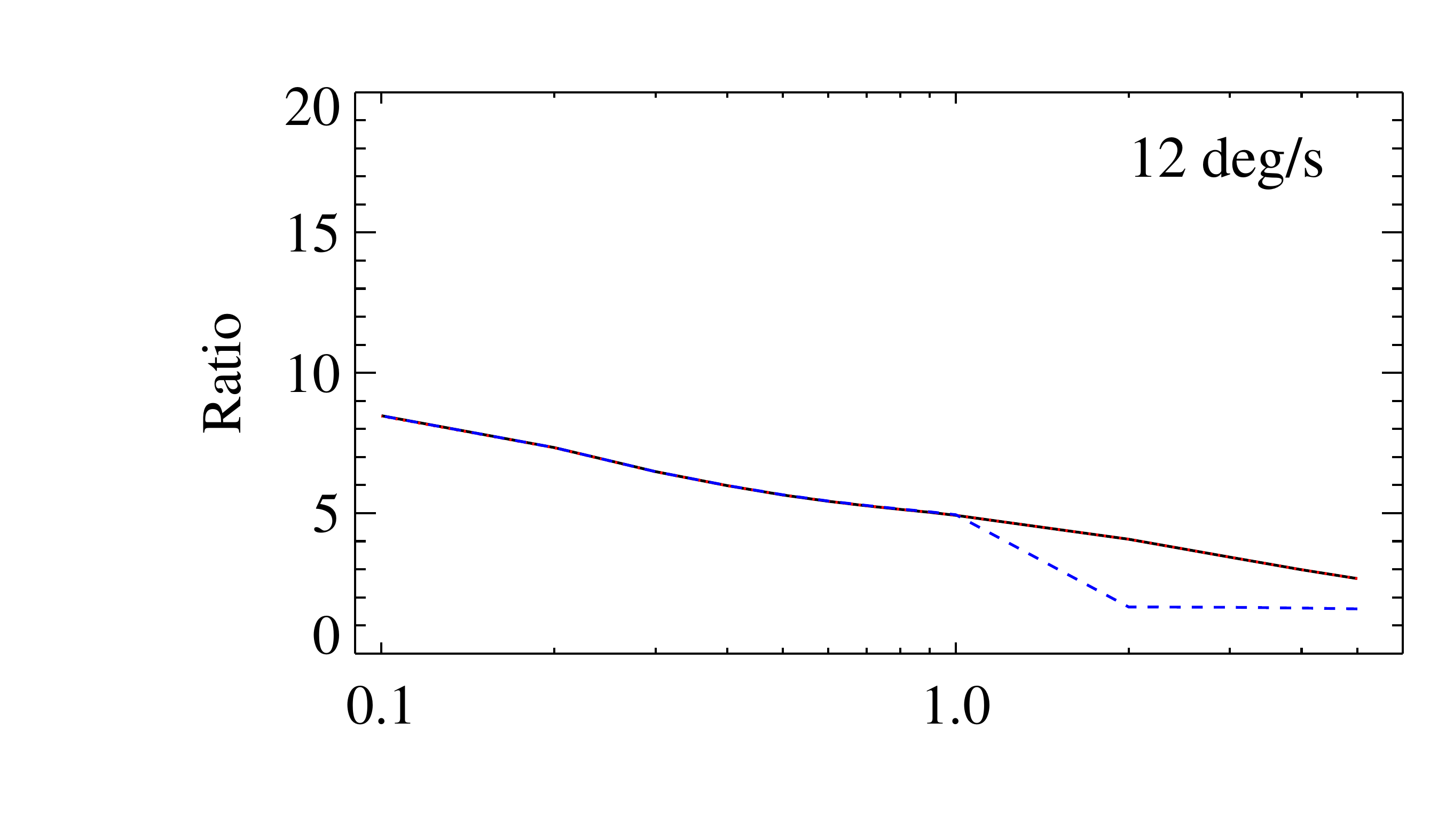}
\hskip 3em  \includegraphics[angle=90, width=0.39\textwidth]{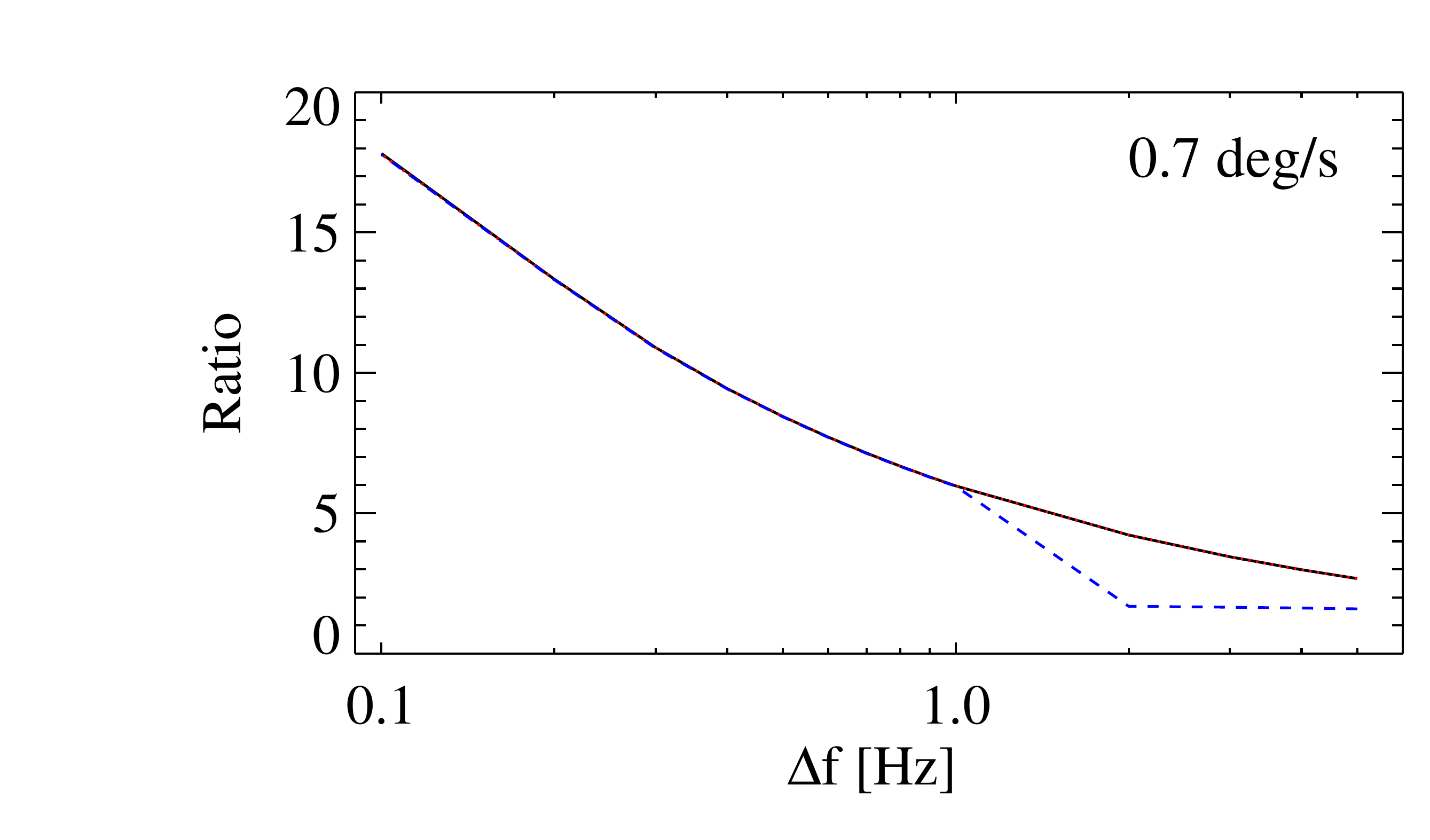}
  \caption{Percentage of polarization intensity power lost when different band-pass filters are applied (first row) and ratios of the polarization signal-to-noise ratio (SNR) between the cases with and without band-pass filters (second row). The bandwidth of the filter $\Delta f=(f_{max}-f_{min})/2$ is set to vary from $0.1~\mathrm{Hz}$ to $5~\mathrm{Hz}$ around the peak frequency (four times the rotation frequency).
  Plots are for a HWP spinning at $5~\mathrm{Hz}$ (in solid black line), $2~\mathrm{Hz}$ (in dotted red line) and $0.5~\mathrm{Hz}$ (in dashed blue line), for a telescope scanning at $12~\mathrm{deg/s}$ and $0.7~\mathrm{deg/s}$.}\label{fig:bandpass}
\end{figure*}
A spinning HWP scheme allows for the interesting possibility of using a band-pass filter. The frequency band is centered around the peak frequency (see Fig.~\ref{fig:periodo}), i.e. at $4f_r$, where $f_r$ is the spinning frequency.

We explore the effect of a set of band-pass filter by calculating the percentage of polarization intensity power lost and the associated SNR as function of the filter bandwidth $\Delta f=(f_{max}-f_{min})/2$, where $\Delta f$ is set to vary from $0.1~\mathrm{Hz}$ to $5~\mathrm{Hz}$. Results are shown in Fig.~\ref{fig:bandpass}. In particular we plot the ratio of the SNR between the cases with and without filters.

While we do not find any relevant dependence on the HWP rotation frequency, the telescope scan-speed has a large effect. We notice a higher SNR when slower telescope scan-speeds are performed: this is because, as shown in Fig.~\ref{fig:periodo}, the bandwidths are narrower in this case. 

For instance, if we want to preserve the 90\% of the cosmological signal, we need a bandwidth of at least $\sim 2~\mathrm{Hz}$ and $0.1~\mathrm{Hz}$ for a telescope scanning at $12~\mathrm{deg/s}$ and $0.7~\mathrm{deg/s}$, respectively. The SNR will consequently increase: it will reach a value of $\sim$4 and $\sim$18 times the reference SNR calculated without any filter, respectively.

\subsection{Minimal observation case}\label{calib}
\begin{figure}
\psfrag{Cl BB res}[c][][3]{Residual $C_{\ell}^{BB}$ $[\si{\mu K^2}]$}
\begin{center}
\hskip -2em \includegraphics[angle=90, width=0.39\textwidth]{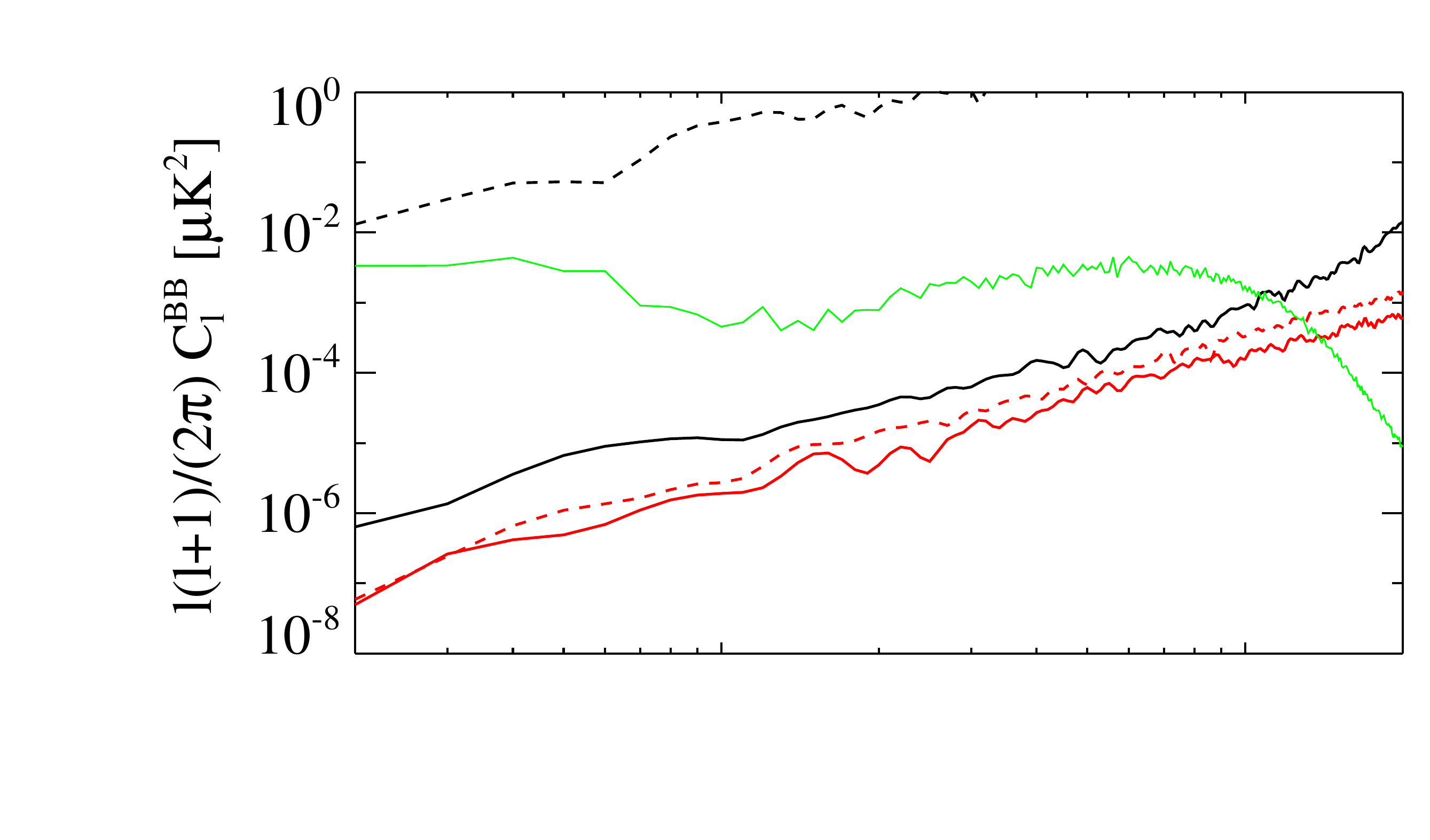}\\ \vskip -3.5em
\hskip -2em \includegraphics[angle=90, width=0.39\textwidth]{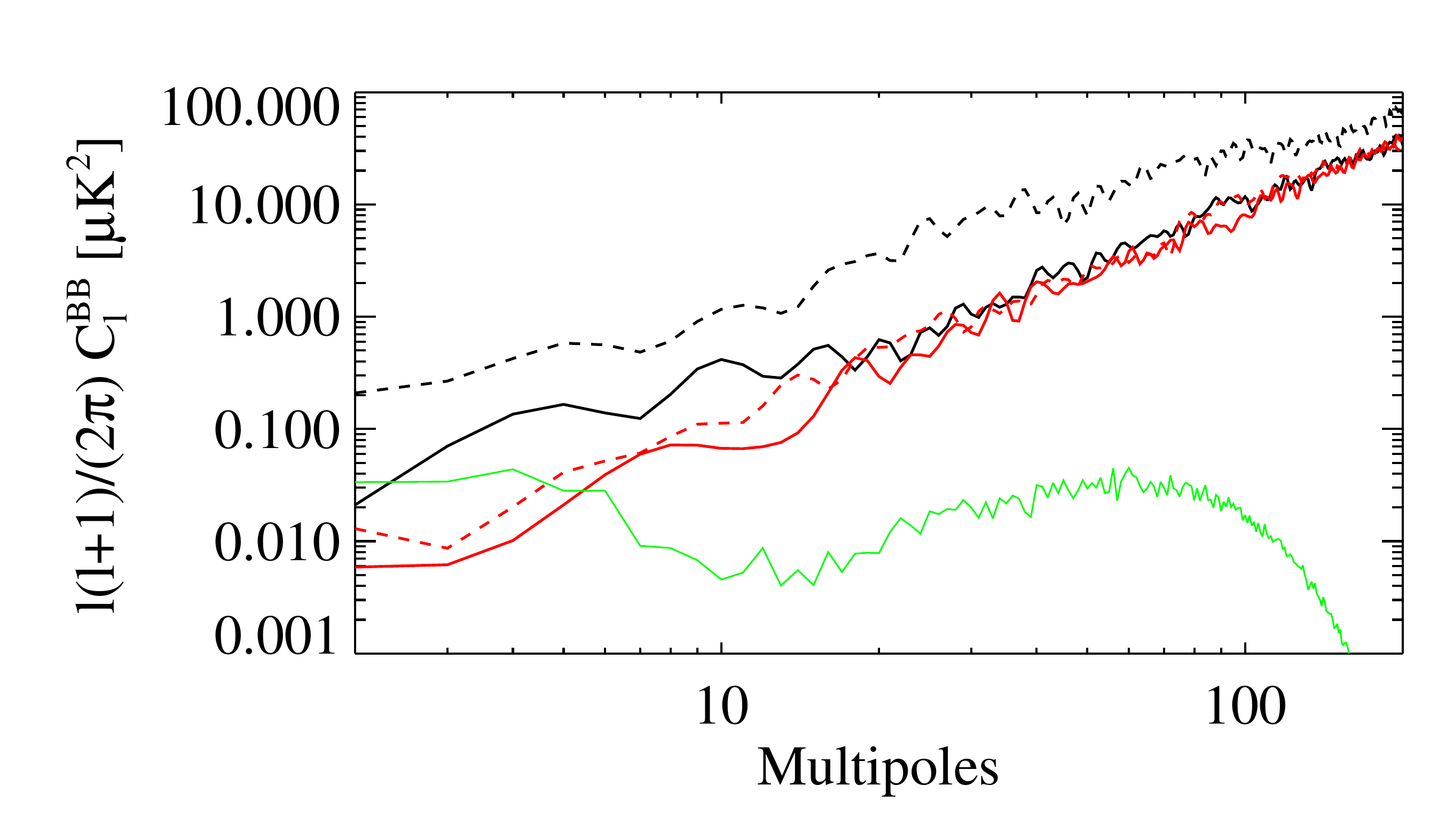}
\end{center}
    \caption{BB angular power spectra from signal-only (on the top) and signal plus noise (on the bottom) residual maps for the minimal case of one detector and one operation day. In solid black line: HWP stepped every $60~\mathrm{s}$ with a telescope scanning rate of $12~\mathrm{deg/s}$; in dashed black line: HWP stepped every $3600~\mathrm{s}$ with a telescope scanning rate of $0.7~\mathrm{deg/s}$; in solid red line: HWP spinning at $5~\mathrm{Hz}$ with a telescope scanning rate of $0.7~\mathrm{deg/s}$; in dashed red line: HWP spinning at $0.5~\mathrm{Hz}$ with a telescope scanning rate of $12~\mathrm{deg/s}$.} The BB spectrum from the input map is shown for comparison in green dot-dashed line (in the signal-only case the spectrum is rescaled down by a factor $10^2$).\label{fig:minimal}
\end{figure}

We compare now the HWP performance in the minimal observation case (one detector for one operation day), which is a crucial test for single-detector calibration purposes.

The basic strategy for LSPE-SWIPE in-flight calibration is described in \citep{2012SPIE.8452E..3FD}. Around one day of the mission will be devoted to calibration scans.

We generate then maps assuming one detector and one day of operation and calculate the residuals. We find that the differences among the various HWP designs are much larger than in the previous two focal plane simulations. 

Results confirm that some stepped HWP configurations are strongly disfavored. For instance, a slowly stepped configuration provides signal-only BB residual spectra with an amplitude orders of magnitude larger than other setups. Furthermore, we find now not negligible differences between fast stepped modes (when combined with a rapid telescope scan-speed) and spinning modes.

As an example, in Fig.~\ref{fig:minimal} we show signal-only and signal plus noise BB residual spectra for the nominal HWP strategy, for two continuously rotating configurations (slowly spinning at $0.5~\mathrm{Hz}$ with a fast gondola rotation of $12~\mathrm{deg/s}$, and rapidly spinning at $5~\mathrm{Hz}$ with a slow telescope scanning rate of $0.7~\mathrm{deg/s}$) and for a slowly stepped HWP ($3600~\mathrm{s}$, combined with a slow gondola scanning frequency of $0.7~\mathrm{deg/s}$).

In the signal-only case, we find that the two spinning modes are slighly more effective in recovering polarization with respect to the nominal configuration at any scale. In particular, the difference increases after $\ell \gtrsim 100$. In the signal plus noise case, we find a large discrepancy between the nominal and the spinning schemes at very low multipoles, while the difference is damped at smaller scales.

\subsection{HWP-induced systematics}
\begin{figure}
\begin{center}
\hskip -2em \includegraphics[angle=90, width=0.39\textwidth]{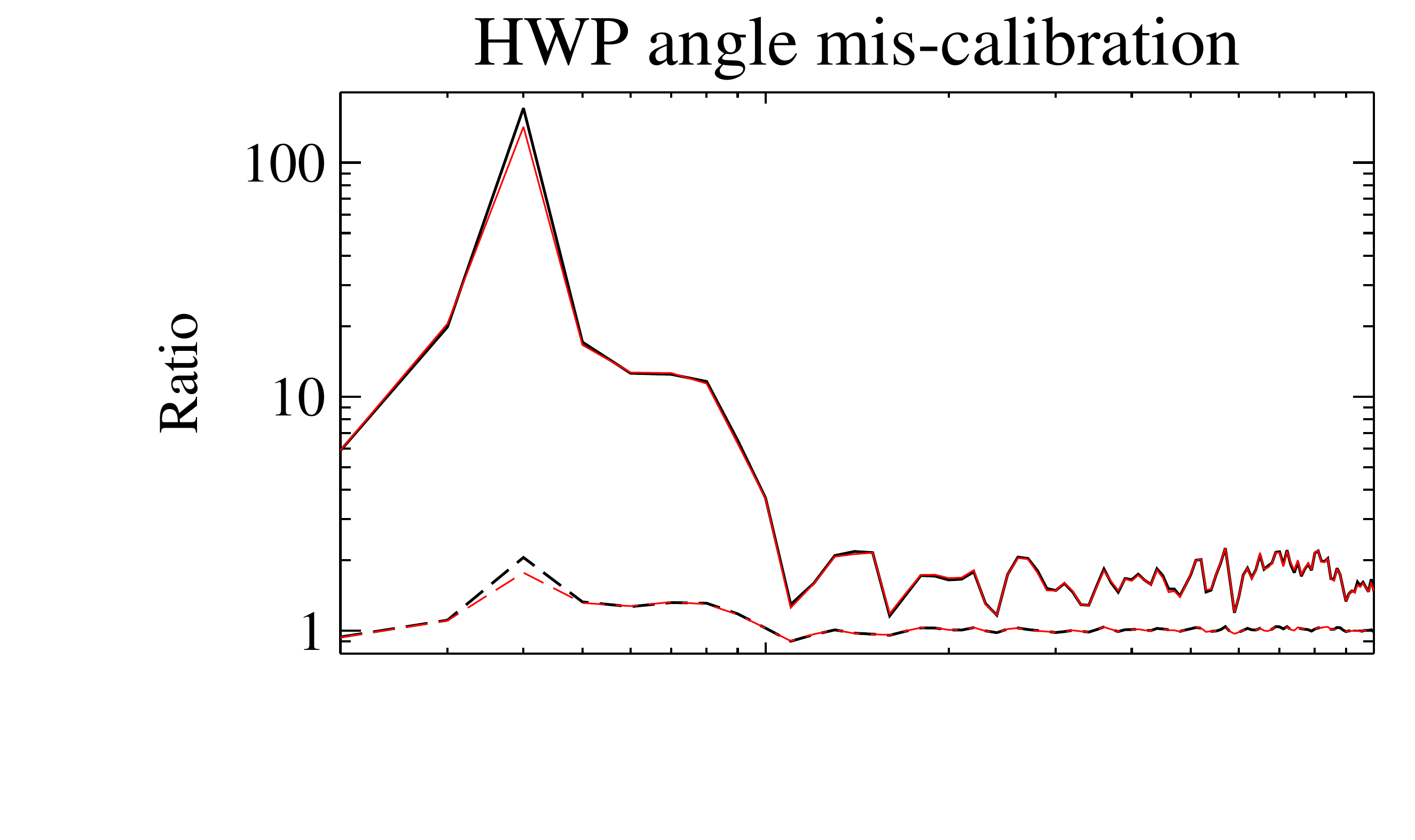}\\ \vskip -2.5em
\hskip -2em \includegraphics[angle=90, width=0.39\textwidth]{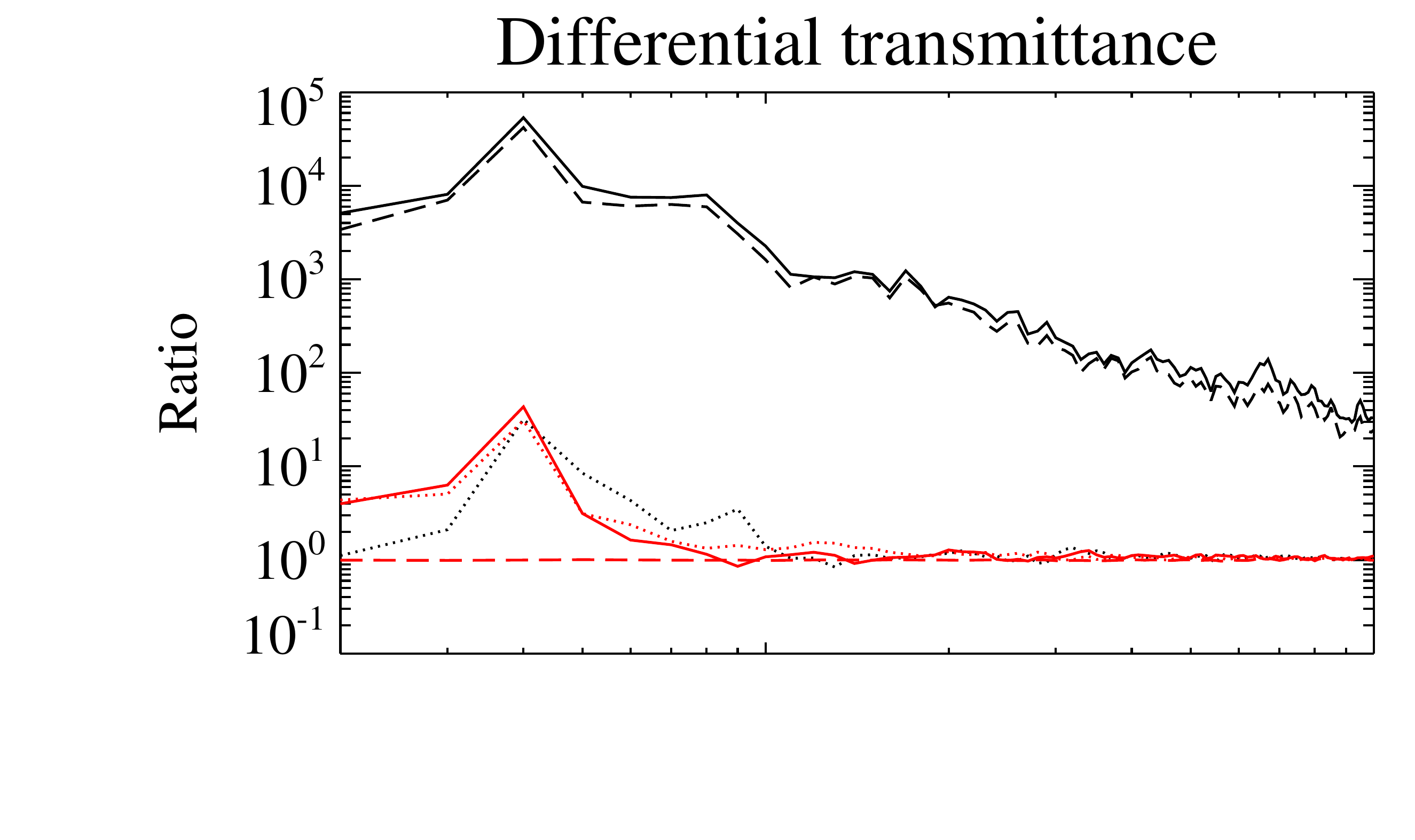}\\ \vskip -2.5em
\hskip -2em \includegraphics[angle=90, width=0.39\textwidth]{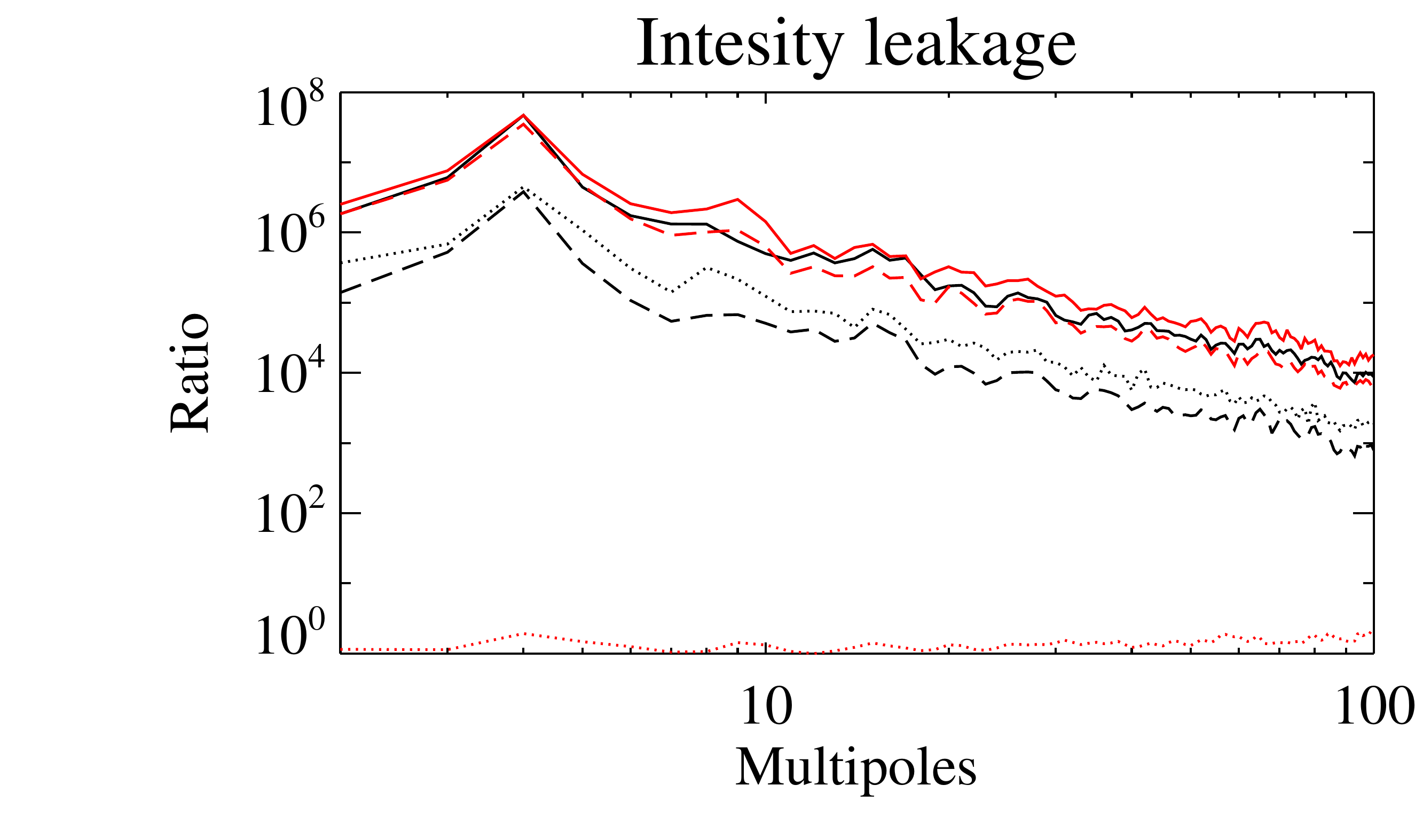}
\end{center}
    \caption{Ratio of signal-only residual BB angular power spectra between the cases with and without HWP-induced systematic effects.  Top panel: HWP angle miscalibration of amplitude 0.1$\degree$ RMS (solid line) and 0.01$\degree$ RMS (dashed line). Middle panel: differential transmittance between the two orthogonal states of amplitude 1\%, producing both a $Q$ and $U$ mis-calibration and an intensity leakage modulated at 2$f_r$ (solid line), only the former (dotted line), only the latter (dashed line). Bottom panel: generic intensity leakage of amplitude 5\% modulated both at 2 and 4$f_r$ (solid line), only at 2$f_r$ (dotted line), only at 4$f_r$ (dashed line). Two HWP configurations are considered: the nominal LSPE-SWIPE stepped setup (in black) and a fast spinning design (rotating at $5~\mathrm{Hz}$ with a slow gondola scanning at $0.7~\mathrm{deg/s}$, in red).}  \label{fig:HWPsys}
\end{figure}
In the simulations above we assumed an ideal HWP, i.e. not introducing systematic effects of its own. In this Section we investigate the impact of typical HWP-induced systematics on the B-mode polarization recovery. In particular, we explore the effects of: i) a mis-estimation of the HWP angles; ii) a possible differential transmittance between the two orthogonal states; iii) a leakage from temperature to polarization, arising, for instance, from an imperfect optical setup.

We limit ourselves to the default LSPE-SWIPE stepped configuration and a fast spinning design (rotating at $5~\mathrm{Hz}$, with a slow gondola scanning rate of $0.7~\mathrm{deg/s}$).

As example of HWP angle mis-estimation, we simulate the effect of both a random error and a systematic offset on the HWP angles. As pessimistic (optimistic) case, we consider both these systematics to have an amplitude of 0.1$\degree$ (0.01$\degree$) RMS. As figure of merit, we use the recovered BB power spectra from signal-only residual maps. In Fig.~\ref{fig:HWPsys} we show the ratio of the BB residuals between the cases with and without the HWP angle mis-calibration, for the two chosen error amplitudes and for the two HWP configurations under exam. 

First, we notice that the impact of the HWP angle error is independent on the HWP configuration: this is in agreement with the results presented in \citet{2009MNRAS.397..634B}. When considering separately the random error and the systematic offset, we find that the impact of the former is negligible with respect the latter. Moreover, as we show, a HWP angle error can potentially cause a large increase in the BB residuals at the largest scales ($\ell \lesssim 10$). An accuracy of better 0.01$\degree$ RMS is therefore needed for the systematic offset to avoid any bias in the large scale B-mode recovery.
In LSPE-SWIPE, around 80 detectors will be devoted to constantly monitor the HWP position, making this requirement achievable.

As second example of HWP systematic effect, we consider a differential transmittance between the two orthogonal states of amplitude $\delta=1\%$. At first order, this implies a miscalibration of $Q$ and $U$ and a leakage from $I$ modulated at $2f_{r}$ (see \citealt{2009MNRAS.397..634B}). In Fig.~\ref{fig:HWPsys} we show the ratio of BB residual spectra between the cases with and without differential transmittance. Confirming the general expectation, we find that the effect of the polarization mis-calibration is sub-dominant and independent on the HWP setup. The intensity leakage is particularly worrisome in the stepped HWP case since it causes an increase of the BB residuals up to 4-5 orders of magnitude (at large scales), while its impact is drastically reduced in a spinning scheme, as the signal is instead modulated at $4f_{r}$.

Finally, we consider a possible leakage from $I$ to $Q$ and $U$ of amplitude $\alpha=5\%$, that can be sourced, for instance, by internal reflections of linearly polarized light between the polarizer and the HWP \citep{2010arXiv1006.3225S}. In this case, the spurious polarization is modulated both at 2 and $4f_{r}$. In Fig.~\ref{fig:HWPsys} we display the ratio of BB residual spectra between the cases with and without leakage, where we also compare the separate effects of the two modulated leakage modes. As expected, we find that only the contribution at $4f_{r}$ impacts on the spinning HWP performance, while both the 2 and $4f_{r}$ contributions affect a stepped HWP. Nonetheless, the overall effect is comparable in the two configurations.

When decreasing the intensity leakage to an amplitude $\alpha=1\%$, we find that the BB residual level is rescaled by only an order of magnitude. Sophisticated techniques aimed at minimizing the intensity-polarization coupling (see, e.g., \citealt{2015MNRAS.453.2058W}) will therefore represent a crucial issue.

\section{Conclusions}\label{Conclusions}

In this work we present preliminary forecasts of the LSPE-SWIPE experiment with the final goal to optimize the HWP polarization modulation strategy. 

We depart from the nominal SWIPE modulation strategy (stepped every $60~\mathrm{s}$ with telescope scan-speed $12~\mathrm{deg/s}$) and perform a detailed investigation of a wide range of possible HWP rotation schemes, either in stepped (every $1~\mathrm{s}$, $60~\mathrm{s}$ and $3600~\mathrm{s}$) or spinning (at rate $5~\mathrm{Hz}$, $2~\mathrm{Hz}$, $0.5~\mathrm{Hz}$) mode, allowing for two different azimuth telescope scan-speeds ($12~\mathrm{deg/s}$ and $0.7~\mathrm{deg/s}$). 
 
We explore the response of the different HWP setups in the frequency, map and angular power spectrum domains. Maps are generated by the optimal ROMA MPI-parallel algorithm. Spectra are estimated using an implementation of the MASTER pseudo-$C_l$ method. Our analysis accounts for a $1/f$ low-frequency noise both self-correlated and cross-correlated among the polarimeters. 

Furthermore, we quantify the effect of high-pass and band-pass filters of the data stream. In particular, we show that a band-pass filter is a very interesting possibility, as it enables to achieve a remarkably higher signal-to-noise ratio with respect to the more common high-pass filters, especially when a slow telescope modulation is performed. We finally analyze the minimal observation case (one detector for one day of operation), critical for the single-detector calibration process, and test the HWP performance against typical HWP-induced systematics.

In terms of pixel angle coverage, map-making residuals (signal-only and signal plus noise) and BB power spectrum standard deviations, we find that a slowly stepped HWP ($\gtrsim 3600~\mathrm{s}$) provides much poorer performance and that the performance of a stepped HWP drastically worsens when slowing down the telescope scan-speed. Therefore, even if easiest to operate, these configurations must be rejected from further investigation. At fast telescope scan-speed, we find no difference between a HWP stepped every $1~\mathrm{s}$ or $60~\mathrm{s}$, making the nominal configuration the best option among the set of stepped schemes.

The performance of a spinning HWP is not very sensitive on both the HWP rotation rate and the gondola scanning frequency. Moreover, the various spinning designs provide comparable results to the default stepped modulation strategy.

However, we find that the nominal SWIPE configuration may not be the most convenient choice when accounting for some specific real-case situations.  

For instance, a rapidly spinning HWP scheme ($5~\mathrm{Hz}$) combined with a slow telescope modulation rate ($0.7~\mathrm{deg/s}$) provides the highest performance in presence of high-pass and band-pass filters since the polarization signal is shifted to very high frequencies into a very narrow band.

In order to preserve 90\% of polarization signal information, any high-pass filter could at most increase the signal-to-noise ratio of less than 20\% and up to 40-50\%, for the nominal modulation case and for a HWP spinning at $5~\mathrm{Hz}$, respectively. Moreover, for any spinning setup combined with a gondola scanning at $0.7~\mathrm{deg/s}$, a band-pass filter could increase the signal-to-noise ratio up to $\sim$18 times, without losing more than 10\% of polarization power.

In addition, this specific spinning design allows a more efficient polarization recovery in the case of one detector-one operation day. The corresponding signal-only BB residuals are lower at any angular scales with respect to the nominal HWP configuration. In the signal plus noise case we find an improvement at the lowest multipoles while the difference is damped at smaller scales.

When including systematics intrinsic to the HWP, we find that any spurious polarization contribution (arising for instance from differential transmittance or imperfect optical setup) modulated at twice the HWP rotation frequency is completely suppressed by a spinning mode, while it can largely affect the performance of a stepped configuration (at large scales). This is expected, as in the spinning case the polarization signal is modulated at four times the spinning rate. Instead, when considering realistic HWP angle errors, $Q$ and $U$ mis-calibration due to differential transmittance, and generic leakage from $I$ to $Q$ and $U$ modulated at four times the rotation frequency, we find no relevant difference between the stepped and spinning mode responses.

The range of HWP-induced systematics analyzed here is not exhaustive. In addition, in this work we did not account for other instrumental systematic effects (e.g. pointing errors and calibration drifts).  
To draw final conclusions, a detailed investigation of these issues is required and left to future work.

Although the simulations presented in this paper are specific to the SWIPE instrument, our results are qualitatively valid for any scanning telescope B-mode mission aiming at large angular scales.

\begin{acknowledgements} 

    The LSPE project is supported by the contract I/022/11/0 of the Italian Space Agency and by INFN CSN2.
    We acknowledge the use of the HEALPix package \citep{2005ApJ...622..759G}, the CAMB software \citep{2002PhRvD..66j3511L} and the FFTW library \citep{FFTW05}.
    We acknowledge the CINECA award under the ISCRA initiative, for the availability of high performance computing resources and support.
    This research used resources of the National Energy Research Scientific Computing Center (NERSC), which is supported by the Office of Science of the U.S. Department of Energy under Contract No. DE-AC02-05CH11231.
    AB thanks Fabio Columbro for useful discussions.
    We finally wish to thank the LSPE collaboration and the anonymous referee for useful suggestions.    
    
\end{acknowledgements}

\bibliographystyle{aa} 
\bibliography{roma} 

\end{document}